\documentclass[twocolumn,twoside]{IEEEtran} 

\usepackage{bm}
\usepackage{graphicx}
\usepackage{epstopdf}
\usepackage{amsmath, amsthm, amsfonts, amssymb, amsbsy}
\usepackage{setspace}
\usepackage{pstricks,arydshln}
\usepackage{multirow}
\usepackage{booktabs}
\usepackage{stfloats}
\usepackage{amscd,textcomp,amssymb,epsfig,graphics,epsf,color,amsmath,balance,cite}
\usepackage{subfigure}
\usepackage{caption}
\usepackage{cite}
\usepackage{xcolor}
\usepackage{makecell}
\usepackage{epstopdf}
\usepackage{dsfont}
\usepackage{color}

\usepackage{cases}

\usepackage[noend]{algpseudocode}
\usepackage{algorithmicx,algorithm}

\newtheorem{thm}{Theorem}

\newtheorem{rem}{Remark}

\newcommand{\RNum}[1]{\uppercase\expandafter{\romannumeral #1\relax}}
\newtheorem{prop}[thm]{Proposition}
\allowdisplaybreaks[4]
\begin{document}

\title{Hybrid Mechanical and Electronic Beam Steering for Maximizing OAM Channel Capacity}

\author{Rui Chen,~\IEEEmembership{Member,~IEEE,} Zhenyang Tian,~\IEEEmembership{Graduate Student Member,~IEEE,} Wen-Xuan Long,\\ \IEEEmembership{Graduate Student Member,~IEEE,} Xiaodong Wang,~\IEEEmembership{Fellow,~IEEE} and Wei Zhang,~\IEEEmembership{Fellow,~IEEE}
\thanks{This work was supported in part by the Natural Science Basic Research Program of Shaanxi under Grant 2021JZ-18 and in part by the Natural Science Foundation of Guangdong Province of China under Grant 2021A1515010812 and in part by the open research fund of National Mobile Communications Research Laboratory, Southeast University under
Grant 2021D04.}
\thanks{R. Chen is with the State Key Laboratory of ISN, Xidian University, Xi'an 710071,
China, and also with the National Mobile Communications Research Laboratory,
Southeast University, Nanjing 210018, China (e-mail: rchen@xidian.edu.cn).}
\thanks{Z. Tian and W.-X. Long are with the State Key Laboratory of Integrated Service Networks (ISN), Xidian University, Shaanxi 710071,
China.}
\thanks{X. Wang is with the Electrical Engineering Department, Columbia University, New York, NY 10027 USA (e-mail: wangx@ee.columbia.edu).}
\thanks{W. Zhang is with the School of Electrical Engineering and Telecommunications, The University of New South Wales, Australia, (e-mail: w.zhang@unsw.edu.au).}
}

\maketitle

\thispagestyle{empty}
\begin{abstract}
Radio frequency-orbital angular momentum (RF-OAM) is a novel approach of multiplexing a set of orthogonal modes on the same frequency channel to achieve high spectrum efficiencies. Since OAM requires precise alignment of the transmit and the receive antennas, the electronic beam steering approach has been proposed for the uniform circular array (UCA)-based OAM communication system to circumvent large performance degradation induced by small antenna misalignment in practical environment.
However, in the case of large-angle misalignment, the OAM channel capacity can not be effectively compensated only by the electronic beam steering. To solve this problem, we propose a hybrid mechanical and electronic beam steering scheme, in which mechanical rotating devices controlled by pulse width modulation (PWM) signals as the execution unit are utilized to eliminate the large misalignment angle, while electronic beam steering is in charge of the remaining small misalignment angle caused by perturbations. Furthermore, due to the interferometry, the receive signal-to-noise ratios (SNRs) are not uniform at the elements of the receive UCA. Therefore, a rotatable UCA structure is proposed for the OAM receiver to maximize the channel capacity, in which the simulated annealing algorithm is adopted to obtain the optimal rotation angle at first, then the servo system performs mechanical rotation, at last the electronic beam steering is adjusted accordingly. Both mathematical analysis and simulation results validate that the proposed hybrid mechanical and electronic beam steering scheme can effectively eliminate the effect of diverse misalignment errors of any practical OAM channel and maximize the OAM channel capacity.
\end{abstract}

\begin{IEEEkeywords}
Orbital angular momentum (OAM), uniform circular array (UCA), hybrid mechanical and electronic, beam steering.
\end{IEEEkeywords}

\section{Introduction}
The rapid development of emerging applications, such as high-definition (HD) video, virtual reality (VR) and auto-pilot driving, results in a never-ending growth in mobile data traffic. It is recognized that the next 6G communication system will be required to increase the capacity of current 5G systems hundred fold \cite{Tataria20216}. To meet the requirement, more and more high frequency bands such as millimeter wave and terahertz bands are being licensed. As radio frequency (RF) spectrum resources are scarce, besides exploiting more frequency bandwidth, innovative techniques to enhance spectrum efficiency (SE) have been explored, such as advanced coding, cognitive radio (CR) and ultra-massive multiple-input multiple-output (MIMO). In essence, all these techniques are based on planar electromagnetic (EM) waves physically.

Since the discovery in 1992 that light beams with helical phase fronts can carry orbital angular momentum (OAM) \cite{Allen1992Orbital}, a significant research effort has also been focused on vortex EM waves as a novel approach to carrying a set of orthogonal OAM modes on the same frequency channel and achieving high-SE mode multiplexing\cite{Tamburini2012Encoding,Yan2014High,Ren2017Line,Zhang2017Mode, Chen2018Beam,Chen2018A,Zhang2019Orbital,Chen2020Orbital,Chen2020Multi-mode,Long2021AoA,Long2021Joint,Zhang2021Orbital,Xiong2020Performance,Trichili2019Communicating,Tian2021Broadband}. Multiple OAM modes could significantly reduce the receivers complexity, compared to a MIMO single mode communication, without affecting the system capacity\cite{Zhang2017Mode,Trichili2019Communicating}. The helical phase wave front of vortex EM is described by a phase term $e^{i\ell\phi}$, where $\phi$ is the transverse azimuthal angle and $\ell$ is an infinite integer, known as topological charge or OAM mode number \cite{Allen1992Orbital}. So far, there have been several methods to generate radio OAM beams such as using spiral phase plate (SPP) \cite{Yan2014High}, solid cylinder dielectric resonator antenna \cite{Liang2016Orbital}, tri-mode concentric circularly polarized patch antenna, metallic traveling-wave ring-slot structure \cite{Zhang2017Four}, time-switched array structure \cite{Tennant2012Generation}, uniform circular array (UCA) \cite{Mohammadi2010system,Chen2018A,Chen2020Multi-mode} and metasurface \cite{Shen2018Generating}, among which UCA is the most popular device because of the ubiquitous application of antenna array on 4G, 5G and even 6G base stations (BSs) and its compatibility with MIMO beam steering\cite{Chen2020Multi-mode}. Therefore, we focus on the UCA rather than other particular antenna structures for OAM waves' generation and reception in this paper.

Since OAM communication requires precise alignment of the transmit and the receive antennas, the electronic beam steering approach has been proposed for the uniform circular array (UCA)-based OAM communication system to circumvent large performance degradation induced by small antenna misalignment in practical environments \cite{Chen2018Beam,Chen2020Multi-mode}. However, in large-angle misalignment case, the OAM channel capacity can not be effectively compensated only by the electronic beam steering, which has not been investigated before as far as we know. Moreover, it is observed that when multiple OAM modes are transmitted simultaneously, due to the interferometry, the electric field intensities of the multi-mode OAM beam are not uniform on the ring of its main lobe. Hence, the receive signal-to-noise ratios (SNRs) are not equal at the elements of the receive UCA, which results in the OAM channel capacities being quite different at different rotation angles of the receive UCA around its axis.
\begin{figure}[t]
\setlength{\abovecaptionskip}{0.1cm}   
\setlength{\belowcaptionskip}{-0.4cm}   
\begin{center}
\includegraphics[width=8.7cm,height=5.1cm]{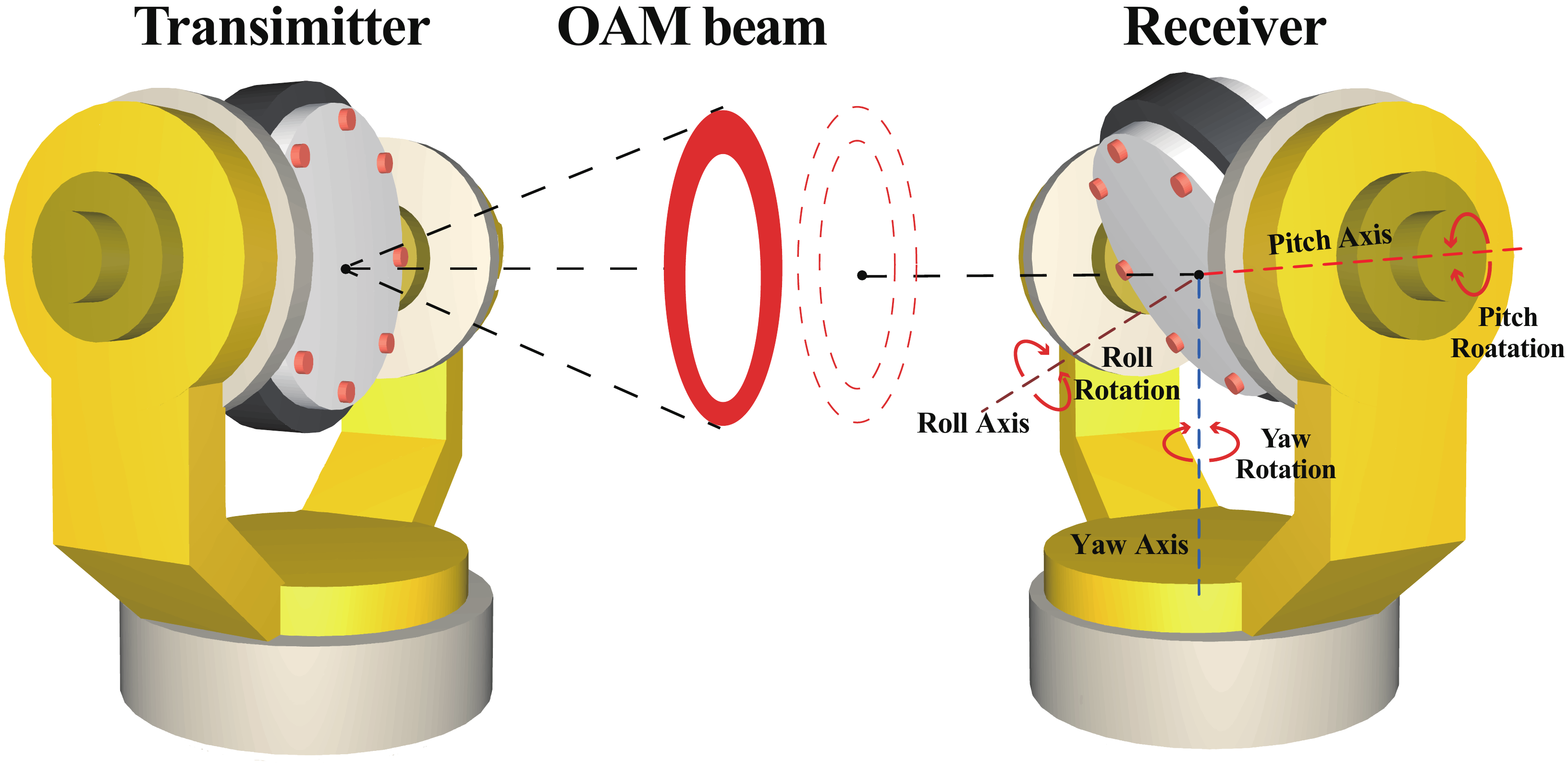}
\end{center}
\caption{The UCA-based LoS MCMM-OAM communication system with hybrid mechanical and electronic beam steerable transceiver.}
\label{fig1}
\end{figure}

In this paper, to deal with the aforementioned problems, we first propose a hybrid mechanical and electronic beam steerable transceiver structure, which consists of a UCA fixed with mechanical rotation devices in pitch, roll and yaw directions, respectively, as shown in Fig. \ref{fig1}. Specifically, the coarse alignment is first achieved by mechanical beam steering, and then electronic beam steering is exploited to compensate the small-scale misalignment error caused by perturbations, such as random wind, rain, or installation errors \cite{Jian2021Non}. The principle of electronic beam steering for UCA-based OAM communication system is compensating the changed phases caused by oblique angle at the transmit or receive UCA through adjusting the delays of antenna elements by phase shifters, then the normal phases used only for OAM piralization/despiralization are tuned to the new phases used for both OAM piralization/despiralization and beam steering towards the direction of departure/arrival of OAM beams \cite{Chen2018Beam}.
Furthermore, due to the interferometry, the receive UCA needs rotating to an optimal angle for maximizing the OAM channel capacity. The mathematical analysis and numerical simulations show that the proposed hybrid mechanical and electronic beam steering scheme outperforms the traditional scheme of only using electronic beam steering, especially in the practical case of large misalignment error, and obtaining the maximum OAM channel capacity. The novelty and major contributions of this paper are summarized as follows:
\begin{itemize}
\item
First, based on the line-of-sight (LoS) OAM channel model, we show mathematically why only using electronic beam steering is not enough for compensating the performance degradation of OAM channel capacity in the large-angle misalignment case.
\item
Second, we propose the hybrid mechanical and electronic beam steering scheme to effectively improve the OAM channel capacity in practical environments, in which mechanical rotating devices controlled by pulse width modulation (PWM) signals as the execution unit are utilized to eliminate the large misalignment angle, while electronic beam steering is in charge of the remaining small misalignment angle caused by perturbations.
\item
Third, considering that the rotation angle of the receive UCA around its axis has a large effect on the receive signal-to-noise ratio (SNR) due to the interferometry, we propose a rotatable UCA structure for the OAM receiver to maximize the channel capacity, for which the simulated annealing algorithm is adopted to obtain the optimal rotation angle at first, then the servo system performs mechanical rotation, at last the electronic beam steering is adjusted accordingly.
\end{itemize}

The remainder of this paper is organized as follows. In Section II, we model the UCA-based LoS OAM communication system with hybrid mechanical and electronic beam steerable transceiver. In Section III, the effect of electronic beam steering in large-angle misalignment case is analyzed. In Section IV, to deal with the performance degradation of electronic beam steering in large-angle misalignment case and maximize the OAM channel capacity, we propose the hybrid mechanical and electronic beam steering scheme. Simulation results are shown in Section V and conclusions are summarized in Section VI.

{\sl Notations}: Bold lowercase letters denote column vectors, and bold uppercase letters denote matrices. $(\cdot)^T$ and $(\cdot)^H$ denote transpose and conjugate transpose. $\mathbf{I}$ is identity matrix. $\odot$ and $|\cdot|$ denote the Hadamard product and modulus. $\mathbb{E}\{\cdot\}$ represents expectation.
\begin{figure}[t]
\begin{center}
\includegraphics[width=8.7cm,height=5cm]{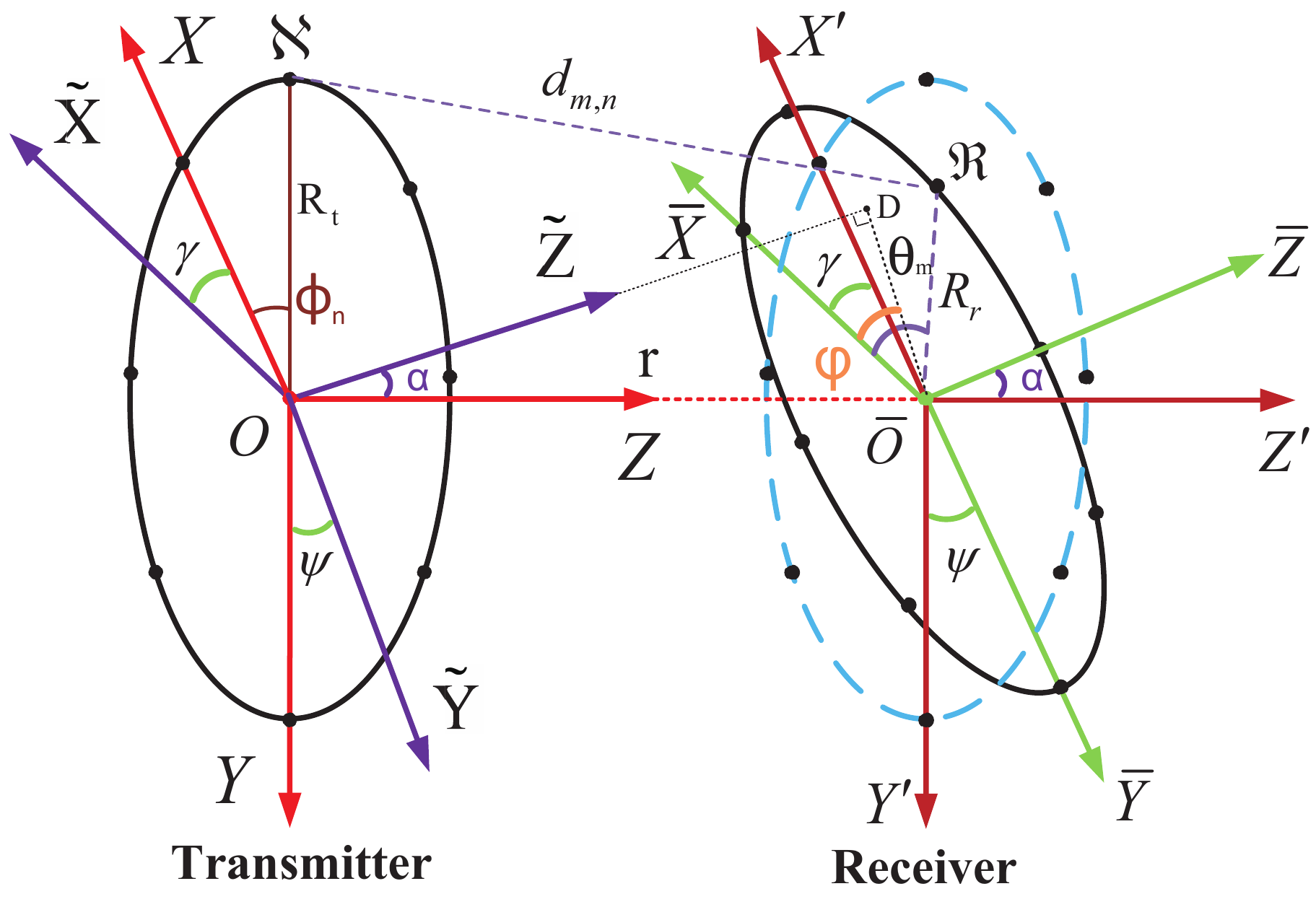}
\end{center}
\caption{Geometrical relationship between the transmit and receive UCAs in the non-parallel misalignment case\cite{Tian2021Broadband}.}
\label{fig2}
\end{figure}
\begin{figure*}[b]
\hrulefill
\setcounter{equation}{0}
\begin{equation}
\left\{
\begin{aligned}
{\varphi  = \frac{\pi }{2} + \arccos \left( {\frac{{2\cos \gamma \sin \psi }}{{\sqrt {3 - 2\cos (2\gamma ){{\cos }^2}\psi  - \cos (2\psi )} }}} \right), 0\le \gamma  < \frac{\pi }{2}},\\
{\varphi  = \frac{\pi }{2} - \arccos \left( {\frac{{2\cos \gamma \sin \psi }}{{\sqrt {3 - 2\cos (2\gamma ){{\cos }^2}\psi  - \cos (2\psi )} }}} \right), - \frac{\pi }{2} < \gamma  < 0}\label{PHI},
\end{aligned}
\right.
\end{equation}
\end{figure*}
\section{UCA-Based OAM Communication Systems}
In this paper, we focus on a UCA-based line-of-sight (LoS) OAM communication system as considered in \cite{Zhang2017Mode,Chen2018A,Chen2018Beam,Chen2020Orbital, Chen2020Multi-mode,Jian2021Non,Long2021AoA,Yagi2021Wireless}. Since the transmit UCA and the receive UCA need to be aligned, beam steering approach is proposed to achieve beam alignment between the UCAs at both sides \cite{Chen2018Beam, Chen2020Multi-mode}. As the functions of the transmit beam steering and the receive beam steering are the same, in order to evaluate the effects of mechanical beam steering and electronic beam steering, we have to separate the effects of transmit and receive beam steering and only consider one side. Therefore, we assume the transmit UCA having been aligned to the center of the receive UCA through mechanical rotation, while the receive UCA will align to the transmit UCA through yaw and pitch mechanical rotations firstly, and then electronic beam steering more accurately in real time, as shown in Fig. \ref{fig1}. Thus, this geometric model falls into the non-parallel misalignment case \cite{Chen2018Beam,Chen2020Multi-mode}, as shown in Fig. \ref{fig2}.

In the considered OAM communication system, a multi-mode OAM beam is generated by an $N$-element UCA at the transmitter and received by another $N$-element UCA at the receiver. For higher data rate transmission, orthogonal frequency-division multiplexing (OFDM)-based LoS multi-carrier and multi-mode OAM (MCMM-OAM) communication scheme proposed in \cite{Chen2020Multi-mode,Long2021AoA} is adopted, and we assume $P$ subcarriers and $U (U\leq N)$ OAM modes for data transmission, $\overline{P}$ subcarriers and $\overline{U}$ OAM modes for coarse angle of arrival (AoA) estimation, and $\widetilde{P}$ subcarriers and $\widetilde{U}$ OAM modes for refined AoA estimation.

\subsection{Channel Model}
The geometrical model of the UCA-based LoS OAM channel is illustrated in Fig. \ref{fig2}. In this model, the transmitter coordinate system $\textrm{Z}-\textrm{X}\textrm{O}\textrm{Y}$ is established using the transmit UCA plane as the $\textrm{X}\textrm{O}\textrm{Y}$ plane and the axis through the transmit UCA center $\textrm{O}$ and perpendicular to the transmit UCA plane as the $\textrm{Z}$-axis, and the receiver coordinate system $\overline{\textrm{Z}}-\overline{\textrm{X}}\overline{\textrm{O}}\overline{\textrm{Y}}$ is established on the plane of the receive UCA by the similar approach, where $\overline {\textrm{X}}$-axis and $\overline {\textrm{Z}}$-axis correspond to the pitch axis and the roll axis in Fig. \ref{fig1}, respectively. Since the transmit UCA is aligned with the center of the receive UCA in the non-parallel misalignment case, the spherical coordinate of the receive UCA center can be denoted as $\overline{\textrm{O}}(r,0,0)$ in $\textrm{Z}-\textrm{X}\textrm{O}\textrm{Y}$ in the non-parallel misalignment case. Suppose the point $\textrm{D}$ is the projection of the point $\textrm{O}$ on the $\overline{\textrm{X}}\overline{\textrm{O}}\overline{\textrm{Y}}$ plane, and the spherical coordinate of the transmit UCA center is denoted as $\textrm{O}(r,\varphi,\alpha)$ in $\overline{\textrm{Z}}-\overline{\textrm{X}}\overline{\textrm{O}}\overline{\textrm{Y}}$, where $r$ is the distance between the transmit and the receive UCA centers, $\varphi$ is the azimuth angle, $\alpha$ is the elevation angle, and $\alpha$ and $\varphi$ are defined as the AoA of the OAM beam\cite{Chen2020Multi-mode,Long2021AoA}.

The coordinate system $\widetilde{\textrm{Z}}-\widetilde{\textrm{X}}{\textrm{O}}\widetilde{\textrm{Y}}$ is built so that its origin is at the point O and $\widetilde{\textrm{X}}{\textrm{O}}\widetilde{\textrm{Y}}$ plane is parallel to $\overline{\textrm{X}}\overline{\textrm{O}}\overline{\textrm{Y}}$. Meanwhile, we build the coordinate system $\textrm{Z}'-\textrm{X}'\overline{\textrm{O}}\textrm{Y}'$ so that its origin is at the point $\overline{\textrm{O}}$ and $\textrm{X}'\overline{\textrm{O}}\textrm{Y}'$ plane is parallel to ${\textrm{X}}{\textrm{O}}{\textrm{Y}}$, where ${\textrm{Y}}'$-axis corresponds to the yaw axis in Fig. \ref{fig1}. According to the geometrical model, the angle between $\widetilde{\textrm{Z}}$-axis and $\textrm{Z}$-axis is $\alpha$, so is that between $\textrm{Z}'$-axis and $\overline{\textrm{Z}}$-axis. Denote the angle between the $\textrm{X}'$-axis and the $\overline{\textrm{X}}$-axis as $\gamma$ ($\gamma\in(-\pi/2,\pi/2)$), so is that between $\textrm{X}$-axis and $\widetilde{\textrm{X}}$-axis, where $\gamma$ $\emph{is the yaw angle}$. Denote the angle between the $\textrm{Y}'$-axis and the $\overline{\textrm{Y}}$-axis as $\psi$ ($\psi\in(-\pi/2,\pi/2)$), so is that between $\textrm{Y}$-axis and $\widetilde{\textrm{Y}}$-axis, where $\psi$ $\emph{is the pitch angle}$, and $\psi$ could be obtained as ${\psi = \arccos \bigg(\frac{{\cos \alpha }}{{\cos \gamma }}\bigg)}$,
$\gamma$ and $\alpha$ can be obtained by the OAM AoA estimation method\cite{Chen2020Multi-mode,Long2021AoA}. The proof of $\psi$ is given in Appendix A. Then, $\varphi$ could be calculated as \eqref{PHI},
whose proof is given in Appendix B. Denote the angle between the line $\textrm{O}\aleph_n$ and $\textrm{X}$-axis as $\phi_n$ and the angle between the line $\overline{\textrm{O}}\Re_m$ and $\overline{\textrm{X}}$-axis as $\theta_m$, where $\aleph_n$ is the position of the $n$th ($1\le n \le N$) element at the transmitter, $\Re_m$ is the position of the $m$th ($1\le m \le N$) element at the receiver, $\phi_n = [2\pi (n - 1)/N + \phi_0]$ and $\theta_m  = [2\pi (m - 1)/N + {\theta _0}]$, $\phi_0$ and $\theta_0$ are respectively the corresponding initial angles of the first reference antenna elements in both UCAs.

According to Fig. \ref{fig2}, the cartesian coordinate of the $n$th antenna element on the transmit UCA is $(R_t \cos\phi_n, R_t \sin\phi_n, 0)$ in $\textrm{Z}-\textrm{X}\textrm{O}\textrm{Y}$ coordinate system, and the cartesian coordinate of the $m$th antenna element on the receive UCA can be written as $(R_r\cos\theta_m, R_r\sin\theta_m, 0)$ in $\overline{\textrm{Z}}-\overline{\textrm{X}}\overline{\textrm{O}}\overline{\textrm{Y}}$ coordinate system, where $R_t$ and $R_r$ are respectively the radii of the transmit and receive UCAs. In order to calculate the distance between the $n$th antenna element on the transmit UCA and the $m$th antenna element on the receive UCA, we transform the coordinate of the $m$th antenna element on the receive UCA in $\overline{\textrm{Z}}-\overline{\textrm{X}}\overline{\textrm{O}}\overline{\textrm{Y}}$ coordinate system to the coordinate in $\textrm{Z}'-\textrm{X}'\overline{\textrm{O}}\textrm{Y}'$ coordinate system. Hence, the cartesian coordinate of the $m$th antenna element on the receive UCA in $\textrm{Z}'-\textrm{X}'\overline{\textrm{O}}\textrm{Y}'$ coordinate system, denoted by $(a_m,b_m,c_m)$, can be obtained as $[a_m, b_m, c_m]^T={\mathbf{R}}_{Y}(\gamma){\mathbf{R}}_{P}(\psi)[R_r\cos\theta_m, R_r\sin\theta_m, 0]^T$ \cite{Rahmat1979Useful},
where ${\mathbf{R}}_{P}(\psi)$ and ${\mathbf{R}}_{Y}(\gamma)$ representing the axis rotation matrixes corresponding to the pitch direction and yaw direction, respectively, take the forms
\begin{equation} \label{Rx}
{\mathbf{R}}_{P}(\psi)=
\begin{bmatrix}1 & 0 & 0 \\ 0 & \cos\psi & -\sin\psi \\ 0 & \sin\psi & \cos\psi \end{bmatrix},\nonumber
\end{equation}
\begin{equation} \label{Ry}
{\mathbf{R}}_{Y}(\gamma)=
\begin{bmatrix} \cos\gamma & 0 & \sin\gamma\\ 0 & 1 & 0 \\ -\sin\gamma & 0 & \cos\gamma  \end{bmatrix}.
\end{equation}
Thus, the cartesian coordinate of the $m$th antenna element on the receive UCA in $\textrm{Z}'-\textrm{X}'\overline{\textrm{O}}\textrm{Y}'$ coordinate system can be written as\cite{Tian2021Broadband}
\begin{equation}\label{aaa}
\left\{
\begin{aligned}
a_m &= R_r\cos\theta_m\cos\gamma + R_r\sin\theta_m\sin\psi\sin\gamma,\\
b_m &= R_r\sin\theta_m\cos\psi,\\
c_m &= R_r\sin\theta_m\sin\psi\cos\gamma - R_r\cos\theta_m\sin\gamma.
\end{aligned}
\right.
\end{equation}
As the coordinate system ${\textrm{Z}'}-{\textrm{X}'}\overline{\textrm{O}}{\textrm{Y}'}$ is parallel to ${\textrm{Z}}-{\textrm{X}} {\textrm{O}}{\textrm{Y}}$, the cartesian coordinate of the $m$th antenna element of the receive UCA in ${\textrm{Z}}-{\textrm{X}}{\textrm{O}}{\textrm{Y}}$ coordinate system, denoted by $({\tilde a_m},{\tilde b_m},{\tilde c_m})$, can be obtained as $[{\tilde a_m}, {\tilde b_m}, {\tilde c_m}]^T = [{a_m}, {b_m}, {c_m}]^T + [0, 0, r]^T$.
With the above coordinates, we can obtain the transmission distance ${d_{m,n}}$ from the $n$th transmit antenna element to the $m$th receive antenna element. By assuming that the transmit and receive UCAs are in far-field distance region of each other, thus $r\gg R_t$ and $r\gg R_r$, we can approximate $d_{m,n}$ as
\begin{align} \label{dmn_appx}
d_{m,n}\overset{(a)}{\approx}& \sqrt{R_r^2+R_t^2+r^2}-\frac{R_tR_r \sin\theta_m \cos\phi_n \sin\psi \sin\gamma}{\sqrt{R_r^2+R_t^2+r^2}}\nonumber\\
& -\frac{R_rR_t\left(\cos\theta_m \cos\phi_n\cos\gamma + \sin\theta_m \sin\phi_n\cos\psi\right)}{\sqrt{R_r^2+R_t^2+r^2}}\nonumber\\
& +\frac{rR_r\left(\sin\theta_m \sin\psi \cos\gamma-\cos\theta_m \sin\gamma\right)}{\sqrt{R_r^2+R_t^2+r^2}}\nonumber\\
\overset{(b)}{\approx}& r - \frac{R_rR_t}{r} \sin\theta_m \cos\phi_n\sin\psi \sin\gamma\nonumber\\
&-\frac{R_rR_t}{r}\big(\cos\theta_m \cos\phi_n\cos\gamma + \sin\theta_m \sin\phi_n\cos\psi \big) \nonumber\\
&+R_r \left(\sin\theta_m \sin\psi \cos\gamma-\cos\theta_m \sin\gamma \right),
\end{align}
where (a) uses the method of completing a square and the condition $r\gg R_t,R_r$ as same as the simple case $\sqrt{a^2-2b}\approx a-\frac{b}{a},a\gg b$, (b) is directly obtained from the condition of $r\gg R_t,R_r$. In LoS communications, propagation through the RF channel leads to attenuation and phase rotation of the transmitted signal. This effect is modelled through multiplying by a complex constant $h_{m,n}(p)$, whose value depends on the frequency and the distance $d_{m,n}$ between the transmit and receive antennas, as
\begin{align}\label{GS9}
&{h_{m,n}}(p)=\frac{\beta}{2k_pd_{m,n}}\textrm{exp}\left(-ik_pd_{m,n}\right)\nonumber\\
&\overset{(a)}{\approx}\frac{\beta}{{2{k_p}r}}\exp\big(iS_{{k_p}}\sin\theta_m \cos\phi_n \sin\psi\sin\gamma  \nonumber\\
&\quad+iS_{{k_p}}\cos\theta_m \cos\phi_n \cos\gamma
+iS_{{k_p}}\sin\theta_m \sin\phi_n\cos\psi \nonumber\\
&\quad-ik_pr-ik_pR_r (\sin\theta_m \sin\psi \cos\gamma-\cos\theta_m \sin\gamma)\big),
\end{align}
where $S_{{k_p}}=k_p\frac{R_rR_t}{r}$, (a) neglects a few minor terms in the denominator of the amplitude term and thus only $2k_pr$ is left, $k_p=2\pi/\lambda_p$ is the wave number at the $p$th subcarrier, and $\lambda_p$ is the wavelength at the $p$th subcarrier, and $1/(2k_pd_{m,n})$ denotes the degradation of amplitude, $\beta=\beta_r\beta_t$, $\beta_r$ models all the constants relative to each receive antenna element, and the complex exponential term is the phase difference due to the propagation distance. In order to mainly analyze the factors we concerned, the transmission loss caused by the polarization mismatch between the transmit and receive UCAs is ignored.

Then, the channel matrix of the UCA-based LoS communication system can be expressed as $\mathbf{H}(p)=[h_{m,n}(p)]_{N\times N}$.
Note that when $\psi=0$ and $\gamma=0$, $\mathbf{H}(p)$ is a circulant matrix that can be decomposed by the $N$-dimentional Fourier matrix $\bm{F}_N$ as $\mathbf{H}(p)=\bm{F}_N^H\mathbf{\Lambda}(p)\bm{F}_N$, where $\mathbf{\Lambda}(p)$ is a diagonal matrix with the eigenvalues of $\mathbf{H}(p)$ on its diagonal. In \cite{Zhang2017Mode,Chen2018Beam,Chen2018A,Chen2020Multi-mode}, the OAM channel is defined as $\mathbf{H}_{\textmd{OAM}}(p)=\bm{F}_N\mathbf{H}(p)\bm{F}_N^H$, which will become $\mathbf{\Lambda}(p)$ in perfect alignment case.

\subsection{Signal Model}
As a UCA can generate multi-mode OAM beams with the baseband (partial) discrete Fourier transform (DFT) matrix $\mathbf{F}_U$, the equivalent baseband signal models of transmitting multi-mode OAM data symbols could be expressed as $\mathbf{F}_U^H\mathbf{s}(p)$, where $\mathbf{F}^H_U$ is used for spiralization, $\mathbf{s}(p)=[s(p,1),s(p,2),\cdots,$ $s(p,U)]^T$ contains the modulation symbols transmitted on $U$ OAM modes at the $p$th $(1\leq p\leq P)$ subcarrier simultaneously, $\ell_u$ is the $u$th OAM mode number. Thus, the received OAM signal $\mathbf{y}(p)$ can be written as
\begin{align} \label{y}
\mathbf{y}(p)=&\mathbf{F}_U\left(\mathbf{H}(p)\mathbf{F}_U^H\mathbf{s}(p) +\mathbf{z}(p)\right)\nonumber\\
=&\mathbf{{H}}_{\rm OAM}(p)\mathbf{s}(p)+\mathbf{\bar{z}}(p),
\end{align}
where ${\mathbf{F}_U}=[{\mathbf{f}^H}({\ell_1}), {\mathbf{f}^H}({\ell_2}), \cdots ,{\mathbf{f}^H}({\ell_U})]^H $ is a $U \times N$ right circularly shifted (partial) Fourier matrix used for despiralization, $\mathbf{f}({\ell_u})=\frac{1}{\sqrt N}\left[1,{e^{-i\frac{{2\pi {\ell_u}}}{N}}},\cdots,{e^{-i\frac{{2\pi{\ell_u}(N-1)}}{N}}}\right]$, $\mathbf{z}(p)=[z(p,1),z(p,2),$ $\cdots,z(p,U)]^T$ is the complex Gaussian noise vector with zero mean and covariance matrix $\sigma _z^2{{\mathbf{{I}}}_N}$, $\bar{\mathbf z}(p)={\mathbf{F}_U}\mathbf{z}(p)$, ${\mathbf{y}}(p)=[{y}(p,1),{y}(p,2),\cdots ,{y}(p,U)]^T$, ${y}(p,u)$ is the received information signal on the $u$th mode OAM and the $p$th subcarrier of receiver, and can be written as
\begin{align}
y(p,u)=& \sum\limits_{v = 1}^U {h_{\textrm{OAM},p}(u,v){s(p,v)} + {\bar{z}(p,u)}} \nonumber\\
=&h_{\textrm{OAM},p}(u,u){s(p,u)}+ {\bar{z}(p,u)} \nonumber\\
&+ \sum_{u\ne v}h_{\textrm{OAM},p}(u,v){s(p,v)},\label{c12}
\end{align}
$h_{\textrm{OAM},p}(u,v)=\mathbf{f}(\ell_u)\mathbf{H}(p){\mathbf{f}^{H}}({\ell_v})$ is the $u$th-row and $v$th-column element of $\mathbf{H}_{\rm OAM}(p)$, and can be expressed as
\begin{align}
h_{\textrm{OAM},p}&(u,v) =\eta(p)\sum_{m = 1}^N\sum_{n = 1}^N
\exp\big(-i{\ell_u}\theta_m +i{\ell_v}\phi_n \nonumber\\
&+iS_{{k_p}}\sin\theta_m \cos\phi_n \sin\psi\sin\gamma\nonumber\\
&+iS_{{k_p}}(\cos\theta_m \cos\phi_n \cos\gamma+\sin\theta_m \sin\phi_n\cos\psi)\nonumber\\
&-ik_pR_r \left(\sin\theta_m \sin\psi \cos\gamma-\cos\theta_m \sin\gamma \right)\big),\label{e}
\end{align}
$\eta(p)=\frac{\beta}{{2{k_p}rN}}\exp(-i{k_p}r)$, $\bar{z}(p,u)={\rm\mathbf{{f}}}({\ell_u})\mathbf{z}(p)$, $u,v = 1,2, \cdots ,U$ and $p = 1,2,\cdots,P$. In \eqref{c12}, $h_{\textrm{OAM},p}(u,u)s(p,u)$ carries useful signals. Once $\gamma\ne0$ or $\psi\ne0$, $\sum_{u \ne v} h_{\textrm{OAM},p}(u,v)s(p,v)\neq 0$, which becomes interferences.

For this model, in order to alleviate the interference that is resulted by the large misalignment error in the practical case, we propose the hybrid mechanical and electronic beam steering scheme, as shown in Fig. \ref{fig1}. It can be observed from (\ref{GS9}) that mechanical rotations of the UCA plane will result in the changes of $\{d_{m,n},m=1,\cdots,M; n=1,\cdots,N\}$, and the amplitudes and phases of $\{h_{m,n},m=1,\cdots,M; n=1,\cdots,N\}$. It is worth noting that the amplitude and phase of $h_{m,n}$ do not linearly change with the rotation angles, which are dependent on $\psi$, $\gamma$ and $\theta_m$. Therefore, the mechanical beam steering is a non-linear transformation for the channel matrix ${\mathbf{H}}(p)$. In order to represent the operation of the mechanical beam steering, we assume a non-linear function $\mathcal{F}(\cdot)$ and then the effect of mechanical beam steering on the channel matrix can be expressed as
\begin{align}
\mathbf{H}_{\mathcal{F}}(p)=\mathcal{F}\left({\mathbf{H}}(p), \hat\psi,\hat\gamma,\theta^\ast\right),
\end{align}
where $\hat\gamma$ is the estimate of $\gamma$ by the OAM AoA estimation technique, $\hat\psi$ is the estimate of $\psi$ obtained with $\hat\gamma$ and the estimate of $\alpha$, i.e., $\hat\alpha$, and $\theta^\ast$ is the optimal value of the rotation angle $\theta^\ast$ in roll direction for maximizing the OAM channel capacity, so $\hat\gamma$, $\hat\psi$ and $\theta^\ast$ are mechanical rotation angles in the pitch, yaw and roll directions, respectively, and $\mathbf{H}_{\mathcal{F}}(p)$ is the channel matrix after mechanical rotations of the receive UCA. In contrast to the mechanical beam steering, the electronic beam steering is linear transformation \cite{Chen2018Beam,Chen2020Multi-mode}. With the hybrid non-linear mechanical beam steering and linear electronic beam steering, the recovered information signal vector $\mathbf{y}_{H}(p)$ becomes
\begin{align}
\mathbf{y}_{H}(p)=\left(\mathbf{F}_U\odot\mathbf{B}_{E}(p)\right) \left(\mathbf{H}_{\mathcal{F}}(p)\mathbf{F}_U^H\mathbf{s}(p)+\mathbf{z}(p)\right), \label{yH14}
\end{align}
where ${\mathbf{B}_{E}(p)}$ is the electronic beam steering matrix at the receiver. The procedure of the mechanical and electronic beam steering for the LoS OAM communication receiver will be detailed in Section IV.

\section{The Effect of Electronic Beam Steering and Solution in Large-Angle Misalignment Case}
\begin{figure}
\setlength{\abovecaptionskip}{-0.2cm}   
\setlength{\belowcaptionskip}{-0.2cm}   
\centering
\subfigure[]{
\includegraphics[scale=0.15]{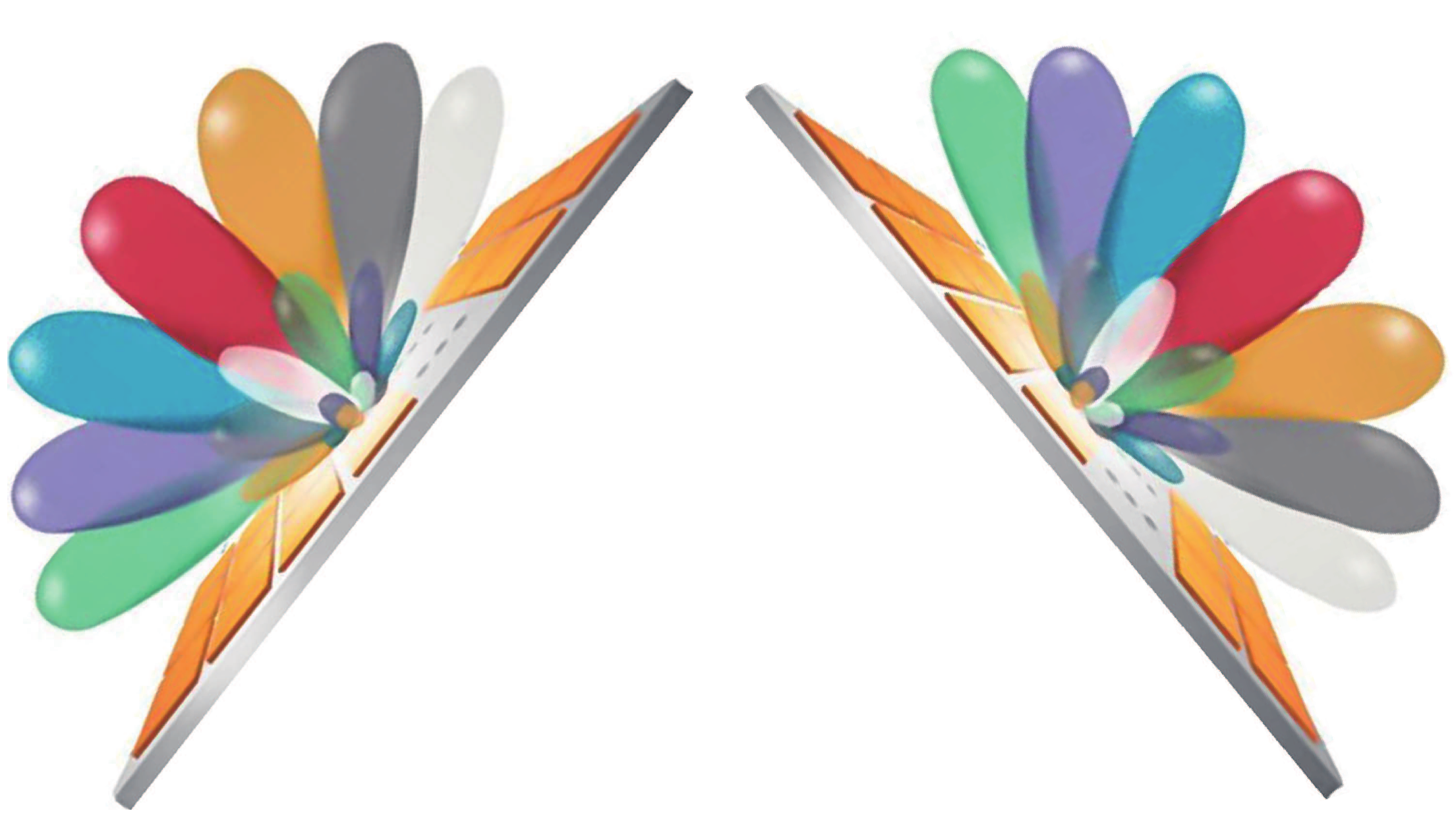}
}
\quad
\subfigure[]{
\includegraphics[scale=0.15]{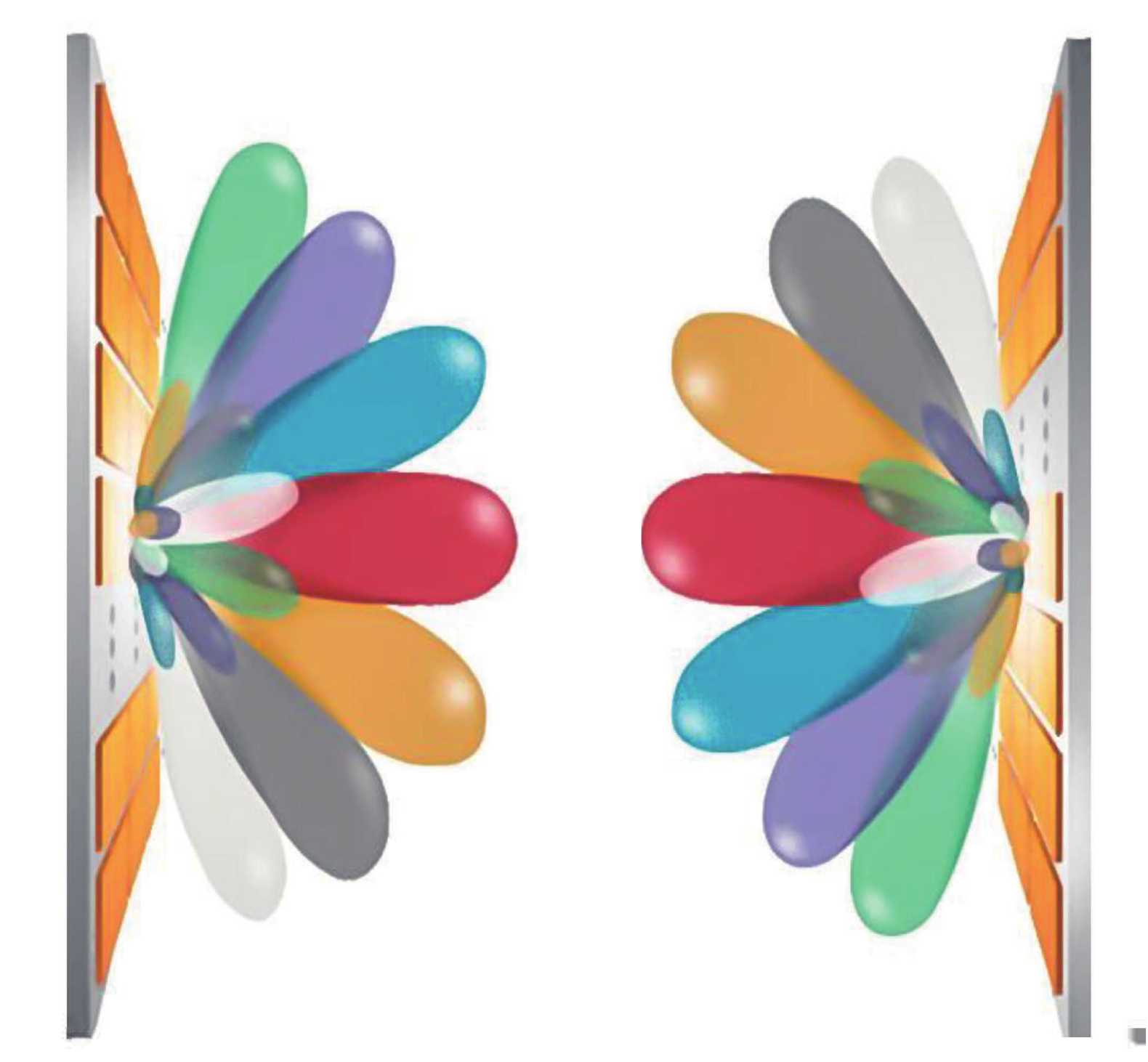}
}
\caption{Two examples of transmit and receive array directions in (a) large angle misalignment case, (b) small angle misalignment case.}
\label{MIMO12}
\end{figure}

The motivation of the proposed hybrid beam steering is from practical implementation of mmWave relays. Let us consider the case in the Fig. \ref{MIMO12}(a), no matter it is OAM array or mmWave massive MIMO array, the performance of the communication is poor. However, with the proposed the hybrid beam steering scheme, which first enables the coarse alignment between the transmit array plane and the receive array plane through mechanical rotation as in Fig. \ref{MIMO12}(b), the good performance can be expected through refined electronic beam steering.
\subsection{The Effect of Electronic Beam Steering}
In this section, we first investigate the effect of electronic beam steering, especially in large-angle misalignment case. With only electronic beam steering, the recovered information signal vector $\mathbf{y}_{E}(p)$ takes the form \cite{Chen2020Multi-mode}
\begin{align}
\mathbf{y}_{E}(p)=&\left({\mathbf{F}_U}\odot{\mathbf{B}_{EO}(p)}\right)\left({\mathbf{H}}(p) \mathbf{F}_U^H\mathbf{s}(p) + \mathbf{z}(p)\right)\nonumber\\
=&\mathbf{\widetilde{H}}_{\rm OAM}(p)\mathbf{s}(p)+\mathbf{\tilde{z}}(p),
\end{align}
where $\mathbf{\widetilde{H}}_{\rm OAM}(p)=\left(\mathbf{F}_U\odot{\mathbf{B}_{EO}(p)}\right)\mathbf{H}(p) \mathbf{F}_U^H$ is defined as the effective OAM channel matrix after electronic beam steering, the electronic beam steering matrix ${\mathbf{B}_{EO}(p)}$ of the receive UCA is designed as ${\mathbf{B}_{EO}(p)} = \mathbf{1}\otimes {\mathbf{\mathfrak{b}}}_{EO}(p)$, ${\mathbf{\mathfrak{b}}}_{EO}(p) = [{e^{i{W_1}(p)}},{e^{i{W_2}(p)}},\cdots,{e^{i{W_N}(p)}}]$ and
${W_m}(p)$ $=k_pR_r\left(\sin\theta_m\sin\psi\cos\gamma-\cos\theta_m\sin\gamma\right)$,
$m=1,\cdots,N$, $p=1,\cdots,P$, $\mathbf{\tilde{z}}(p)=\left(\mathbf{F}_U\odot{\mathbf{B}_{EO}(p)}\right) \mathbf{z}(p)=[\tilde{z}(p,1),\tilde{z}(p,2),\cdots,\tilde{z}(p,U)]^T$, $\mathbf{y}_{E}(p)=[{y}_{E}(p,1), $ ${y}_{E}(p,2),\cdots ,{y}_{E}(p,U)]^T$, ${y}_{E}(p,u)$ is the information symbol on the $u$th mode OAM and the $p$th subcarrier after electronic beam steering, and can be written as
\begin{align}
{y}_{E}(p,u)=&\sum\limits_{v = 1}^U {\tilde{h}_{\textrm{OAM},p}(u,v){s(p,v)} + {\tilde{z}(p,u)}}\nonumber\\
=&\tilde{h}_{\textrm{OAM},p}(u,u){s(p,u)}+ {\tilde{z}(p,u)} \nonumber\\
&+\sum\limits_{u \ne v} \tilde{h}_{\textrm{OAM},p}(u,v){s(p,v)}\label{GS15},
\end{align}
$\tilde{h}_{\textrm{OAM},p}(u,v)=\big({\mathbf{\mathfrak{b}}}_{EO}(p)\odot\mathbf{f}(\ell_u)\big)\mathbf{H}(p){\mathbf{f}^{H}}({\ell_v})$  is the $u$th-row and $v$th-column element of $\mathbf{\widetilde{H}}_{\rm OAM}(p)$, and can be expressed as
\begin{align}
&\tilde{h}_{\textrm{OAM},p}(u,v)
=\eta (p)\sum\limits_{m = 1}^N\sum\limits_{n = 1}^N\exp\bigg(-i{\ell_u}\theta_m +i{\ell_v}\phi_n\nonumber\\
&+iS_{{k_p}}\sin\theta_m \cos\phi_n \sin\psi \sin\gamma\nonumber\\
&+ iS_{{k_p}}\left(\cos\theta_m \cos\phi_n \cos\gamma + \sin\theta_m \sin\phi_n\cos\psi \right)\bigg)\label{g11}.
\end{align}
\begin{figure}[t]
\setlength{\abovecaptionskip}{-0.2cm}   
\setlength{\belowcaptionskip}{-0.2cm}   
\centering
\subfigure[]{
\includegraphics[width=3.9cm,height=3cm]{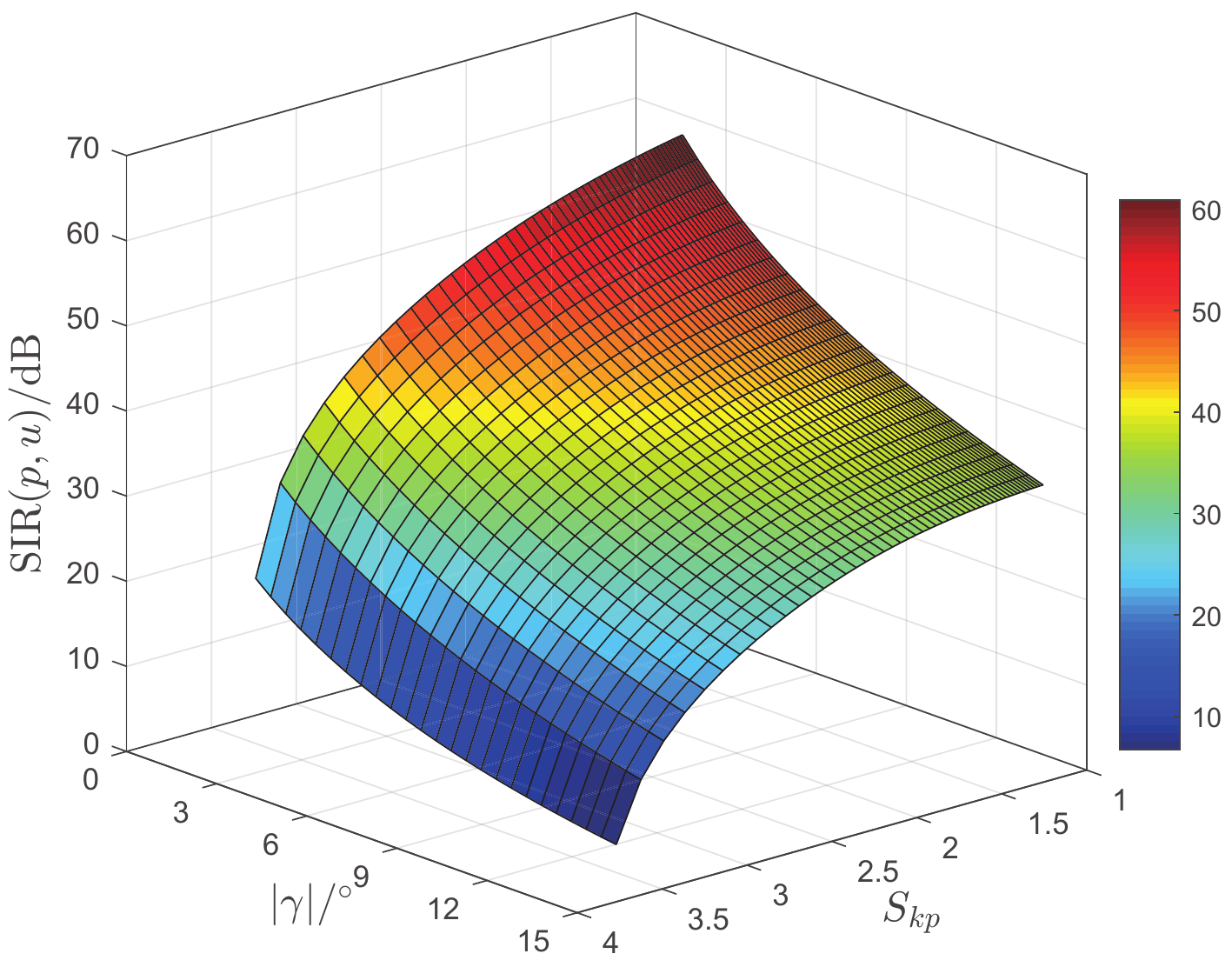}
}
\quad
\subfigure[]{
\includegraphics[width=3.9cm,height=3cm]{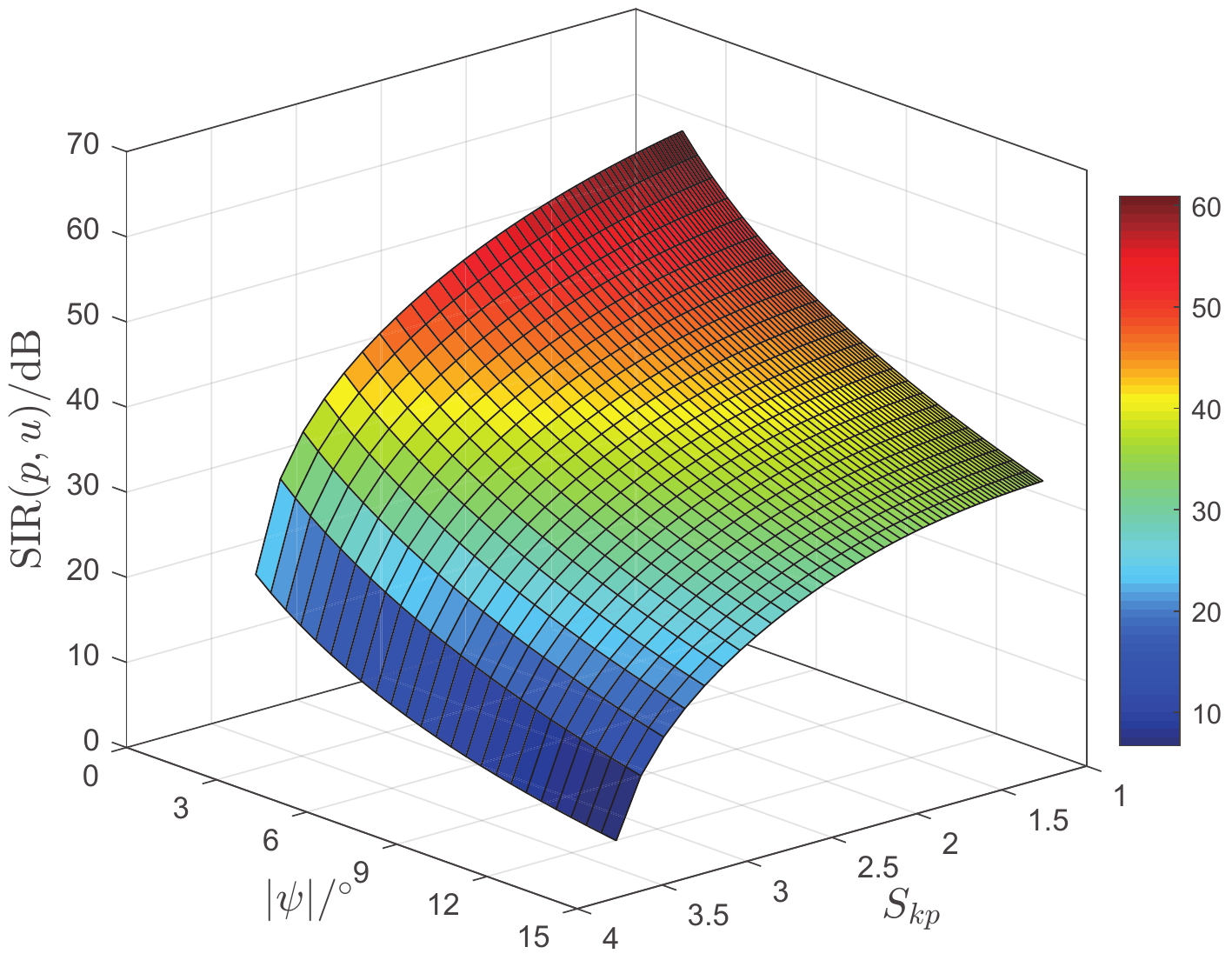}
}
\quad
\subfigure[]{
\includegraphics[width=8cm,height=6cm]{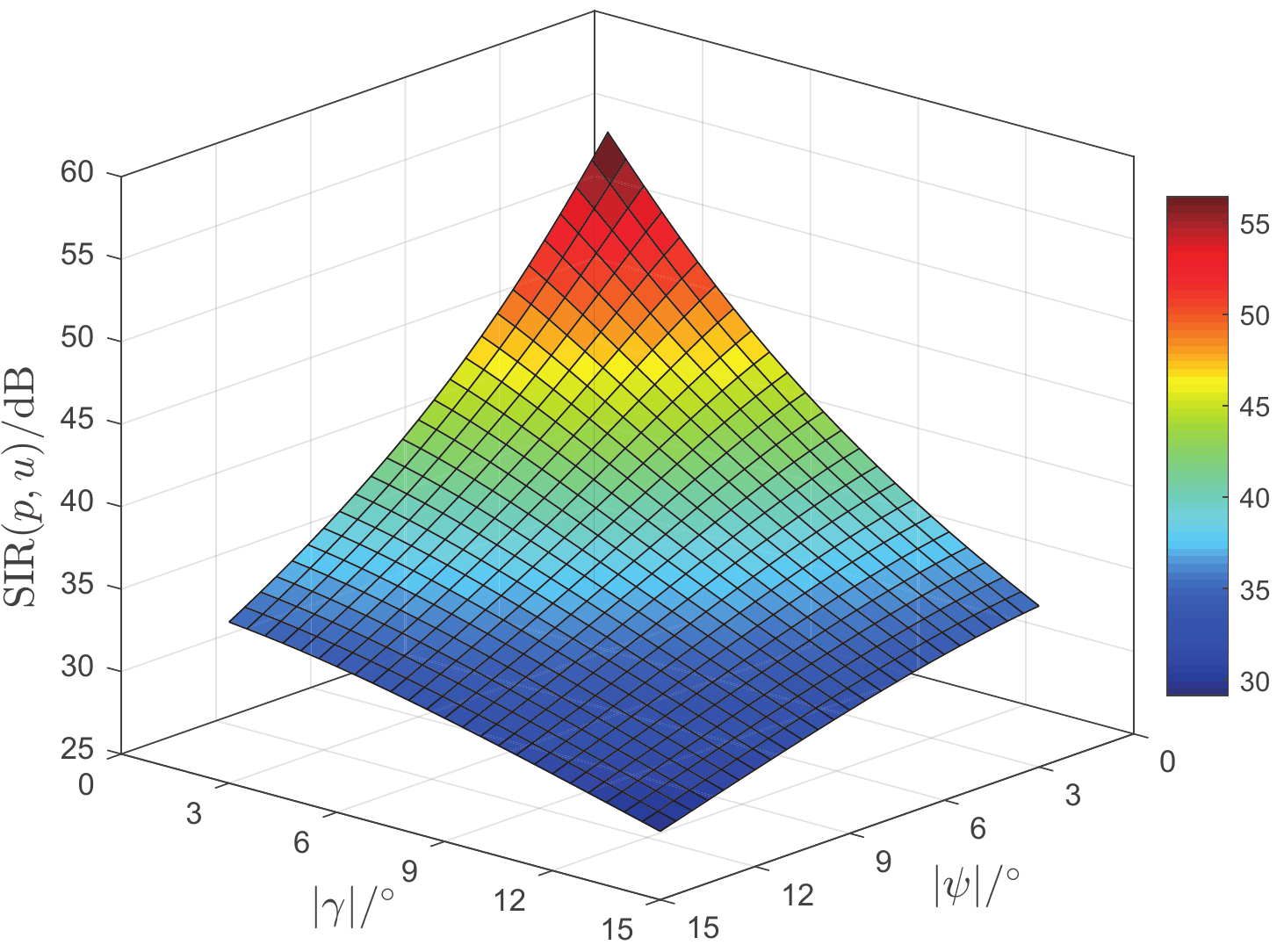}
}
\caption{The effects of (a) yaw angle $\gamma$, (b) pitch angle $\psi$, (c) yaw angle $\gamma$ and pitch angle $\psi$ with $S_{{k_p}}=1$ on SIR. $U = 9$, ${\ell_v} \in [- 4,4]$, ${\ell_u}=1$, $N=10$, ${R_r} = {R_t} = 20\lambda_1$.}
\label{FDJX}
\end{figure}
\begin{figure}
\setlength{\abovecaptionskip}{-0.2cm}   
\setlength{\belowcaptionskip}{-0.2cm}   
\centering
\subfigure[]{
\includegraphics[width=8cm,height=6.5cm]{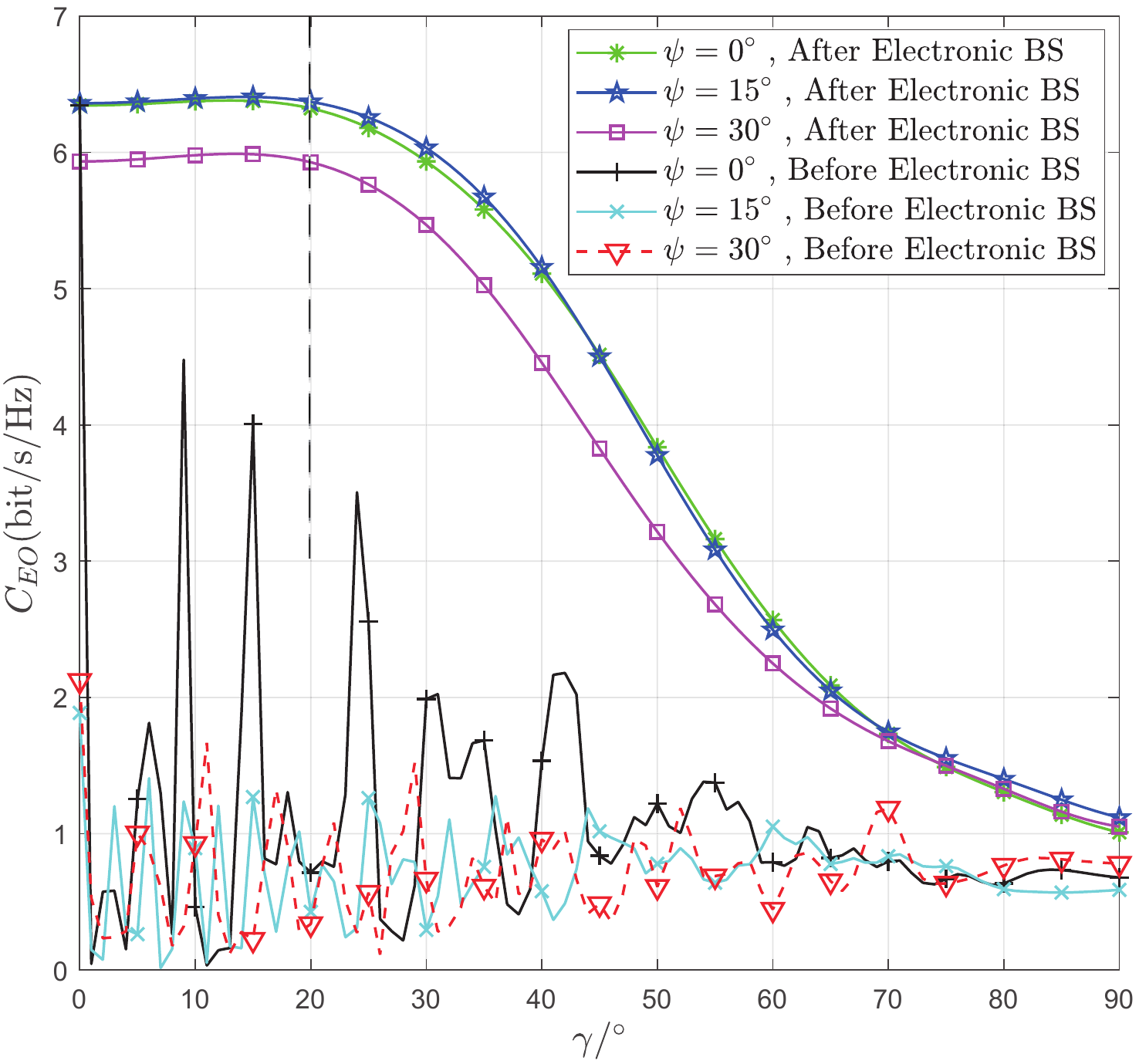}
}
\quad
\subfigure[]{
\includegraphics[width=8cm,height=6.5cm]{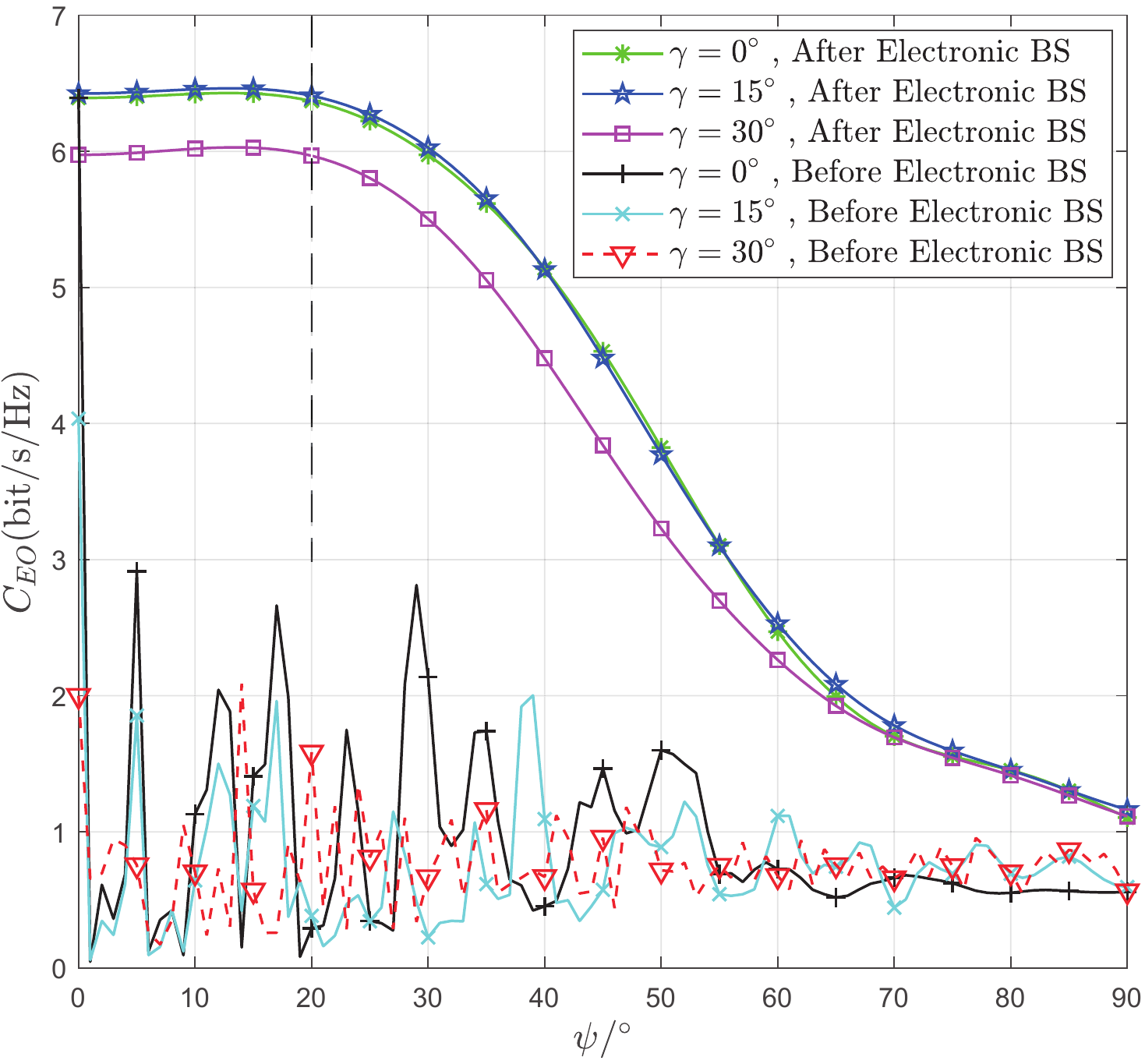}
}
\caption{The effects of (a) yaw angle $\gamma$, (b) pitch angle $\psi$ on $C_{EO}$, $P = 6$ subcarriers from $3.9982$ GHz to $4.2387$ GHz, $U = 9$ with ${\ell_v} \in [- 4,4]$, ${R_r} = {R_t} = 20\lambda_1$, $r=450\lambda_1$. BS: Beam steering.}
\label{PY}
\end{figure}
Hence, after electronic beam steering, the signal-to-interference-plus-noise ratio (SINR) on the $u$th mode and the $p$th subcarrier can be formulated as
\begin{align}\label{SINREO}
&\textrm{SINR}_{EO}(p,u)\nonumber\\
&=\frac{{{{\left| \tilde{h}_{\textrm{OAM},p}(u,u) \right|}^2}\mathbb{E}\left({{\left| {s(p,u})\right|}^2}\right)}}{{\sum\limits_{u \ne v}{{{\left| \tilde{h}_{\textrm{OAM},p}(u,v)\right|}^2}\mathbb{E}\left({{\left| {s(p,v}) \right|}^2}\right) + \sigma_z^2} }},
\end{align}
where $\mathbb{E}\big({{\left| {s(p,v}) \right|}^2}\big)$ is the average power of modulation symbols on the $v$th mode and $p$th subcarrier. We assume that the average power of modulation symbols on all modes and subcarriers are the same, and then the SINR approaches to the signal-to-interference ratio (SIR) at high SNR as
\begin{align}
&\lim_{\rho\rightarrow \infty}\textrm{SINR}_{EO}(p,u)=\textrm{SIR}_{EO}(p,u) \nonumber\\
&=\frac{{\left| \tilde{h}_{\textrm{OAM},p}(u,u)\right|}^2}{\sum\limits_{u \ne v} {{\left| \tilde{h}_{\textrm{OAM},p}(u,v)\right|}^2} }\label{GS19},
\end{align}
where $\rho=\mathbb{E}\left({{\left| {s(p,u})\right|}^2}\right)/\sigma_z^2$. Then, we evaluate how the SINRs change with the increase of $|\gamma|$ and $|\psi|$ of the electronic beam steering. If we substitute \eqref{g11} into \eqref{SINREO}, the obtained two-dimensional function of $\gamma$ and $\psi$ is too complex to analyze. Therefore, we turn to check how $\textrm{SIR}_{EO}(p,u)$ changes with the increase of $|\gamma|$ when $\psi=0$, and how $\textrm{SIR}_{EO}(p,u)$ changes with the increase of $|\psi|$ when $\gamma=0$, and obtain the following results.
\begin{prop}
When $\psi=0$, the SIR after electronic beam steering on the $u$th mode and $p$th subcarrier decreases with the increase of $|\gamma|$ as $S_{{k_p}}$ goes to zero.
\label{PropI}
\end{prop}
\begin{IEEEproof}
The proof is given in Appendix C.
\end{IEEEproof}
\begin{prop}
When $\gamma=0$, the SIR after electronic beam steering on the $u$th mode and $p$th subcarrier decreases with the increase of $|\psi|$ as $S_{{k_p}}$ goes to zero.
\label{PropII}
\end{prop}
\begin{IEEEproof}
The proof is completely similar to Proposition 1.
\end{IEEEproof}

The Proposition 1 and Proposition 2 show that $\textrm{SIR}_{EO}(p,u)$ is monotonically decreasing for $|\gamma|$ and $|\psi|$ in two one-dimensional directions respectively on the condition that $S_{{k_p}}$ approaches to zero. Moreover, we show through simulation results in Fig. \ref{FDJX}(a) and Fig. \ref{FDJX}(b) that $\textrm{SIR}_{EO}(p,u)$ is also monotonically decreasing for $|\gamma|$ and $|\psi|$ in the case $S_{{k_p}}$ is larger. Furthermore, Fig. \ref{FDJX}(c) shows that when $S_{k_p}$ is fixed to be a value other than zero, $\textrm{SIR}_{EO}(p,u)$ is still monotonically decreasing for $|\gamma|$ and $|\psi|$. It follows that with the increase of oblique angles the SINR after applying only electronic beam steering will get worse.

In order to further figure out the total effect of all the subcarriers, the MCMM-OAM system with only electronic beam steering can be evaluated according to the channel capacity
\begin{align}
C_{EO} = \frac{1}{P}\sum\limits_{p = 1}^P \sum\limits_{u = 1}^U {{{\log }_2}\left( {1 +\textrm{SINR}_{EO}(p,u)} \right)} \label{ceo}.
\end{align}
In Fig. \ref{PY}, we illustrate what is the condition for ``small-angle misalignment" so that electronic steering is sufficient, and what is the condition for ``large-angle misalignment" so that it is insufficient. It can be seen apparently from the figure that under this set of parameters the threshhold between ``small-angle" and ``large-angle" is about 20 degree. When $|\gamma|$ and $|\psi|$ are smaller than the threshold, the electronic beam steering can well compensate for the performance loss caused by misalignment, while $|\gamma|$ and $|\psi|$ are larger than this threshold, the electronic beam steering is not good enough especially when $|\gamma|$ and $|\psi|$ approach $\pi/2$.
\begin{figure*}[t]
\begin{center}
\includegraphics[scale=0.8]{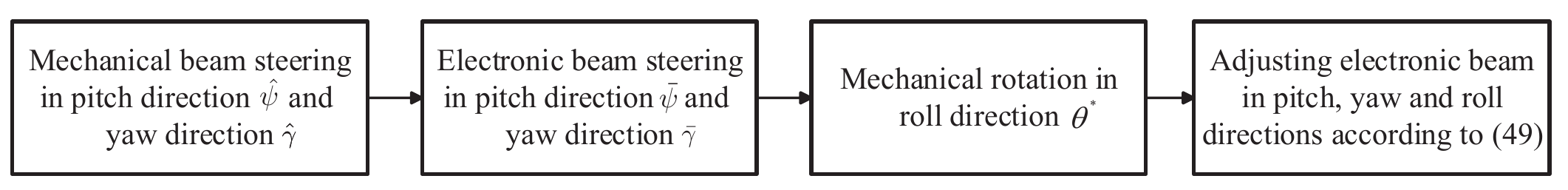}
\end{center}
\caption{The procedure of hybrid mechanical and electronic beam steering for the LoS MCMM-OAM receiver.}
\label{liucheng}
\end{figure*}
\begin{figure*}[b]
\hrulefill
\begin{align}\label{yH}
\setcounter{equation}{17}
\mathbf{y}_{H}(p)=&\Big({\mathbf{F}_U}\odot{\mathbf{B}_{E_2}(p)} \odot{\mathbf{B}_{E_1}(p)}\Big)\left(\mathcal{F}_2\left( {\mathcal{F}}_1\big(\mathbf{H}(p),\hat \psi ,\hat \gamma \big),\theta^* \right)\mathbf{F}_U^H\mathbf{s}(p) + \mathbf{z}(p)\right)\nonumber\\
=&{\mathbf{H}}^{{\rm OAM}}_{H}(p)\mathbf{s}(p)+\mathbf{\hat{z}}(p),
\end{align}
\end{figure*}
\subsection{Solution in Large-Angle Misalignment Case}
The reason for the performance degradation of electronic beam steering is that with the increase of misalignment angles, the inter-mode interferences increase. In order to minimize the value of the interference term $\sum_{u \ne v}| \tilde{h}_{\textrm{OAM},p}(u,v)|^2$, we observe from \eqref{g11} that the smaller the $|\gamma|$ and $|\psi|$, the smaller the value of the interference term. As long as the $\gamma$ and $\psi$ are equal to zero, $\tilde{h}_{\textrm{OAM},p}(u,v)=0$ and interferences disappear. It follows that when the receive UCA is in the large-angle misalignment with the transmit UCA, it is the best choice to firstly steer the yaw and pitch directions of the receive UCA physically in order that $|\gamma|$ and $|\psi|$ are as small as possible. Furthermore, besides minimizing the interference, maximizing the channel capacity needs also maximizing the received signal power, that is, maximizing the term $|\tilde{h}_{\textrm{OAM},p}(u,u)|^2$. Note that when $\gamma$ and $\psi$ are fixed, for the receive UCA only $\theta_0$ has effect on $\tilde{h}_{\textrm{OAM},p}(u,v)$, which means the roll rotation angle $\theta$ needs to be optimized. The above analysis could be formulated in the follows.
\begin{align}\label{maxSINR}
\setcounter{equation}{16}
\max C_{EO}\Leftrightarrow& \max \textrm{SINR}_{EO} \nonumber\\
\Leftrightarrow& \left\{
\begin{aligned}
&{\min\sum\limits_{u \ne v}{{\left| {{\tilde{h}_{{\rm{OAM}},p}}(u,v)} \right|}^2} }\\
&{\max {{\left| {{\tilde{h}_{{\rm{OAM}},p}}(u,u)} \right|}^2}}
\end{aligned}
\right.\nonumber\\
\Leftrightarrow& \left\{
\begin{aligned}
&{\min \ (|\gamma | + |\psi |) \ \ \textrm{in yaw and pitch directions}}\\
&{\textrm{Search for} \ \ \theta^\star \ \textrm{in roll direction}.}
\end{aligned}
\right.
\end{align}
This is the motivation of proposing hybrid mechanical and electronic beam steering. Mechanical beam steering enables the misalignment angles fall into the range of ``small angle" and leave them for electronic beam steering, which happens when the receiver starts running or moves to another place.
\section{Hybrid Mechanical and Electronic Beam Steering of OAM Communication Systems}
According to \eqref{maxSINR}, the procedure of the hybrid mechanical and electronic beam steering scheme can be designed as the four successive steps shown in Fig. \ref{liucheng}. Then, the hybrid beam steering can be formulated as \eqref{yH}, where $\mathbf{y}_{H}(p)$ is the recovered information signal vector with hybrid beam steering, ${\mathbf{H}}^{{\rm OAM}}_{H}(p)=({\mathbf{F}_U}\odot{\mathbf{B}_{E_2}(p)} \odot{\mathbf{B}_{E_1}(p)})(\mathcal{F}_2( {\mathcal{F}}_1\big(\mathbf{H}(p),\hat \psi ,\hat \gamma \big),\theta^*)\mathbf{F}_U^H)$ is defined as the effective OAM channel matrix after the hybrid steering, $\mathbf{\hat{z}}(p)=\left({\mathbf{F}_U}\odot{\mathbf{B}_{E_2}(p)} \odot{\mathbf{B}_{E_1}(p)}\right)\mathbf{z}(p)=[\hat{z}(p,1),\hat{z}(p,2),\cdots,\hat{z}(p,U)]^T$, ${\mathbf{B}_{E_1}(p)}$ is the first electronic beam steering matrix in pitch and yaw directions, ${\mathbf{B}_{E_2}(p)}$ is the second beam steering matrix that represents adjusting the electronic beam in roll direction, ${{\mathcal{F}}_1}$ represents the operation of the first mechanical beam steering in pitch and yaw directions, and ${{\mathcal{F}}_2}$ represents the operation of the second mechanical beam steering in roll direction.

\begin{figure}[t]
\setlength{\abovecaptionskip}{0.0cm}   
\setlength{\belowcaptionskip}{-0.0cm}   
\begin{center}
\includegraphics[scale=0.6]{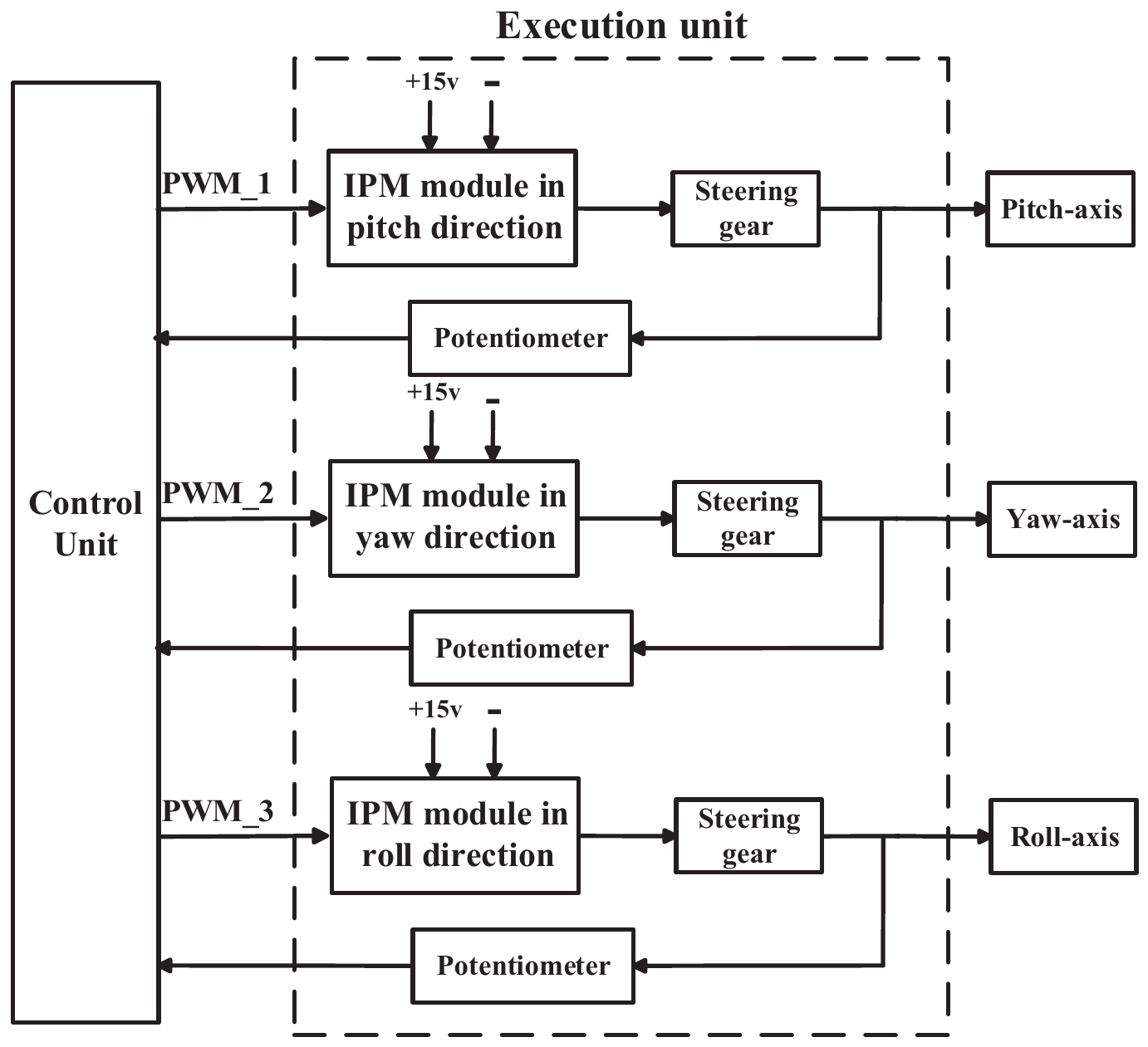}
\end{center}
\caption{The schematic diagram of the servo system of the mechanical rotating devices.}
\label{002}
\end{figure}

The implementation of mechanical beam steering relies on the servo system of mechanical rotating devices, which is controlled by PWM signals \cite{Tajuddin2009TMS320F2812}. The schematic diagram of the servo system corresponding to the hybrid OAM transceiver in Fig. \ref{fig1} is shown in Fig. \ref{002}, where the control
unit generates three tunable PWM signals for pitch-axis, yaw-axis and roll-axis, respectively. In the execution unit, intelligent power modules (IPMs) are used to drive steering gears in the three directions respectively, and potentiometer is used to convert the motion of steering gear to voltage and feedback to control unit. We denote the accuracy of the potentiometer as $\nu$. The relationship between the duty cycle $D$ of the PWM signal and the expected rotation angle $\vartheta$ can be obtained as $\vartheta =\frac{\pi}{p_e-p_s}(DK - {p_0})$,
where $K$ is the signal period of PWM signal, $p_s$, $p_0$ and $p_e$ are the pulse widths of PWM signal corresponding to the minimum angle, middle angle and the maximum angle of the mechanical rotating device \cite{Pinckney2006Pulse}. Here, $\vartheta$ represents $\hat\psi$, $\hat\gamma$ and $\theta^*$ in pitch-axis, yaw-axis and roll-axis, respectively.
\subsection{Mechanical Beam Steering in Pitch and Yaw Directions}
Firstly, the coarse alignment is achieved by mechanical beam steering. From the Fig. \ref{fig1} we can see that only with rotations in pitch and yaw directions can the receive UCA align to the transmit UCA. Then, according to the Fig. \ref{fig2}, we can see that the mechanical beam steering in yaw and pitch directions are performed around $\textrm{Y}'$-axis and $\overline{\textrm{X}}$-axis, respectively. It is assumed that $\hat\gamma$ and $\hat\psi$ are the coarse estimates of $\gamma$ and $\psi$ in this step, and $\bar\gamma=\gamma-\hat{\gamma}$ and $\bar\psi=\psi-\hat{\psi}$ are the remaining small misalignment errors in yaw and pitch directions, respectively. In the first-stage yaw direction, the mechanical beam steering makes the receive UCA rotate around the $\textrm{Y}'$-axis of the $\textrm{Z}'-\textrm{X}'\overline{\textrm{O}}\textrm{Y}'$ coordinate system so that the angle between the $\overline{\textrm{X}}$-axis and the $\textrm{X}'$-axis changes from $\gamma$ to $\bar\gamma$. Then, in the second-stage pitch direction, the receive UCA rotates around the rotated $\overline{\textrm{X}}$-axis so that the angle between the $\overline{\textrm{Y}}$-axis and $\textrm{Y}'$-axis changes from $\psi$ to $\bar\psi$. With the pitch and yaw rotations, the cartesian coordinate of the $m$th antenna element on the receive UCA in ${\textrm{Z}'}-{\textrm{X}'}\overline{\textrm{O}}{\textrm{Y}'}$, denoted by $({\bar{a}_m},{\bar{b}_m},{\bar{c}_m})$, can be obtained as $[{\bar{a}_m}, {\bar{b}_m}, {\bar{c}_m}]^T ={\mathbf{R}}_{Y} (\bar\gamma)$ ${\mathbf{R}}_{P}(-\hat\psi)$${\mathbf{R}}_{Y}(-\bar\gamma){\mathbf{R}}_{Y}(-\hat\gamma)
[{a_m},{b_m},{c_m}]^T$,
where ${\mathbf{R}}_{P}(-\hat\psi)$ and ${\mathbf{R}}_{Y}(-\hat\gamma)$ represent the coordinate rotation matrixes in pitch and yaw directions
and considering the second-stage pitch rotation around the rotated $\overline{\textrm{X}}$-axis not any axis of the $\textrm{Z}'-\textrm{X}'\overline{\textrm{O}}\textrm{Y}'$ coordinate system, the axis rotation matrix ${\mathbf{R}}_{Y}(-\bar\gamma)$ is required after the first-stage yaw rotation for the coordinate transformation from $\textrm{X}'$-axis to $\overline{\textrm{X}}$-axis, in the end, ${\mathbf{R}}_{Y}(\bar\gamma)$ performs the opposite coordinate transformation from $\overline{\textrm{X}}$-axis to $\textrm{X}'$-axis.
Thus, after the pitch and yaw rotations the transmission distance from the $n$th transmit antenna element to the $m$th receive antenna element becomes
\begin{align}
\setcounter{equation}{18}
{\bar d_{m,n}}
=&r - \frac{R_rR_t}{r} \sin\theta_m \cos\phi_n\sin\bar\psi \sin\bar\gamma\nonumber\\
&-\frac{R_rR_t}{r}\big(\cos\theta_m \cos\phi_n\cos\bar\gamma + \sin\theta_m \sin\phi_n\cos\bar\psi \big)  \nonumber\\
&+R_r \left(\sin\theta_m \sin\bar\psi \cos\bar\gamma-\cos\theta_m \sin\bar\gamma\right)\label{c}.
\end{align}
Hence, the channel matrix after the mechanical beam steering in pitch and yaw directions can be expressed as ${\mathbf{H}}_{{\mathcal{F}}_1}(p)=[{\bar h_{m,n}}(p)]_{N\times N}$, where the $m$th-row $n$th-column element ${\bar h_{m,n}}(p)$ can be written as
\vspace{-0.2cm}
\begin{align}
&{\bar h_{m,n}}(p)=\frac{\beta}{2k_p\bar{d}_{m,n}}\textrm{exp}\left(-ik_p\bar{d}_{m,n}\right)\nonumber\\
&\qquad\mathop \approx \limits^{(a)}\frac{\beta }{{2{k_p}r}}\exp\bigg(iS_{{k_p}}\sin\theta_m \cos\phi_n \sin\bar\psi\sin\bar\gamma  \nonumber\\
&\qquad+ iS_{{k_p}}\cos\theta_m \cos\phi_n \cos\bar\gamma\nonumber\\
&\qquad+ iS_{{k_p}}\sin\theta_m \sin\phi_n \cos\bar\psi-ik_pr\nonumber\\
&\qquad- ik_pR_r \left(\sin\theta_m \sin\bar\psi \cos\bar\gamma -\cos\theta_m \sin\bar\gamma \right)\bigg)\label{GS28},
\end{align}
(a) neglects a few small terms in the denominator and only $2k_pr$ is left. If only with the mechanical beam steering, the received OAM signal $\mathbf{{y}}_{{\mathcal{F}}_1}(p)$ takes the form
\begin{align} \label{y}
\mathbf{{y}}_{{\mathcal{F}}_1}(p)=&\mathbf{F}_U\left(\mathbf{{H}}_{{\mathcal{F}}_1}(p)\mathbf{F}_U^H\mathbf{s}(p) +\mathbf{z}(p)\right)\nonumber\\
=&\mathbf{{H}}^{\rm OAM}_{{\mathcal{F}}_1}(p)\mathbf{s}(p)+\mathbf{\bar{z}}(p),
\end{align}
where $\mathbf{{H}}^{\textmd{OAM}}_{{\mathcal{F}}_1}(p)=\mathbf{F}_U\mathbf{{H}}_{{\mathcal{F}}_1}(p)\mathbf{F}_U^H$ is defined as the effective multi-mode OAM channel matrix after the first mechanical beam steering. ${h}^{\textrm{OAM}}_{{\mathcal{F}}_1,p}(u,v)=\mathbf{f}(\ell_u)\mathbf{{H}}_{{\mathcal{F}}_1}(p){\mathbf{f}^{H}}({\ell_v})$ is the $u$th-row and $v$th-column element of $\mathbf{{H}}^{\textmd{OAM}}_{{\mathcal{F}}_1}(p)$, and can be calculated as
\begin{align}
&{h}^{\textrm{OAM}}_{{\mathcal{F}}_1,p}(u,v)=\eta ({p})\sum\limits_{m = 1}^N\sum\limits_{n = 1}^N
\exp\bigg(-i{\ell_u}\theta_m +i{\ell_v}\phi_n \nonumber\\
&+iS_{{k_p}}\sin\theta_m \cos\phi_n \sin\bar\psi\sin\bar\gamma\nonumber\\
&+iS_{{k_p}}(\cos\theta_m \cos\phi_n \cos\bar\gamma+\sin\theta_m \sin\phi_n\cos\bar\psi)\nonumber\\
&-ik_pR_r \left(\sin\theta_m \sin\bar\psi \cos\bar\gamma-\cos\theta_m \sin\bar\gamma \right)\bigg)\nonumber\\
&\mathop \approx \limits^{(a)}\eta(p)\sum\limits_{\delta = 1}^N\exp\left(i\frac{2\pi \delta}{N}\ell_v+iS_{{k_p}}\cos\frac{2\pi \delta}{N}\right)\times\nonumber\\
&\sum\limits_{m = 1}^N\sum\limits_{n = 1}^N\exp \bigg(-i\frac{2\pi (m-1)}{N}t+ik_pR_r\cos\theta_m \sin\bar\gamma\nonumber\\
&-ik_pR_r\sin\theta_m \sin\bar\psi \cos\bar\gamma \bigg),
\end{align}
where $\delta=n-m\in \mathds{Z}$, (a) applies the approximation $\cos a\approx1-\frac{a^2}{2}$ for $\cos\bar\psi$ and $\cos\bar\gamma$ in the case that $\bar\psi$ and $\bar\gamma$ are relatively small, and neglects a few small terms under the condition $r\gg R_r$. As $\mathbf{{H}}^{\textmd{OAM}}_{{\mathcal{F}}_1}(p)$ is a non-diagonal matrix, we can conclude that there is still inter-mode interferences after the first mechanical beam steering.

\subsection{Electronic Beam Steering in Pitch and Yaw Directions}
In order to eliminate the inter-mode interferences induced by the remaining small perturbations $\bar\psi$ and $\bar\gamma$ after previous mechanical beam steering, periodic refined AoA estimation are necessary. It is assumed that $\bar\psi$ and $\bar\gamma$ varies slower than the duration of AoA estimation, and $\bar\psi$ and $\bar\gamma$ can be perfectly estimated. With the estimates of $\bar\psi$ and $\bar\gamma$, electronic beam steering can enable the receive beam to align with the direction of the arrived OAM beam. The electronic beam steering matrix can be designed as ${\mathbf{B}}_{E_1}(p) = \mathbf{1}\otimes {\mathbf{\mathfrak{b}}}_{E_1}(p)$, where ${\mathbf{\mathfrak{b}}}_{E_1}(p) = [{e^{i{\overline{W}_1}(p)}},{e^{i{\overline{W}_2}(p)}}, \cdots ,{e^{i{\overline{W}_N}(p)}}]$ and $\overline{W}_m(p) =  k_pR_r\left( \sin\theta_m\sin\bar\psi\cos\bar\gamma-\cos\theta_m\sin\bar\gamma\right)$,
$m = 1,\cdots,N$, $p = 1,\cdots,P$. After involving these phases into the original phases in ${\mathbf{F}_U}$ at the receive UCA, the effective multi-mode OAM channel matrix at the $p$th subcarrier becomes
\begin{align}
{\mathbf{{H}}}^{\rm OAM}_{E_1{\mathcal{F}}_1}(p) = ({\mathbf{F}_U} \odot {\mathbf{B}}_{E_1}(p)){\mathbf{{H}}}_{{\mathcal{F}}_1}(p){\mathbf{F}_U^H}.
\end{align}
Denote ${h}_{E_1{\mathcal{F}}_1,p}^{\textrm{OAM}}(u,v)=\big({\mathbf{\mathfrak{{b}}}}_{E_1}(p)\odot\mathbf{f}(\ell_u)\big){\mathbf{{H}}}_{{\mathcal{F}}_1}(p){\mathbf{f}^{H}}({\ell_v})$  as the $u$th-row and $v$th-column element of ${\mathbf{{H}}}^{\rm OAM}_{E_1{\mathcal{F}}_1}(p)$, which can be expressed as
\begin{align}
{h}_{E_1{\mathcal{F}}_1,p}^{\textrm{OAM}}(u,v)
\approx & \eta (p)\sum\limits_{m = 1}^N \exp (-i\theta_m t)\times \nonumber\\
&\sum\limits_{\delta=1}^N {\exp \left( {i\frac{{2\pi\delta}}{N}{\ell_v} + iS_{{k_p}}\cos \frac{{2\pi\delta}}{N}} \right)}\label{E_1F1},
\end{align}
$t={\ell_u}-{\ell_v}\in \mathds{Z}$. Note that in \eqref{E_1F1}, we have
\begin{align}
\left\{ {\begin{array}{*{20}{c}}
{\sum\limits_{m = 1}^N {\exp } \left( { - i{\theta _m}t} \right) = 0, \quad t \ne 0},\\
{\sum\limits_{m = 1}^N {\exp } \left( { - i{\theta _m}t} \right) = N, \quad t = 0}.
\end{array}} \right.\label{GS33}
\end{align}
That is to say, the non-diagonal elements of ${\mathbf{{H}}}^{\rm OAM}_{E_1{\mathcal{F}}_1}(p)$ equal to zero approximately. It follows that after the mechanical and electronic beam steering in pitch and yaw directions, the inter-mode interferences are almost completely eliminated in the case of large-angle misalignment. At the same time, according to the conclusion in Section III, steering the OAM beam with only a small angle in electronic way does not result in large performance degradation. Therefore, the proposed hybrid mechanical and electronic beam steering outperforms only using electronic beam steering in a wide variety of practical scenarios.
\begin{figure}[t]
\setlength{\abovecaptionskip}{-0.1cm}   
\setlength{\belowcaptionskip}{-0.8cm}   
\centering
\subfigure[]{
\includegraphics[width=3cm]{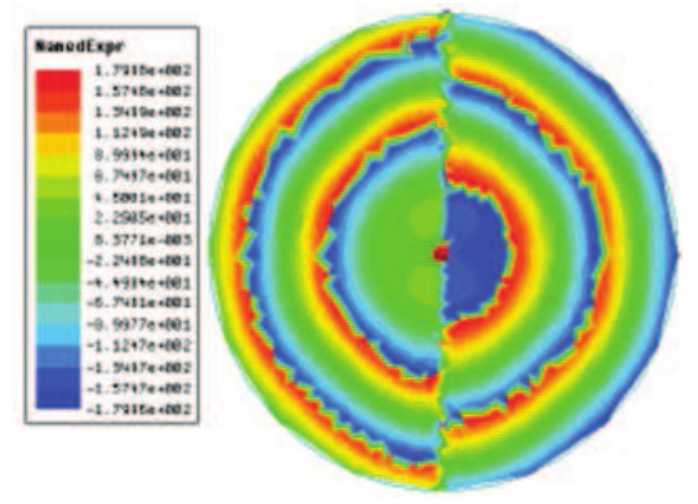}}
\subfigure[]{
\includegraphics[width=3cm]{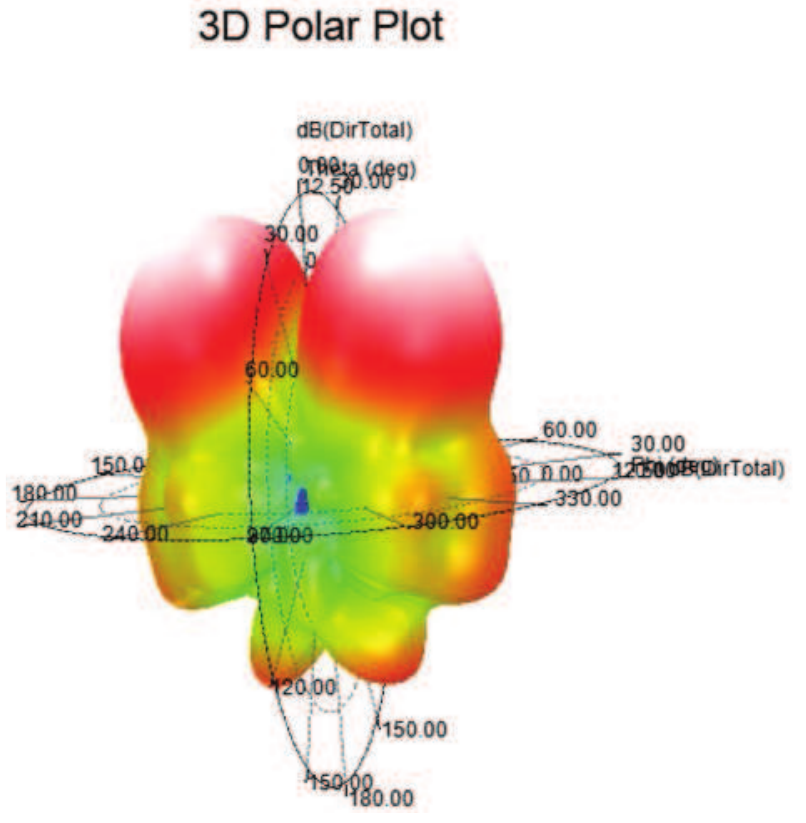}}

\subfigure[]{
\includegraphics[width=3cm]{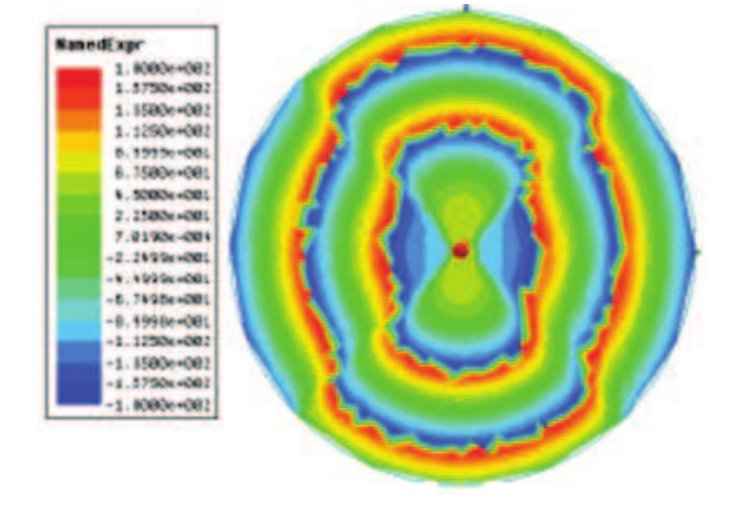}}
\subfigure[]{
\includegraphics[width=3cm]{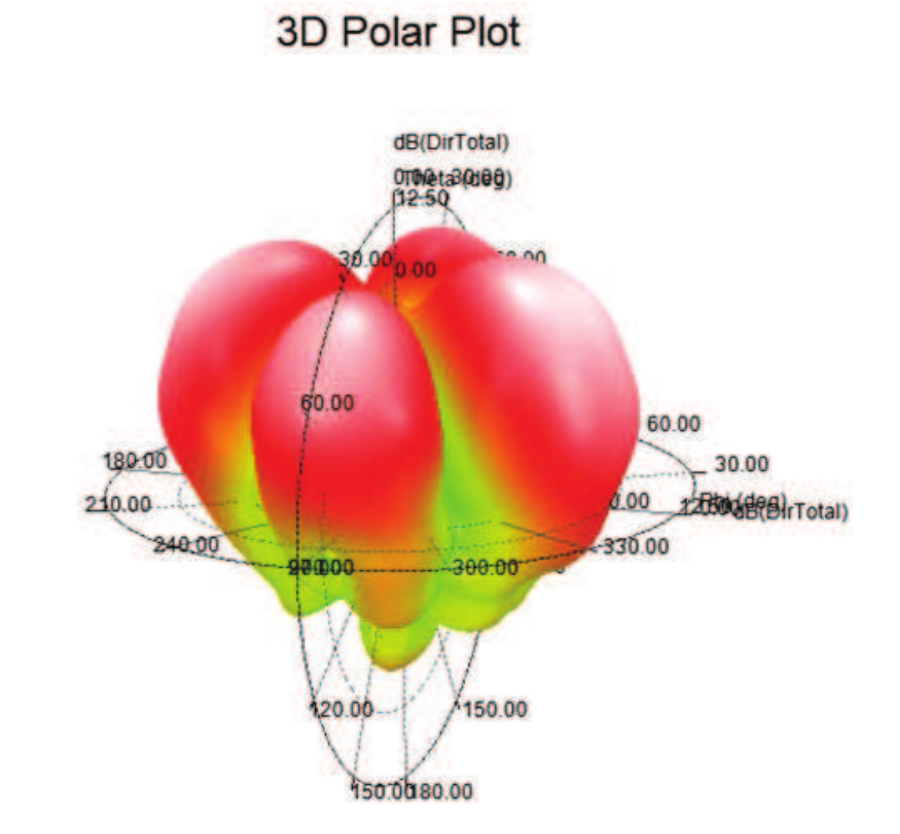}}

\subfigure[]{
\includegraphics[width=3cm]{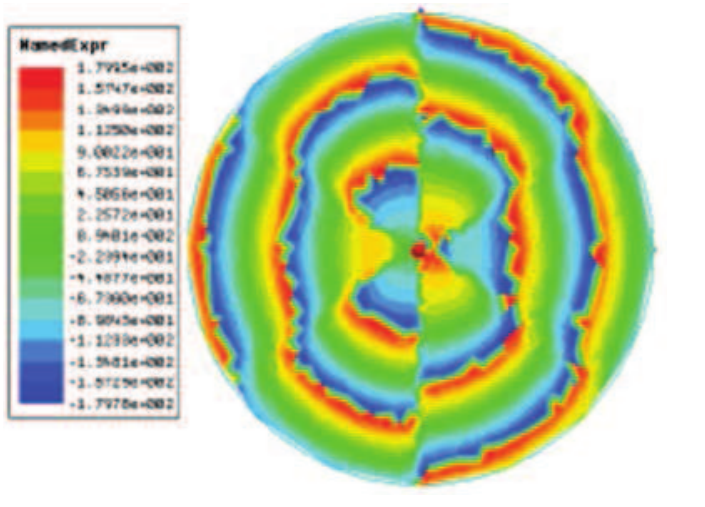}}
\subfigure[]{
\includegraphics[width=3cm]{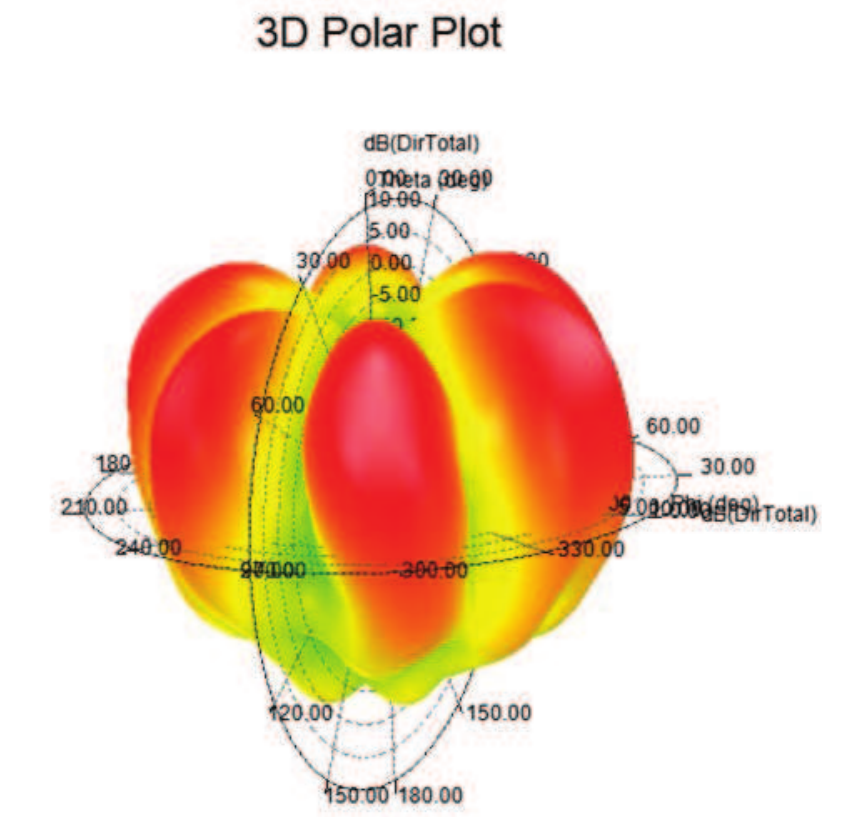}}

\caption{The phase distributions (left) and radiation patterns (right) of two-mode OAM beam simulated by HFSS with mode numbers: (a)(b) $\ell_1=+1$, $\ell_2=-1$, (c)(d) $\ell_1=+2$, $\ell_2=-2$, (e)(f) $\ell_1=+3$, $\ell_2=-3$.}
\label{Pattern}
\end{figure}
\vspace{-0.2cm}
\subsection{Mechanical Rotation in Roll Direction for OAM Channel Capacity Maximization}

When a multi-mode OAM beam that can be deemed as the superposition of multiple OAM modes is transmitted, we can observe the phenomenon of interferometry as shown in the simulation results of HFSS in Fig. \ref{Pattern}, that is, the electric field intensities of one multi-mode OAM beam are not uniform on the ring of its main lobe. Therefore, the receive antenna elements need to be set in the maximum SNR positions on the ring of the main lobe, which means that another rotation angle in roll direction within the plane of the receive UCA should be optimized.
In order to obtain the optimal rotation angle in roll direction, we first formulate the optimization problem and then apply SA algorithm to solve the problem.
\begin{algorithm}[t]
\caption{Solving The Optimization Problem \eqref{h}}
\hspace*{0.02in} {\bf Input:}
${\theta} \in [-\frac{\pi }{N},\frac{\pi }{N}]$, ${T}_{\rm {min}}\in\mathbb{R}$, ${T_{\rm{init}}}\in\mathbb{R}$, $\zeta\in\mathbb{R}$\\
\hspace*{0.02in} {\bf Output:}
$\theta^*\in\mathbb{R}$
\begin{algorithmic}[1]
\State \textbf{procedure} \textbf{SA} $({\theta},{T}_{\rm{min}},{T_{\rm{init}}},\zeta)$
\State \quad ${C}\leftarrow $use (\ref{C27})
\State \quad\textbf{while}  ${T_{\rm{init}}}$ $ > $ ${T}_{\rm{min}}$
\State \quad\quad\quad $\theta^*\leftarrow {\theta} $, $C_{\rm{max}}\leftarrow {C} $, ${L}\leftarrow 0$
\State \quad\quad\textbf{for} $j=1:J$ \textbf{do}
\State \quad\quad\quad\quad ${e}\leftarrow $ a small random value$ $
\State \quad\quad\quad\quad $\theta_{\rm{new}}\leftarrow \left\{
{\begin{array}{*{2}{ll}}
{\theta+ e}, \ \textrm{if} \ -{\frac{\pi}{N}}<\theta+ e < {\frac{\pi}{N}}\\
{\theta- e}, \ \textrm{otherwise}
\end{array}} \right.$
\State \quad\quad\quad\quad$\theta\leftarrow\theta_{\rm{new}}$, $C_{\rm{new}}\leftarrow$ use (\ref{C27})
\State \quad\quad\quad\textbf{if}\;\;{$C_{\rm{new}}>{C}$ or meet the Metropolis criterion }
\State \quad\quad\quad\quad\quad ${L}\leftarrow 1 $, ${\theta}\leftarrow \theta_{\rm{new}}$, ${C}\leftarrow C_{\rm{new}}$
\State \quad\quad\quad\quad\textbf{if}\;\;{${C}>C_{\rm{max}}$}
\State \quad\quad\quad\quad\quad\quad $\theta^*\leftarrow {\theta}$, $C_{\rm{max}}\leftarrow {C}$
\State \quad\quad\quad\quad\textbf{end}
\State \quad\quad\quad\textbf{end}
\State \quad\quad\textbf{end}
\State \quad\quad\textbf{if}  {$L \ne 0 $}
\State \quad\quad\quad\quad ${\theta}\leftarrow \theta^*$, ${C}\leftarrow C_{\rm{max}}$
\State \quad\quad\textbf{end}
\State \quad\quad\quad ${T_{\rm{init}}}\leftarrow {T_{\rm{init}}} \zeta$
\State \quad\textbf{end}
\State \quad\Return ${\theta^*}$
\State \textbf{end procedure}
\end{algorithmic}
\end{algorithm}

\subsubsection{Optimization Model}
\begin{figure}[t]
\setlength{\abovecaptionskip}{0cm}   
\setlength{\belowcaptionskip}{-0.5cm}   
\begin{center}
\includegraphics[width=8cm,height=6.5cm]{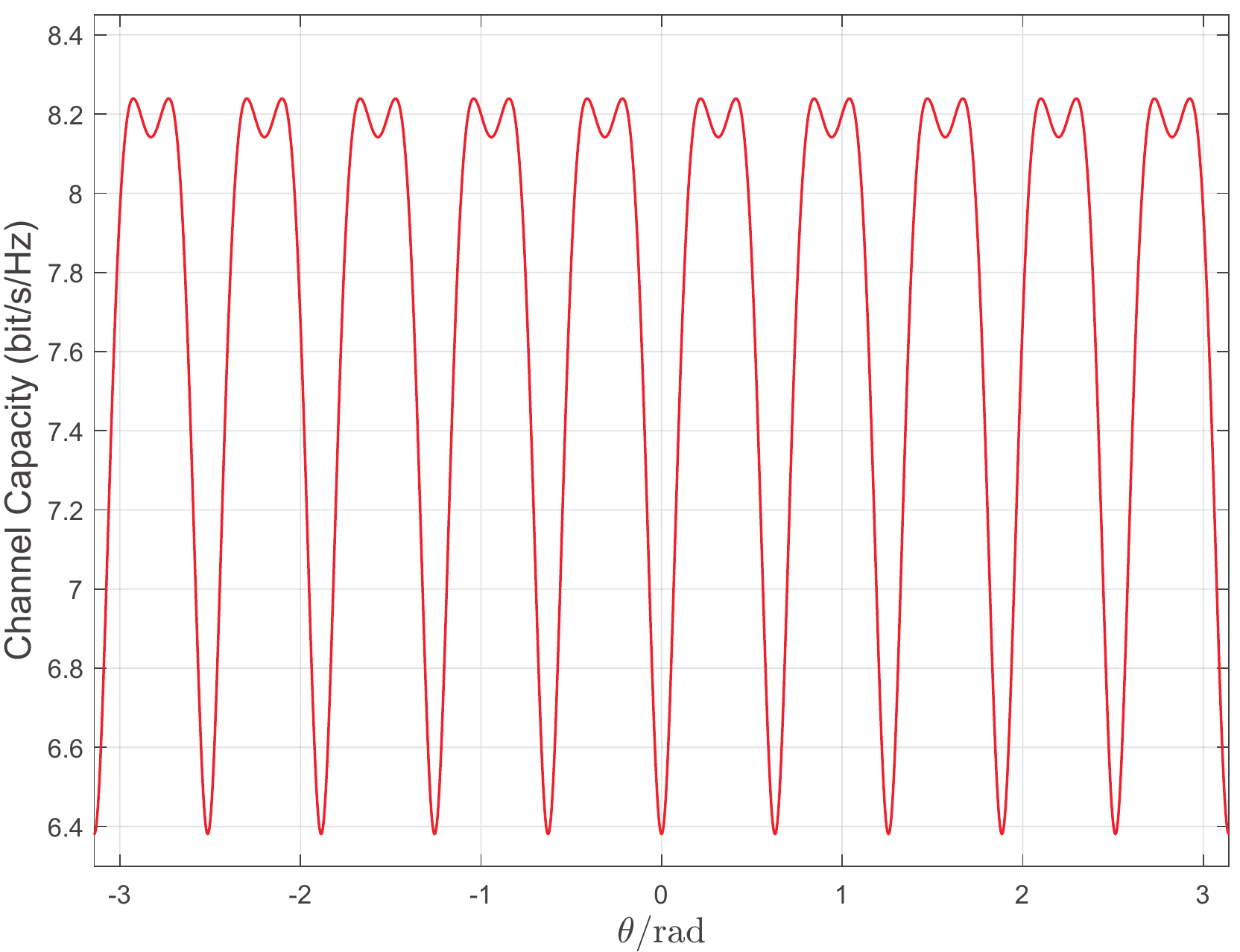}
\end{center}
\caption{The impact of $\theta$ on the capacity of the effective channel ${\mathbf{{H}}}^{\rm OAM}_{E_2\mathcal{F}_2E_1\mathcal{F}_1}$, $P = 6$ subcarriers from $3.9982$ GHz to $4.2387$ GHz, $U = 9$ with ${\ell_u}=-4,-3,\cdots,4$, $N=10$, ${R_r} = $ ${R_t} = 20\lambda_1$, and $r=450\lambda_1$.}
\label{fig5}
\end{figure}

First, we denote the effective OAM channel matrix with twice mechanical and once electronic beam steering as ${\mathbf{{H}}}^{\rm OAM}_{\mathcal{F}_2E_1\mathcal{F}_1}(p)$. Since the roll rotation will result in the misalignment between the transmit and receive beams and deteriorate the performance of the OAM communication systems, we assume the second-step roll rotation can be well compensated by the forth-step electronic beam steering. Only in this way, can the effective OAM channel capacity be maximized. For this purpose, we denote the effective OAM channel matrix with twice mechanical and twice electronic beam steering as ${\mathbf{{H}}}^{\rm OAM}_{E_2\mathcal{F}_2E_1\mathcal{F}_1}(p)$. Then, based on \eqref{E_1F1} the $u$th diagonal element of ${\mathbf{{H}}}^{\rm OAM}_{E_2\mathcal{F}_2E_1\mathcal{F}_1}(p)$ can be obtained as
\begin{align}
&{h}_{E_2\mathcal{F}_2E_1\mathcal{F}_1,p}^{\textrm{OAM}}(u,u) = \nonumber\\
&N\eta (p)\sum\limits_{\delta=1}^N \exp \bigg(i\left(\frac{{2\pi\delta}}{N} -\theta\right){\ell_u}
+ iS_{{k_p}}\cos \left(\frac{{2\pi\delta}}{N} -\theta\right) \bigg)\label{C26},
\end{align}
where $\theta$ is the rotation angle in roll direction. Therefore, the SINR at the $u$th mode and $p$th subcarrier can be formulated as ${\textrm{SINR}}_{E_2\mathcal{F}_2E_1\mathcal{F}_1}(p,u)=\rho{{\left| {{h}_{E_2\mathcal{F}_2E_1\mathcal{F}_1,p}^{\textrm{OAM}}(u,u)} \right|}^2}$.
The channel capacity of the OAM communication system with hybrid beam steering can be expressed as
\begin{align}
C = \frac{1}{P}\sum\limits_{p = 1}^P {\sum\limits_{u = 1}^U {{{\log }_2}\bigg( {1 +{\textrm{SINR}}_{E_2\mathcal{F}_2E_1\mathcal{F}_1}(p,u)} \bigg)} }\label{C27}.
\end{align}
In order to see the impact of $\theta$ on OAM channel capacity intuitively, we plot the relationship between $C$ and $\theta$ in Fig. \ref{fig5}. It can be seen from the figure that the OAM channel capacity varies periodically with $\theta$ in the range of $[-\pi, \pi ]$. That is because the antenna elements on the receive UCA are uniformly arranged around the center of the circle. In Fig. \ref{fig5}, $N=10$ leads to 10 cycles. Therefore, we can limit $\theta$ to only any circle of the range to obtain the optimal rotation angle $\theta^*$ from the following optimization problem.
\begin{align}
\begin{array}{ccccc}
\mathop {\max}\limits & C\\
s.t. 
&-\frac{\pi }{N}\le{\theta}\le\frac{\pi }{N}\label{h}\\
\end{array}
\end{align}
Since the target function $C$ is not guaranteed to be differentiable, gradient-based procedures cannot be applied. A large number of global optimization techniques were developed to provide alternative solution methods for challenging problems, e.g., simulated
annealing (SA) \cite{Xavier2010Coupled}.
SA algorithm is a derivative-free promising algorithm, whose final solution quality does not strongly depend on the choice of the initial solution \cite{Abido2000obust}. So, SA is adopted here to obtain the optimal rotation angle $\theta^*$ within the range $[-\frac{\pi}{N},\frac{\pi}{N}]$. The solving procedure of the problem \eqref{h} is given in Algorithm 1, where ${T_{\rm{init}}}$ is the initial temperature, ${T_{\rm{min}}}$ is the lowest temperature, $\zeta$ is the cooling coefficient, $J$ is the number of inner-layer iterations, $L$ is used to identify whether a new value is accepted at ${T_{\rm{init}}}$, $C_{\rm{max}}$ is the optimal channel capacity, $C_{\rm{new}}$ represents the new solution obtained in the algorithm, and $\theta_{\rm{new}}$ represents the roll angle corresponding to $C_{\rm{new}}$. In the end, the optimal rotation angle $\theta^*$ that maximizes the OAM channel capacity can be obtained.

\begin{rem}
We can observe from \eqref{C26} and \eqref{h} that $\theta^*$ is only decided by \{$\eta(p)$, $\ell_u$, $u=1,\cdots,U$, $p=1,\cdots,P$\}. That is to say, the mechanical beam steering $\mathcal{F}_2$ in roll direction does not dependent on the electronic beam steering $E_1$ in pitch and yaw directions. Therefore, the procedure of hybrid mechanical and electronic beam steering can change its order from $\mathcal{F}_1$, $E_1$, $\mathcal{F}_2$, $E_2$ to $\mathcal{F}_1$, $\mathcal{F}_2$, $E_1$, $E_2$. Thus, we have $\mathbf{H}_{\mathcal{F}}(p)=\mathcal{F}_2( {\mathcal{F}}_1(\mathbf{H}(p),\hat \psi ,\hat \gamma),\theta^*)$, which unifies \eqref{yH14} and \eqref{yH}. Therefore, the relationship between $\mathbf{B}_{E_1}(p)$, $\mathbf{B}_{E_2}(p)$ and $\mathbf{B}_{E}(p)$ is $\mathbf{B}_{E}(p)=\mathbf{B}_{E_2}(p)\odot \mathbf{B}_{E_1}(p)$.
\end{rem}

\subsubsection{Mechanical Rotation in Roll Direction}
Having the optimal rotation angle $\theta^*$, the servo system make the receive UCA rotate around the roll-axis. Since the mechanical beam steering in roll direction is just for maximizing the OAM channel capacity rather than aligning the transmit and receive beams, we don't consider the rotation error in $\theta^*$ for easier analysis.
According to the Fig. \ref{fig2}, the cartesian coordinate of the $m$th antenna element on the receive UCA in $\overline{\textrm{Z}}-\overline{\textrm{X}}\overline{\textrm{O}}\overline{\textrm{Y}}$ coordinate system after the mechanical rotation in roll direction, denoted by $({\check{a}_m},{\check{b}_m},{\check{c}_m})$, can be expressed as $[{\check{a}_m},{\check{b}_m},{\check{c}_m}]^T= {{\bf{R}}_R}({\theta ^*})[{{R_r}\cos {\theta _m}},{{R_r}\sin {\theta _m}},0]^T$,
where ${\mathbf{R}}_{R}(\theta^*)$ represents the coordinate rotation matrix in roll direction and can be expressed as
\begin{align}
{\mathbf{R}}_{R}(\theta^*)=
\left[ {\begin{array}{*{20}{c}}
{\cos \theta^* }&{-\sin \theta^*}&0\\
{\sin \theta^* }&{\cos \theta^*}&0\\
0&0&1
\end{array}} \right].
\end{align}
Thus, the cartesian coordinate of the $m$th antenna element on the receive UCA in ${\textrm{Z}'}-{\textrm{X}'}\overline{\textrm{O}}{\textrm{Y}'}$ coordinate system, denoted by $({\hat{a}_m},{\hat{b}_m},{\hat{c}_m})$, can be expressed as $[{\hat{a}_m},{\hat{b}_m},{\hat{c}_m}]^T= {\mathbf{R}}_{Y}(\bar\gamma){\mathbf{R}}_{P}(\bar\psi)$ $[{\check{a}_m},{\check{b}_m},{\check{c}_m}]^T$,
where ${\mathbf{R}}_{P}(\bar\psi)$ and ${\mathbf{R}}_{Y}(\bar\gamma)$ represent the axis rotation matrixes corresponding to the pitch direction and the yaw direction after the mechanical rotation in roll direction, respectively.
Hence, after roll rotation the transmission distance ${\hat d_{m,n}}$ from the $n$th $(1 \le n \le N)$ transmit antenna element to the $m$th $(1 \le m \le N)$ receive antenna element can be calculated as
\begin{align}
{\hat d_{m,n}}
=&r - \frac{R_rR_t}{r} \sin\tilde\theta_m \cos\phi_n\sin\bar\psi \sin\bar\gamma\nonumber\\
&-\frac{R_rR_t}{r}\Big(\cos\tilde\theta_m \cos\phi_n\cos\bar\gamma + \sin\tilde\theta_m \sin\phi_n\cos\bar\psi \Big)  \nonumber\\
&+R_r \left(\sin\tilde\theta_m \sin\bar\psi \cos\bar\gamma-\cos\tilde\theta_m \sin\bar\gamma \right)\label{c},
\end{align}
where $\tilde\theta_m=\theta_m+\theta^*$. Correspondingly, the channel matrix can be expressed as ${\mathbf{H}}_{{\mathcal{F}}}(p)=[{\hat h_{m,n}}(p)]_{N\times N}$, where the $m$th-row $n$th-column element ${\hat h_{m,n}}(p)$ can be written as
\begin{align}
&\hat h_{m,n}(p)= \frac{\beta}{2k_p\hat{d}_{m,n}}\textrm{exp}\left(-ik_p\hat{d}_{m,n}\right)\nonumber\\
&\mathop \approx \limits^{(a)}\frac{\beta }{{2{k_p}r}}\exp\bigg(iS_{{k_p}}\sin\tilde\theta_m \cos\phi_n \sin\bar\psi\sin\bar\gamma  \nonumber\\
&+ iS_{{k_p}}\cos\tilde\theta_m \cos\phi_n \cos\bar\gamma\nonumber\\
&+ iS_{{k_p}}\sin\tilde\theta_m \sin\phi_n \cos\bar\psi-ik_pr\nonumber\\
&- ik_pR_r \left(\sin\tilde\theta_m \sin\bar\psi \cos\bar\gamma -\cos\tilde\theta_m \sin\bar\gamma \right)\bigg)\label{GS45},
\end{align}
(a) neglects a few small terms in the denominator and thus only $2k_pr$ is left. Thus, the effective multi-mode OAM channel matrix after the roll rotation can be written as
\begin{align} \label{y}
\mathbf{{H}}^{\rm OAM}_{\mathcal{F}_2E_1\mathcal{F}_1}(p)&=\mathbf{{H}}^{\rm OAM}_{E_1\mathcal{F}_2\mathcal{F}_1}(p)\nonumber\\
&=({\mathbf{F}_U} \odot {\mathbf{B}}_{E_1}(p)){\mathbf{H}}_{{\mathcal{F}}}(p)\mathbf{F}_U^H,
\end{align}
where ${h}_{\mathcal{F}_2E_1\mathcal{F}_1,p}^{\textrm{OAM}}(u,v)=\big({\mathbf{\mathfrak{{b}}}}_{E_1}(p)\odot\mathbf{f}(\ell_u)\big){\mathbf{H}}_{\mathcal{F}}(p){\mathbf{f}^{H}}({\ell_v})$ is the $u$th-row and $v$th-column element of $\mathbf{{H}}^{\rm OAM}_{\mathcal{F}_2E_1\mathcal{F}_1}(p)$, and can be calculated as
\begin{align}
&{h}_{\mathcal{F}_2E_1\mathcal{F}_1,p}^{\textrm{OAM}}(u,v)=\eta (p)\sum\limits_{m = 1}^N\sum\limits_{n = 1}^N
\exp\bigg(-i{\ell_u}\tilde\theta_m +i{\ell_v}\phi_n \nonumber\\
&+iS_{{k_p}}\sin\tilde\theta_m \cos\phi_n \sin\bar\psi\sin\bar\gamma\nonumber\\
&+iS_{{k_p}}(\cos\tilde\theta_m \cos\phi_n \cos\bar\gamma+\sin\tilde\theta_m \sin\phi_n\cos\bar\psi)\nonumber\\
&-ik_pR_r \left(\sin\tilde\theta_m \sin\bar\psi \cos\bar\gamma-\cos\tilde\theta_m \sin\bar\gamma \right)\nonumber\\
&+ik_pR_r \left(\sin\theta_m\sin\bar\psi\cos\bar\gamma-\cos\theta_m\sin\bar\gamma\right)\bigg)\nonumber\\
&\mathop \approx \limits^{(a)}\eta(p)\sum\limits_{m = 1}^N\sum\limits_{n = 1}^N\exp \bigg(-i\tilde\theta_mt+ik_pR_r\cos\tilde\theta_m \sin\bar\gamma\nonumber\\
&-ik_pR_r\sin\tilde\theta_m \sin\bar\psi \cos\bar\gamma + ik_pR_r\sin\theta_m \sin\bar\psi \cos\bar\gamma\nonumber\\
&-ik_pR_r\cos\theta_m \sin\bar\gamma\bigg)\times \sum\limits_{\delta = 1}^N\exp\bigg(i\left(\frac{2\pi \delta}{N}-\theta^*\right)\ell_v\nonumber\\
&+iS_{{k_p}}\cos\left(\frac{2\pi \delta}{N}-\theta^*\right)\bigg)
\label{GS46},
\end{align}
(a) applies the approximation $\cos a\approx1-\frac{a^2}{2}$ for $\cos\bar\psi$ and $\cos\bar\gamma$ in the case that $\bar\psi$ and $\bar\gamma$ are relatively small, and neglects a few small terms under the condition $r\gg R_r$.
\subsection{Adjusting Electronic Beam in Three Directions}
It can be seen from \eqref{GS46} that the effective multi-mode OAM channel matrix after roll rotation $\mathbf{{H}}^{\rm OAM}_{\mathcal{F}_2E_1\mathcal{F}_1}(p)$ is not a diagonal matrix any more, because the transmit and receive beam directions are not aligned after the roll rotation. In order to solve the reemerging inter-mode interferences, the electronic beam has to be adjusted in pitch, yaw and roll directions. The beam adjustment matrix ${\mathbf{B}}_{E_2}(p)$ could be designed as ${\mathbf{B}}_{E_2}(p) = \mathbf{1}\otimes {\mathbf{\mathfrak{{b}}}}_{E_2}(p)$, where ${\mathbf{\mathfrak{{b}}}}_{E_2}(p) = [{e^{i{\widehat{W}_1}(p)}},{e^{i{\widehat{W}_2}(p)}}, \cdots ,{e^{i{\widehat{W}_N}(p)}}]$ and ${\widehat{W}_m}(p)$ has the form
\begin{align}
&{\widehat{W}_m}(p)=2k_pR_r\sin\frac{\theta ^*}{2} \nonumber\\
&\times\left(\cos \bar \gamma\cos \left(\frac{\theta ^*}{2}+\theta_m\right)\sin\bar\psi+\sin\bar\gamma\sin\left(\frac{\theta^*}{2} +\theta_m\right)\right),
\end{align}
$m = 1, \cdots ,N,p = 1, \cdots ,P$.
After involving these phases into the original phases in ${\mathbf{F}_U}$ at the receive UCA, the effective multi-mode OAM channel matrix at the $p$th subcarrier becomes
\begin{align}
&{\mathbf{H}}_{E_2\mathcal{F}_2E_1\mathcal{F}_1}^{\rm OAM}(p)= {\mathbf{H}}_{E_2E_1\mathcal{F}_2\mathcal{F}_1}^{\rm OAM}(p) \nonumber\\
&=\left({\mathbf{F}_U}\odot{\mathbf{B}_{E_2}(p)} \odot{\mathbf{B}_{E_1}(p)}\right){\mathbf{H}}_{{\mathcal{F}}}(p){\mathbf{F}_U^H} \nonumber\\
&=\left({\mathbf{F}_U}\odot{\mathbf{B}_{E}(p)}\right){\mathbf{H}}_{{\mathcal{F}}}(p){\mathbf{F}_U^H}
={\mathbf{H}}_{H}^{\rm OAM}(p).
\end{align}
\begin{prop}
The successive implementation of the electronic beam steering matrices $\mathbf{B}_{E_1}(p)$ and $\mathbf{B}_{E_2}(p)$ can eliminate the inter-mode interferences in the channel matrix after mechanical beam steering $\mathbf{H}_{\mathcal{F}}$.
\label{PropI}
\end{prop}
\begin{IEEEproof}
Successive implementing $\mathbf{B}_{E_1}(p)$ and $\mathbf{B}_{E_2}(p)$ to $\mathbf{H}_{\mathcal{F}}$ leads to the $u$th-row and $v$th-column element of ${\mathbf{H}}_{H}^{\rm OAM}(p)$ ${h}_{E_2E_1\mathcal{F}_2\mathcal{F}_1,p}^{\textrm{OAM}}(u,v)={h}_{H,p}^{\textrm{OAM}}(u,v) =\big({\mathbf{\mathfrak{{b}}}}_{E_2}(p)\odot{\mathbf{\mathfrak{{b}}}}_{E_1}(p) \odot\mathbf{f}(\ell_u)\big){\mathbf{H}}_{{\mathcal{F}}}(p){\mathbf{f}^{H}}({\ell_v})$ being expressed as
\begin{align}
&{h}_{H,p}^{\textrm{OAM}}(u,v)=\eta (p)\sum\limits_{m = 1}^N\sum\limits_{n = 1}^N
\exp\bigg(-i{\ell_u}\tilde\theta_m +i{\ell_v}\phi_n \nonumber\\
&+iS_{{k_p}}\sin\tilde\theta_m \cos\phi_n \sin\bar\psi\sin\bar\gamma\nonumber\\
&+iS_{{k_p}}(\cos\tilde\theta_m \cos\phi_n \cos\bar\gamma+\sin\tilde\theta_m \sin\phi_n\cos\bar\psi)\nonumber\\
&-ik_pR_r \left(\sin\tilde\theta_m \sin\bar\psi \cos\bar\gamma-\cos\tilde\theta_m \sin\bar\gamma \right)\nonumber\\
&+ik_pR_r \left(\sin\theta_m\sin\bar\psi\cos\bar\gamma-\cos\theta_m\sin\bar\gamma\right)\nonumber\\
&+2ik_pR_r\sin\frac{\theta ^*}{2}\bigg(\cos \bar \gamma\cos \left(\frac{\theta ^*}{2}+\theta_m\right)\sin\bar\psi\nonumber\\
&+\sin\bar\gamma\sin\left(\frac{\theta^*}{2} +\theta_m\right)\bigg)\bigg)\nonumber\\
&\mathop \approx \limits^{(a)}\eta(p)\sum\limits_{m = 1}^N\exp \left(-i\tilde\theta_m t\right)\nonumber\\
&\times\sum\limits_{\delta = 1}^N\exp\left(i\left(\frac{2\pi \delta}{N}-\theta^*\right)\ell_v+iS_{{k_p}}\cos\left(\frac{2\pi \delta}{N}-\theta^*\right)\right),
\end{align}
where (a) applies the approximation $\cos a\approx1-\frac{a^2}{2}$ for $\cos\bar\psi$ and $\cos\bar\gamma$ in the case that, $\bar\psi$ and $\bar\gamma$ are relatively small, and neglects a few small terms under the condition $r\gg R_r$.
According to (\ref{GS33}), it is apparent that the $u$th diagonal element ${h}_{H,p}^{\textrm{OAM}}(u,u)$ of ${\mathbf{H}}_{H}^{\rm OAM}(p)$ can be expressed as
\vspace{1cm}
\begin{align}
&{h}_{H,p}^{\textrm{OAM}}(u,u)=N\eta (p) \nonumber\\
&\times\sum\limits_{\delta=1}^N {\exp\bigg({i\left(\frac{{2\pi\delta}}{N}-\theta^*\right){\ell_u}+ iS_{{k_p}}\cos\left(\frac{{2\pi\delta}}{N}-\theta^*\right)}\bigg)},
\end{align}
and the non-diagonal elements of ${\mathbf{H}}_{H}^{\rm OAM}(p)$ approximately equal to zero. It follows that the successive implementation of the electronic beam steering matrices $\mathbf{B}_{E_1}(p)$ and $\mathbf{B}_{E_2}(p)$ can eliminate the inter-mode interferences.
\end{IEEEproof}

\begin{rem}
According to the Proposition 3, the procedure of the hybrid mechanical and electronic beam steering for the LoS MCMM-OAM receiver can be simplified from the four steps $\mathcal{F}_1$, $E_1$, $\mathcal{F}_2$, $E_2$ illustrated in Fig. \ref{liucheng} to only the two steps: mechanical beam steering in pitch, yaw and roll directions $\mathcal{F}=\mathcal{F}_2\left(\mathcal{F}_1(\cdot)\right)$ and electronic beam steering in pitch, yaw and roll directions $\mathbf{B}_{E}(p)=\mathbf{B}_{E_2}(p)\odot \mathbf{B}_{E_1}(p)$.
\end{rem}
\subsection{Complexity Analysis}
\begin{figure}[t] 
\begin{center}
\includegraphics[scale=0.5]{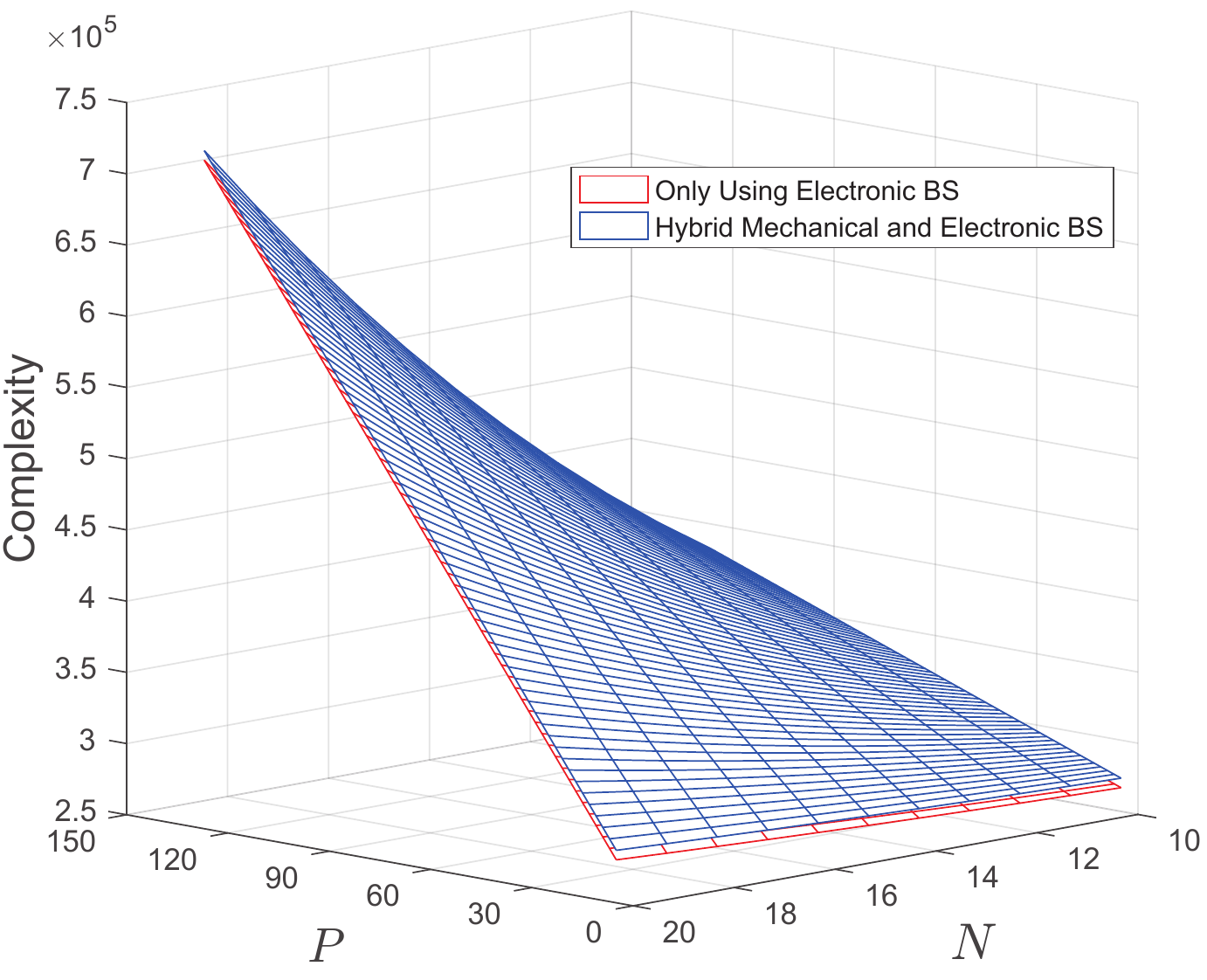}
\end{center}
\caption{The complexities of the hybrid mechanical and electronic beam steering and electronic beam steering. vs. $N$ and $P$ at $\overline{P}=4$, $\widetilde{P}=8$, $U=9$, $\overline{U}=4$, $\widetilde{U}=8$, $J=20$, $\zeta=0.9$, ${T_{\rm{min}}}=0.001$ and ${T_{\rm{init}}}=100$. BS: Beam steering.}
\label{Complexity}
\end{figure}

The specific computational complexity of the proposed hybrid mechanical and electronic beam steering scheme is compared with that of traditional electronic beam steering in the Table \ref{Table1}. For LoS MCMM-OAM communication systems, whether the coarse distance and AoA estimation method or the refined distance and AoA estimation method, their complexities are both determined by the complexity of the EVD in (16) of reference \cite{Chen2020Multi-mode}. However, the complexity of the coarse distance and AoA estimation method is affected by the values of $\overline P$ and $\overline U$, and factors that affect the complexity of the refined distance and AoA estimation method are the values of $\widetilde P$ and $\widetilde U$. Moreover, the complexity of the electronic beam steering is determined by the complexity of the multiplying the received signal vector by $\left(\mathbf{F}_U\odot\mathbf{B}_{E}(p)\right)$. It follows that the values of $P$, $U$ and $N$ decide the complexity of the electronic beam steering. The complexity of ${\mathcal{F}}_1$ is determined by the complexity of the mechanical rotation in pitch and yaw directions, and the value of $\hat\psi$, $\hat\gamma$ and $\nu$ decide the complexity of the ${\mathcal{F}}_1$. Meanwhile, the complexity of the ${\mathcal{F}}_2$ is determined by the complexity of the mechanical rotation in roll direction corresponding to the value of $\theta^*$ and $\nu$. The complexity of the optimization algorithm is determined by the values of $J$, $\zeta$, ${T_{\rm{init}}}$ and ${T_{\rm{min}}}$. The total computational complexity comparison between the hybrid mechanical and electronic beam steering and electronic beam steering scheme is shown in the Fig. \ref{Complexity}. It can be seen from the Fig. \ref{Complexity} that due to the coarse AoA estimation using fewer OAM modes and subcarriers, mechanical beam steering and optimization having relatively low complexity, the total complexity of the proposed hybrid beam steering scheme is only slightly higher than that of traditional electronic beam steering. Furthermore, we calculate the relative complexity from the proposed hybrid beam steering to electronic beam steering as $1 + \mathcal{O}\left(\frac{{{\overline P}^3}{{\overline U}^3}}{{{\widetilde P}^3}{{\widetilde U}^3}}\right)$. To show their difference, we select parameters as $\overline{P}=4$, $\widetilde{P}=8$, $U=9$, $\overline{U}=4$, $\widetilde{U}=8$, $J=20$, $\zeta=0.9$, ${T_{\rm{min}}}=0.001$, ${T_{\rm{init}}}=100$, $\nu=0.3^\circ$, $\psi=\gamma=60^\circ$, $N=10$, and $\theta^{*}=10^\circ$. With these parameters, the complexity of the hybrid beam steering is 1.009 to 1.026 times that of the electronic beam steering.
\begin{table}[t]
\small
\setlength{\abovecaptionskip}{0cm}   
\setlength{\belowcaptionskip}{0cm}   
\caption{The complexity comparison between hybrid mechanical and electronic beam steering and electronic beam steering. BS: Beam steering.}
\begin{center}
\setlength{\tabcolsep}{2mm}{
\begin{tabular}{cll}
  \toprule
  \textbf{Scheme}                                                 &\multicolumn{2}{c}{\textbf{Complexity}}\\
  \midrule
  \multirow{6}{*}{\makecell[c]{Hybrid BS}}
                                                                  &Coarse AoA estimation                                                                   &$\mathcal{O}\left(\overline{P}^3\overline{U}^3\right)$\\
                                                                  \cline{2-3}
                                                                  &Mechanical beam steering ${\mathcal{F}}_1$
                                                                  &$\mathcal{O}\left(\frac{\hat\psi+\hat\gamma}{\nu}\right)$\\
                                                                  \cline{2-3}
                                                                  &Optimization algorithm
                                                                  &$\mathcal{O}\left(J{\log _\zeta}\frac{T_{\min }}{T_{\max}}\right)$\\
                                                                  \cline{2-3}
                                                                  &Mechanical beam steering ${\mathcal{F}}_2$                                                                   &$\mathcal{O}\left(\frac{\theta^*}{\nu}\right)$\\
                                                                  \cline{2-3}
                                                                  &Refined AoA estimation                                                                   &$\mathcal{O}\left(\widetilde{P}^3\widetilde{U}^3\right)$\\
                                                                  \cline{2-3}
                                                                  &Electronic beam steering       &$\mathcal{O}\left(PUN^2\right)$\\
  \midrule
  \multirow{2}{*}{Electronic BS}                                  &AoA estimation
                                                                  &$\mathcal{O}\left(\widetilde{P}^3\widetilde{U}^3\right)$\\
                                                                  \cline{2-3}
                                                                  &Electronic beam steering      &$\mathcal{O}\left(PUN^2\right)$\\
  \bottomrule
  \label{Table1}
\end{tabular}}
\end{center}
\vspace{-1em}
\end{table}
\begin{figure}[t] 
\begin{center}
\includegraphics[width=8cm,height=7cm]{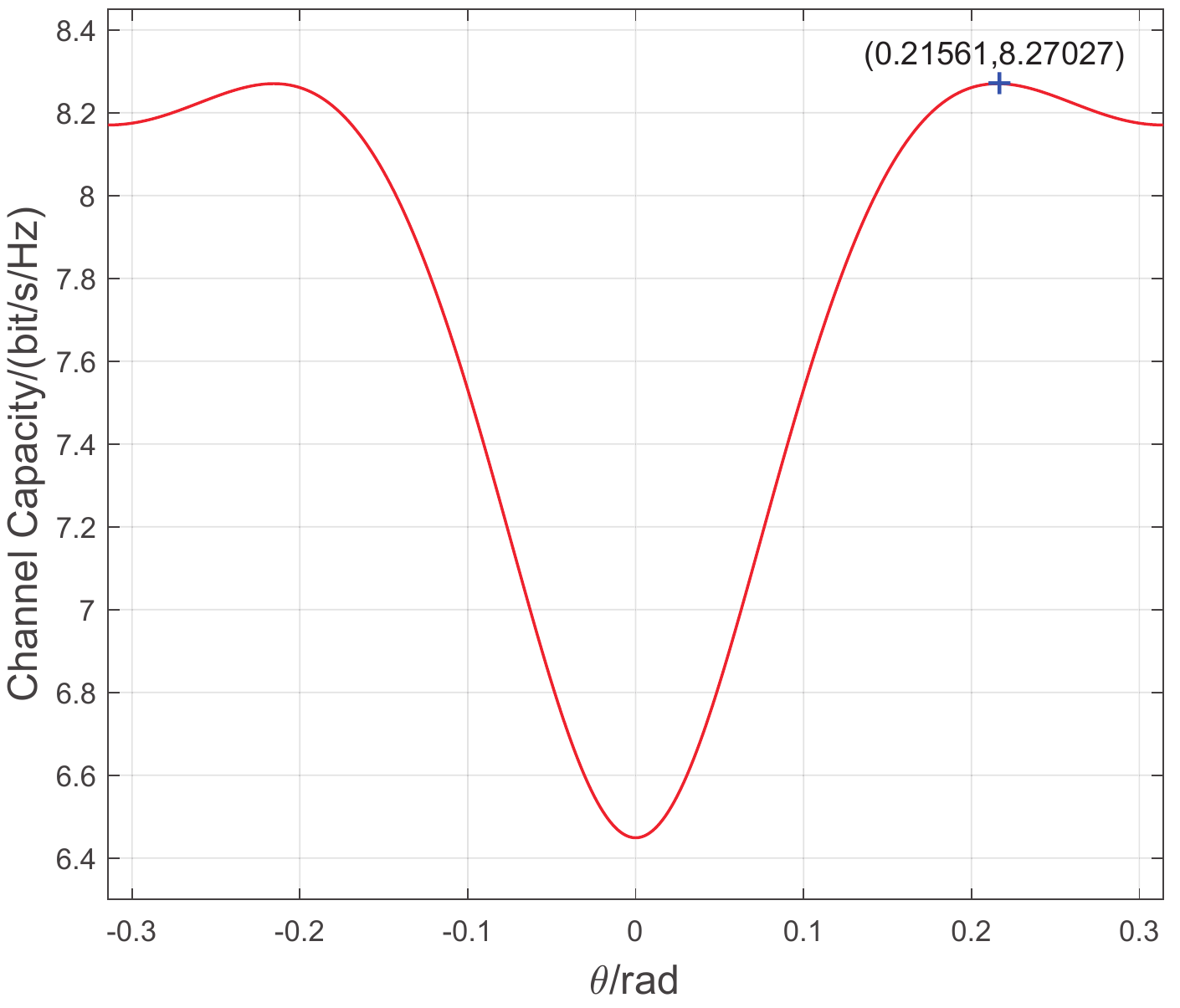}
\end{center}
\caption{The obtained optimal rotation angle $\theta^*$ by Algorithm 1 in the range of $[-\pi/10,\pi/10]$.}
\label{fig6}
\end{figure}
\begin{figure}[t] 
\begin{center}
\includegraphics[width=8cm,height=7cm]{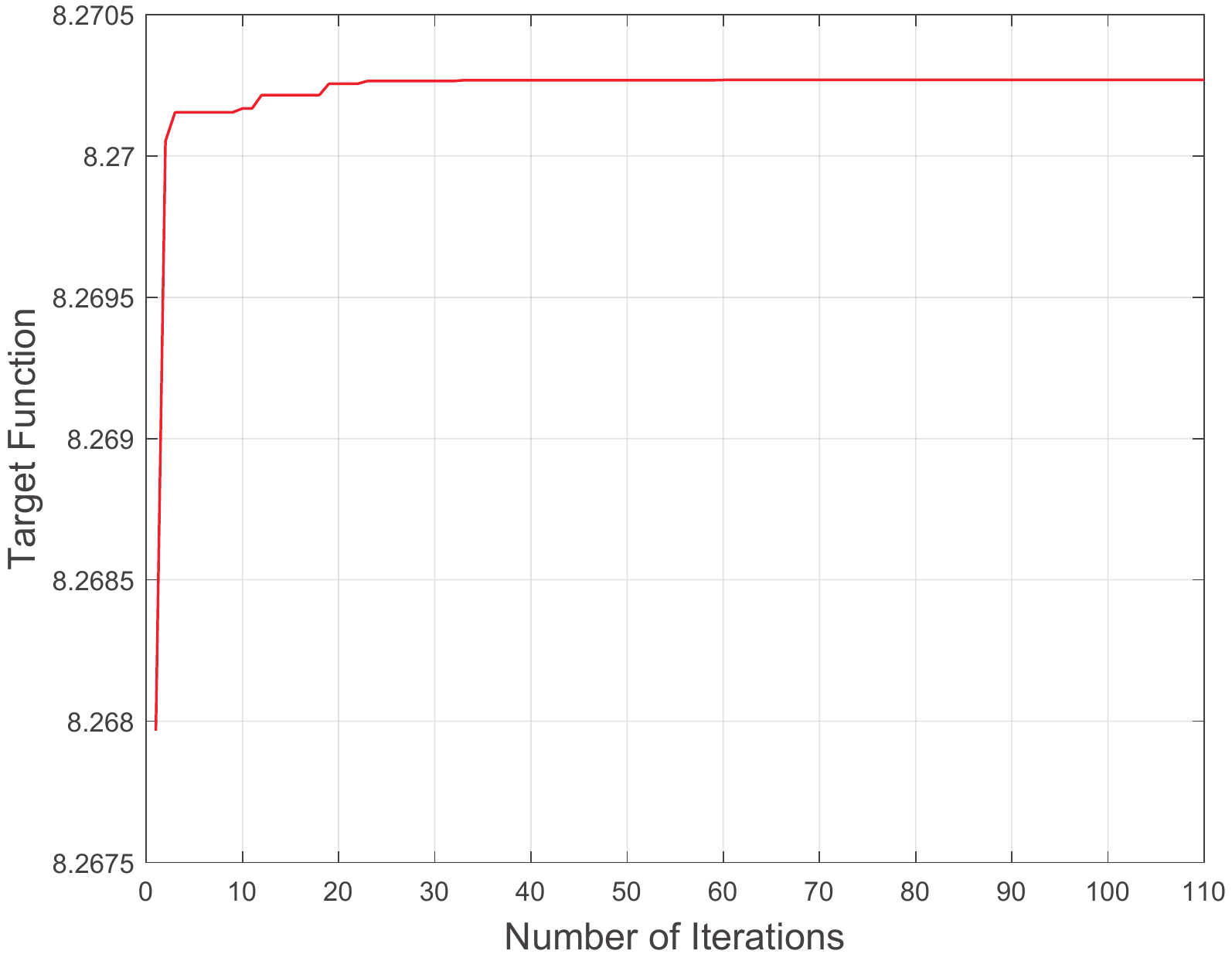}
\end{center}
\caption{The convergence rate of the Algorithm 1.}
\label{diedai}
\end{figure}

\section{Numerical Simulations And Results}
In this section we show the performance of the proposed hybrid mechanical and electronic beam steering method by numerical simulations. We first verify the effectiveness of SA algorithm in solving the optimal solution $\theta^\ast$, and show the convergence rate of the algorithm. Then, we compare the channel capacities of the UCA-based LoS MCMM-OAM system with the proposed hybrid beam steering and that with only electronic beam steering. Unless otherwise stated, the SNRs in all the figures are defined as the ratio of the received signal power versus the noise power.
In the simulations, we choose $P = 8$ subcarriers from $3.9982$ GHz to $4.2387$ GHz, $\overline{P}=4$, $\widetilde{P}=8$, $U = 9$ with ${\ell_v} \in [ - 4,4]$, $\overline{U}=4$, $\widetilde{U}=8$, ${R_r} = {R_t} = 20\lambda_1$, $r = 450\lambda_1$, $\lambda_1=2\pi/k_1$, $N=10$, $\nu=0.3^\circ$.

Fig.\ref{fig6} shows the solution of the optimization problem \eqref{h}. It can be seen from the figure that the channel capacity with obtained rotation angle $\theta$ by Algorithm 1 reaches its optimum.
In Fig.\ref{diedai}, we show the convergence rate of the Algorithm 1 with the number of outer-layer iterations, where ${T_{\rm{init}}}=100$, ${T_{\rm{min}}}=10^{-3}$ and $\zeta=0.9$. It can be clearly seen that through reasonably setting the parameters of SA algorithm, the target function has a fast convergence rate.

\begin{figure}
\begin{center}
\includegraphics[width=8cm,height=7cm]{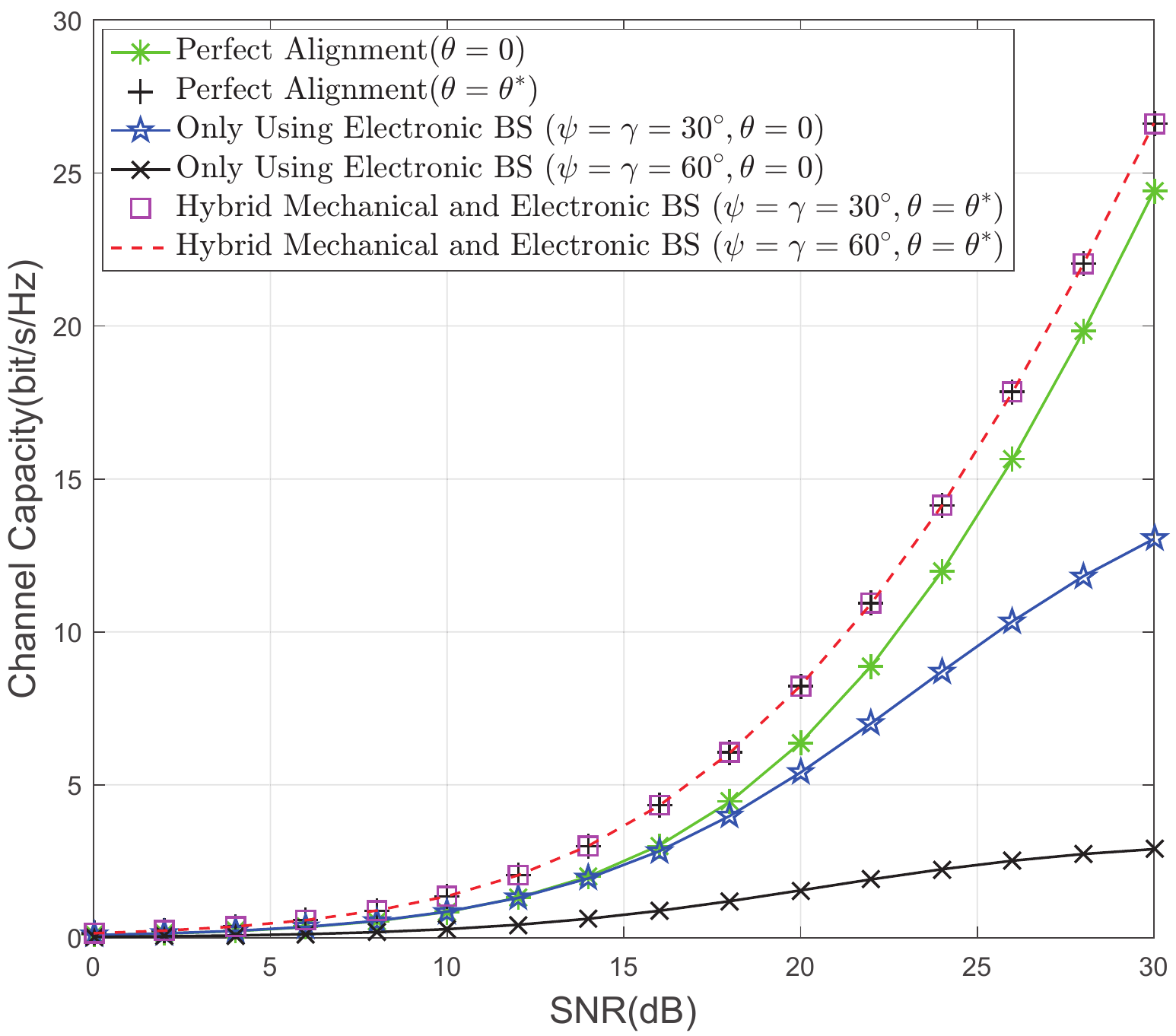}
\end{center}
\caption{The channel capacity of the UCA-based LoS MCMM-OAM system. BS: Beam steering.}
\label{fig4}
\end{figure}

In Fig.\ref{fig4}, we illustrate the channel capacities of the proposed hybrid mechanical and electronic beam steering scheme and the channel capacities of only using electronic beam steering method under different pitch, yaw and roll rotation angles. Also, the channel capacities in the case of perfect alignment is given for clearer comparison. It can be seen from the results that for the electronic beam steering method, the larger the oblique angles, the lower the channel capacity, while the proposed hybrid mechanical and electronic beam steering scheme can effectively eliminate the effect of large misalignment errors of any practical OAM channel and approaches the performance of perfect aligned OAM channel.

Although we performed simulations at 4GHz, the proposed hybrid mechanical and electronic beam steering scheme is applicable to all frequency bands due to that the channel capacity in \eqref{C27} is dependent on the two terms $\eta (p)$ and $S_{k_p}$ related to frequency in \eqref{C26}. Recall that $\eta(p)=\frac{\beta}{{2{k_p}rN}}\exp(-i{k_p}r)$, $S_{k_p}=k_p\frac{R_rR_t}{r}$ and $k_p=\frac{2\pi}{\lambda_p}$, we set $R_t$, $R_r$ and $r$ to be different times of $\lambda_1$ in the simulations. As the difference of $\lambda_1$ and $\lambda_p$ is so small within a certain bandwidth that the values of $\eta(p)$ and $S_{k_p}$ do not vary much with frequency. The simulation results in higher frequencies used in 5G and 6G are similar to those at 4GHz, which are omitted here due to limited space.

\section{Conclusions}
In this paper, a hybrid mechanical and electronic beam steering scheme is proposed for LoS multi-carrier and multi-mode OAM communication systems to deal with the capacity loss encountered by electronic beam steering in large misalignment condition. To intuitively reveal the motivation of the proposed hybrid beam steering scheme, we show from both mathematical proof and numerical simulations that only using the electronic beam steering method, the larger the oblique angles, the lower the receive SINR and OAM channel capacity.
In our proposed hybrid beam steering scheme, mechanical rotating devices controlled by PWM signals are utilized to eliminate the large misalignment angle, while electronic beam steering is in charge of the remaining small misalignment angle caused by perturbations. The whole procedure is designed as four consecutive steps including mechanical beam steering in pitch and yaw directions, electronic beam steering in pitch and yaw directions, mechanical rotation in roll direction and adjusting electronic beam in pitch, yaw and roll directions. As the physical beam steering in mechanical way is nonlinear operation for channel matrix, while electronic beam steering in digital way is linear operation, we show that the four steps of hybrid beam steering can be simplified to only two steps, i.e., mechanical beam steering in pitch, yaw and roll directions and electronic beam steering in these three directions. Numerical simulations show that searching for the optimal roll rotation angle only requires no more than 30 iterations with the SA algorithm, and the proposed hybrid beam steering scheme can effectively eliminate the effect of large misalignment errors of any practical OAM channel approaching the performance of perfect aligned OAM channel.

It is worth noting that for the OAM communication system in the large off-axis and more general misalignment cases, the proposed hybrid mechanical and electronic beam steering scheme is still applicable with the same idea in \cite{Chen2018Beam} that beam steering is performed at both receiver and transmitter. However, the specific procedure needs to be further developed, which can also be used in mmWave massive MIMO systems.
Moreover, the directivity of the antenna units will affect the rotation angle $\theta$. Although in this manuscript the antenna elements are assumed to be ideally omnidirectional, we believe that it is of practical significance to study how the directivity of the antenna units affect the rotation angle $\theta$, which is left for our future work.

\renewcommand\thefigure{\Alph{section}\arabic{figure}}
\renewcommand\thefigure{A\arabic{figure}}
\setcounter{figure}{0}
\begin{figure}
\setlength{\abovecaptionskip}{-0cm}   
\setlength{\belowcaptionskip}{-0.2cm}   
\begin{center}
\includegraphics[scale=0.89]{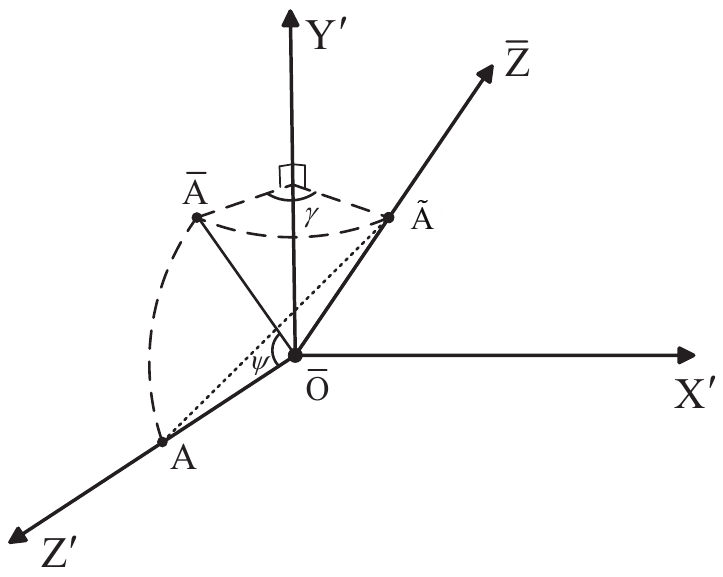}
\end{center}
\caption{Geometrical relationship between $\alpha$, $\psi$ and $\gamma$.}
\label{A1}
\end{figure}
\setcounter{figure}{0}
\renewcommand{\theequation}{B.\arabic{equation}}
\begin{figure*}
\setcounter{equation}{2}
\begin{align}
|\textrm{DB}|=\sqrt{|\textrm{OB}|^2-|\textrm{OD}|^2} =\frac{1}{2}\sqrt {4 + 3{r^2} - r\bigg( {2r\cos (2\gamma ){{\cos }^2}\psi  + r\cos (2\psi ) + 8\cos \gamma \sin \psi } \bigg)}.\label{B3}
\end{align}
\hrulefill
\setcounter{equation}{3}
\end{figure*}
\renewcommand{\appendixname}{Appendix A}
\appendix
\setcounter{equation}{0}
\renewcommand\theequation{\Alph{section}\arabic{equation}}

In order to prove the relationship between $\psi$, $\gamma$ and $\alpha$, we illustrate the geometrical relationship between $\alpha$, $\psi$ and $\gamma$ in Fig. \ref{A1}. We assume that A is the point on the ${\textrm{Z}'}$-axis with its cartesian coordinate $(0,0,1)$ in the coordinate system ${\textrm{Z}'}-{\textrm{X}'}\overline{\textrm{O}}{\textrm{Y}'}$, $\overline{\textrm A}$ is the point that A rotates around $\textrm{X}'$-axis the angel $\psi$ to. Thus, the cartesian coordinate of $\overline{\textrm A}$ is ${\mathbf{R}}_{P}(\psi)
[0,0,1]^T$.
$\widetilde{\textrm A}$ is the point that $\overline{\textrm A}$ rotates around $\textrm{Y}'$-axis the angel $\gamma$ to. Thus, the cartesian coordinate of $\widetilde{\textrm A}$ is ${\mathbf{R}}_{Y}(\gamma){\mathbf{R}}_{P}(\psi)
[0,0,1]^T$.
Therefore, the distance between A and $\widetilde{\textrm A}$ can be obtained as
\renewcommand{\theequation}{A.\arabic{equation}}
\begin{align}
\left|\textrm{A}\widetilde{\textrm{A}}\right|&=\sqrt{\left(\widetilde{A}_x\right)^2+\left(\widetilde{A}_y\right)^2+\left(\widetilde{A}_z-1\right)^2} \nonumber\\
&=\sqrt{2(1-\cos\psi\cos\gamma)}.
\end{align}
According to the law of cosines, $\left|\textrm{A}\widetilde{\textrm{A}}\right|^2=\left|\textrm{A}\overline{\textrm{O}}\right|^2+ \left|\widetilde{\textrm{A}}\overline{\textrm{O}}\right|^2-2\left|\textrm{A}\overline{\textrm{O}}\right| \left|\widetilde{\textrm{A}}\overline{\textrm{O}}\right|\cos\angle\textrm{A}\overline{\textrm{O}}\widetilde{\textrm{A}}$. Thus, $\angle\textrm{A}\overline{\textrm{O}}\widetilde{\textrm{A}}$ can be expressed as
\begin{align}
\angle\textrm{A}\overline{\textrm{O}}\widetilde{\textrm{A}}&=\arccos\left(\frac{2-|\textrm{A}\widetilde{\textrm{A}}|^2}{2}\right) \nonumber\\
&=\arccos\left(\cos \psi \cos \gamma\right).
\end{align}
Because the cartesian coordinate of the point $\widetilde{\textrm A}$ in the coordinate system $\overline{\textrm{Z}}-\overline{\textrm{X}}\overline{\textrm{O}}\overline{\textrm{Y}}$ can be easily obtained as $(0,0,1)$, $\widetilde{\textrm A}$ is on the $\overline{\textrm Z}$-axis. Therefore, the angle $\alpha$ between $\overline{\textrm Z}$-axis and ${\textrm Z}$-axis is $\angle\textrm{A}\overline{\textrm{O}}\widetilde{\textrm{A}}$. Hence, the relationship between the $\psi$, $\alpha$ and $\gamma$ can be obtained as $\cos \alpha  = \cos \psi \cos \gamma$.
\vspace{-0.4cm}
\renewcommand{\appendixname}{Appendix B}
\setcounter{equation}{0}
\renewcommand{\theequation}{B.\arabic{equation}}
\appendix
In order to prove \eqref{PHI}, we first magnify the geometrical model of the receive UCA of Fig. \ref{fig2} when $- \frac{\pi}{2}\!\!<\gamma\!\!< 0$ as Fig. \ref{B1}(a). In the coordinate system ${\textrm{Z}}-{\textrm{X}}{\textrm{O}}{\textrm{Y}}$, the cartesian coordinates of the point $\overline{\textrm{O}}$ and ${\textrm{O}}$ are $(0,0,r)$ and $(0,0,0)$, respectively. We assume that B is the point on the $\overline{\textrm Y}$-axis, $\overline{\textrm{B}}$ is the point on the $\overline{\textrm X}$-axis, and their cartesian coordinates in the coordinate system $\overline{\textrm{Z}}-\overline{\textrm{X}}\overline{\textrm{O}}\overline{\textrm{Y}}$ are $(0,1,0)$ and $(1,0,0)$, respectively. So, the cartesian coordinates of B and $\overline{\textrm{B}}$ are $[{\sin\psi \sin \gamma},{\cos \psi},{r +\sin \psi\cos \gamma}]^T$ and $[{\cos\gamma}, 0, {r - \sin\gamma}]^T$ in the coordinate system ${\textrm{Z}}-{\textrm{X}}{\textrm{O}}{\textrm{Y}}$, respectively.
With the cartesian coordinates of $\overline{\textrm{O}}$, $\overline{\textrm{B}}$ and ${\textrm{B}}$, we can obtain the plane equation of the receive UCA ${ax+by+cz+d=0}$, where $a=-\cos \psi \sin \gamma$, $b=\sin\psi$, $c=-\cos \gamma \cos \psi$, and $d=r\cos \gamma \cos \psi$.
Thus, the distance from the point O to the plane of receive UCA can be written as
\begin{align}
|\textrm{OD}| =\frac{|d|}{\sqrt{a^2 + b^2 + c^2}}=r\cos\psi\cos\gamma.\label{B4}
\end{align}
With the cartesian coordinates of B and ${\textrm{O}}$, the distance between O and B can be obtained as
\begin{align}
|\textrm{OB}|&=\sqrt {B_x^2 + B_y^2 + B_z^2}=\sqrt {1 + {r^2} - 2r\cos \gamma \sin \psi }.\label{B4}
\end{align}
Since the line OD is perpendicular to the plane of the receive UCA, it is perpendicular to the line DB. Hence, OD, DB and OB form a right triangle, where OD and DB are two legs and OB is hypotenuse.
Therefore, the distance from D to B can be expressed as (\ref{B3}). With the cartesian coordinates of $\overline{\textrm{O}}$ and ${\textrm{B}}$, we can calculate $|\overline{\textrm{O}}{\textrm{B}}|$ as
\begin{align}
\setcounter{equation}{2}
&|\overline{\textrm{O}}\textrm{B}|=\sqrt {B_x^2 + B_y^2 + (B_z-r)^2}=1.
\setcounter{equation}{3}
\end{align}
\renewcommand\thefigure{\Alph{section}\arabic{figure}}
\renewcommand\thefigure{B\arabic{figure}}
\setcounter{figure}{0}
\begin{figure}
\setlength{\abovecaptionskip}{-0cm}   
\setlength{\belowcaptionskip}{-0.2cm}   
\centering
\subfigure[]{
\includegraphics[scale=0.4]{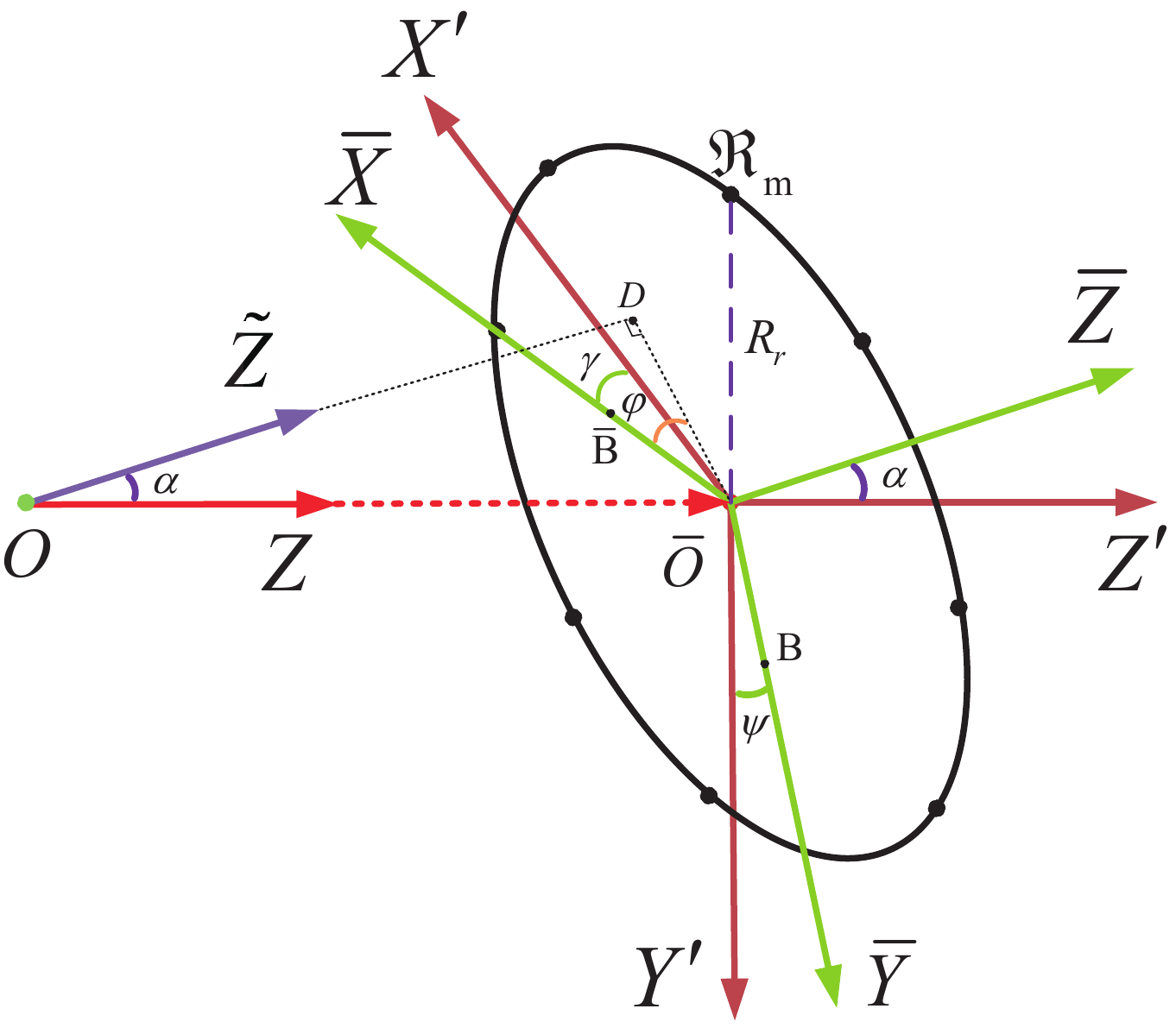}
}
\quad
\subfigure[]{
\includegraphics[scale=0.4]{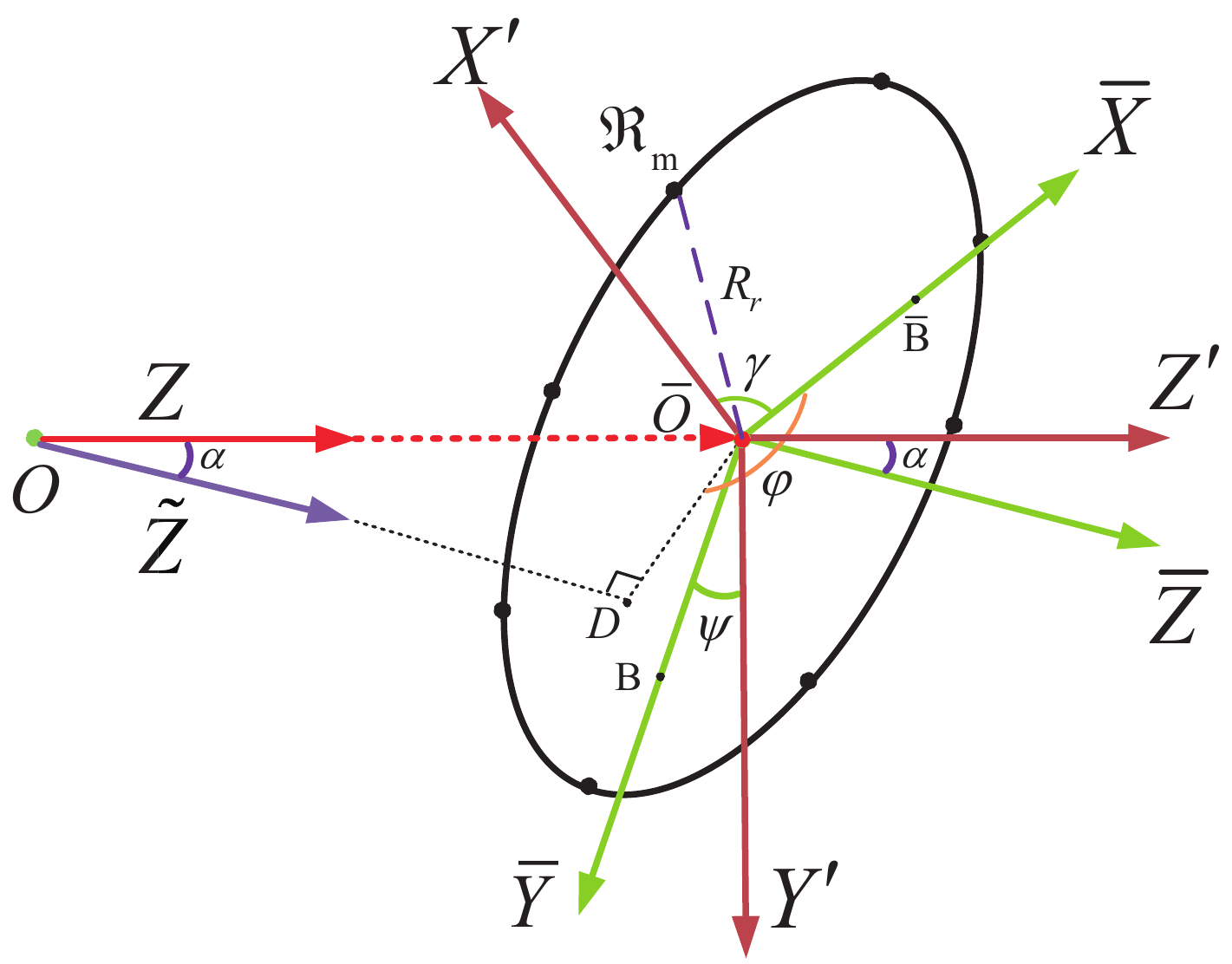}
}
\caption{The geometrical model of the receive UCA before the pitch and yaw rotation when (a) $-\frac{\pi}{2}\!\!<\gamma\!\!<0$, (b) $0\!\!<\gamma\!\!<\frac{\pi}{2}$.}
\label{B1}
\end{figure}

Similarly, the line OD is perpendicular to $\overline{\textrm{O}}\textrm{D}$ due to OD being perpendicular to the plane of the receive UCA. Hence, OD, $\overline{\textrm{O}}\textrm{D}$ and $\overline{\textrm{O}}\textrm{O}$ form a right triangle, where OD and $\overline{\textrm{O}}\textrm{D}$ are two legs and $\overline{\textrm{O}}\textrm{O}$ is hypotenuse. Hence, $|\overline{\textrm{O}}{\textrm{D}}|$ can be written as
\begin{align}
&|\overline{\textrm{O}}{\textrm{D}}|=\sqrt {|\overline{\textrm{O}}\textrm{O}{|^2} - |\textrm{O}\textrm{D}{|^2}}=r\sin\alpha.
\end{align}
According to the law of cosines, $\left|\textrm{D}\textrm{B}\right|^2 = \left|\overline{\textrm{O}}\textrm{D}\right|^2 + \left|\overline{\textrm{O}}\textrm{B}\right|^2 - 2\left|\overline{\textrm{O}}\textrm{D}\right|\left|\overline{\textrm{O}}\textrm{B}\right|\cos\angle \textrm{D}\overline{\textrm{O}}\textrm{B}$. Therefore,
\begin{align}
\setcounter{equation}{4}
\angle \textrm{D}\overline{\textrm{O}}{\textrm{B}} = \arccos \left( {\frac{{2\cos \gamma \sin \psi }}{{\sqrt {3 - 2\cos (2\gamma ){{\cos }^2}\psi  - \cos (2\psi )} }}} \right).
\setcounter{equation}{5}
\end{align}
Since B is on the $\overline{\textrm{Y}}$-axis while $\overline{\textrm{B}}$ is on the $\overline{\textrm{X}}$-axis, $\angle\overline{\textrm{B}}\overline{\textrm{O}}{\textrm{B}}=\frac{\pi}{2}$. Meanwhile, $\angle \overline{\textrm{B}}\overline{\textrm{O}}{\textrm{B}}=\angle {\textrm{D}}\overline{\textrm{O}}{\textrm{B}}+\overline{\textrm{B}}\overline{\textrm{O}}{\textrm{D}}$ and $\angle\overline{\textrm{B}}\overline{\textrm{O}}{\textrm{D}}=\varphi$, Therefore, $\varphi$ can be obtained as $\varphi=\frac{\pi }{2} - \angle \textrm{D}\overline{\textrm{O}}{\textrm{B}}$ when $- \frac{\pi}{2}\!\!<\gamma\!\!< 0$.

Similarly, when $0\!\!<\gamma\!\!<\frac{\pi}{2}$, the geometrical model of the receive UCA is shown in Fig. \ref{B1}(b). We can prove in the same way that $\varphi=\frac{\pi}{2}+\angle \textrm{D}\overline{\textrm{O}}{\textrm{B}}$ when $0\!\!<\gamma\!\!<\frac{\pi}{2}$.

\vspace{-0.6cm}
\renewcommand{\appendixname}{Appendix C}
\appendix
\setcounter{equation}{0}
\renewcommand{\theequation}{C.\arabic{equation}}
On one hand, when $\psi=0$ and $\gamma\ne0$, the $u$th diagonal element $\tilde{h}_{\textrm{OAM},p}(u,u)$ of $\widetilde{\mathbf{H}}_{\rm OAM}(p)$ can be obtained as
\begin{align}
\tilde{h}&_{\textrm{OAM},p}(u,u)=\eta(p)\nonumber\\
&\times\sum\limits_{\delta= 1}^N \exp\bigg(i\frac{{2\pi \delta}}{N}{\ell _u} + \frac{1}{2}i{{S_{{k_p}}}}(1 + \cos \gamma )\cos \frac{{2\pi \delta}}{N}\bigg)\varsigma(m),
\end{align}
where $\varsigma(m)$ is defined as
\begin{align}
\setcounter{equation}{1}
\varsigma(m)=\sum\limits_{m = 1}^N \exp\bigg(\frac{1}{2}i{{S_{{k_p}}}}(\cos \gamma-1)\cos \frac{{2\pi (2m + \delta - 2)}}{N}\bigg).\label{C2}
\setcounter{equation}{1}
\end{align}
When $S_{{k_p}}$ goes to zero, $\varsigma(m)\approx N$.
Therefore, $\tilde{h}_{\textrm{OAM},p}(u,u)$ can be approximated as
\begin{align}
\tilde{h}&_{\textrm{OAM},p}(u,u)\nonumber\\
\approx &N\eta (p)\sum\limits_{\delta= 1}^N {\exp\bigg(i\frac{{2\pi \delta}}{N}{\ell _u} + \frac{1}{2}i{{S_{{k_p}}}}(1 + \cos \gamma )\cos \frac{{2\pi \delta}}{N}\bigg)}.\label{C3}
\end{align}
According to \cite{Chen2020Multi-mode}, we can substitute the Taylor series expansion of the exponential function ${e^x} = \sum\nolimits_{g = 0}^{ + \infty} {({x^g})/g!}$ into \eqref{C3} and eventually have $\tilde{h}_{\textrm{OAM},p}(u,u)\approx \eta (p)\frac{{{N^2}}}{{{2^\tau }}}\cdot \frac{{{i^\tau }}}{{\tau !}}$ ${\bigg(\frac{1}{2}{S_{{k_p}}}(1 + \cos \gamma )\bigg)^\tau }$,
where $\tau={\min\left\{|\ell_u|, N-|\ell_u|\right\}}$.
\renewcommand{\theequation}{C.\arabic{equation}}
\begin{figure*}
\begin{align}
\setcounter{equation}{4}
\bar \varsigma(m)&= \sum\limits_{m = 1}^N {\exp \bigg( - i\frac{{2\pi (m - 1)}}{N}t\bigg)\exp \bigg(\frac{1}{2}i{S_{{k_p}}}(\cos\gamma-1)\cos \frac{{2\pi (2m + \delta  - 2)}}{N}\bigg)} \nonumber\\
&\!=\!\sum\limits_{g = 0}^{ + \infty }\!{\frac{{{i^g}}}{{{2^g}g!}}}\!\sum\limits_{l = 0}^{\rm{g}}\!\bigg(
\begin{matrix}
g\\
l
\end{matrix}
\bigg)\exp \bigg(i\frac{{2\pi t}}{N}\!+\!i(2l\!-\!g)\frac{{2\pi \left( {\delta\!-\!2} \right)}}{N}\bigg)\!\!\sum\limits_{m = 1}^N\!{\exp \bigg(i(2l\!-\!g\!-\!\frac{t}{2})\frac{{2\pi (2m)}}{N}\bigg)} {\bigg(\frac{1}{2}{S_{{k_p}}}( \cos\gamma\!-\!1)\bigg)^g}.\label{C7}
\setcounter{equation}{5}
\end{align}
\hrulefill
\end{figure*}

On the other hand, when $\psi=0$ and $\gamma\ne0$, the $u$th-row and $v$th-column element $\tilde{h}_{\textrm{OAM},p}(u,v)$ of $\widetilde{\mathbf{H}}_{\rm OAM}(p)$ can be obtained as
\begin{align}
&\tilde{h}_{\textrm{OAM},p}(u,v)=\eta (p)\nonumber\\
&\times\sum\limits_{\delta= 1}^N {\exp\bigg(i\frac{{2\pi \delta}}{N}{\ell _v} + \frac{1}{2}i{{S_{{k_p}}}}(1 + \cos \gamma )\cos \frac{{2\pi \delta}}{N}\bigg)}\bar\varsigma(m),\label{C5}
\setcounter{equation}{3}
\end{align}
where $\bar\varsigma(m)$ is defined as
\begin{align}
\bar\varsigma(m)=&\sum\limits_{m = 1}^N \exp\bigg(\frac{1}{2}i{{S_{{k_p}}}}\cos \gamma\cos \frac{{2\pi (2m + \delta - 2)}}{N}\nonumber
\end{align}
\begin{align}
&-\frac{1}{2}i{{S_{{k_p}}}}\cos \frac{{2\pi (2m + \delta- 2)}}{N}-i\frac{{2\pi (m - 1)}t}{N}\bigg).\label{C6}
\end{align}
We substitute ${e^x} = \sum\nolimits_{g = 0}^{ + \infty} {({x^g})/g!}$ into $\bar\varsigma(m)$ and obtain \eqref{C7}.
For the last summation in (\ref{C7}), it is easy to verify that
\begin{align}
\setcounter{equation}{5}
&\sum\limits_{m = 1}^N\exp\bigg(i(2l-g-\frac{t}{2})\frac{4\pi m}{N}\bigg)=\left\{
\begin{smallmatrix}
0, &\textrm{Mod}_N(2l-g-\frac{t}{2})\neq0;\\
N, &\textrm{Mod}_N(2l-g-\frac{t}{2})=0.
\end{smallmatrix}\right.\label{C8}
\setcounter{equation}{6}
\end{align}
When $0\leq g\leq\min\left\{|\frac{t}{2}|, N-|\frac{t}{2}|\right\}$, we always have (recall that $0 \leq l \leq g$) $-N<2l-g-\frac{t}{2}< N$ according to (34) of \cite{Chen2020Multi-mode}.
When $0\leq g< \min\left\{|\frac{t}{2}|, N-|\frac{t}{2}|\right\}$, (\ref{C8}) is always zero, i.e., the Taylor series of $\tilde{h}_{\textrm{OAM},p}(u,v)$ does not contain the terms $S^g_{k_p}$, and when $g=\min\left\{|\frac{t}{2}|, N-|\frac{t}{2}|\right\}$, (\ref{C8}) equals to $N$ if and only if $l=g (t\geq0)$; $l=0 (t<0)$.
Furthermore, when $g>\min\left\{|\frac{t}{2}|, N-|\frac{t}{2}|\right\}$, the coefficient of $S^g_{k_p}$ is much smaller due to the term $\frac{1}{2^gg!}$, and hence the term $\{S^g_{k_p}|g>\min\left\{|\frac{t}{2}|, N-|\frac{t}{2}|\right\}$ is counted into the higher order infinitesimal of $S^g_{k_p}$ denoted as $o\{S^g_{k_p}\}$ as $S_{k_p}$ goes to zero. Thus, (\ref{C6}) can be approximated as
\begin{align}
\bar \varsigma(m)\approx \frac{{{N}}}{{{2^{\bar\tau} }}} \cdot \frac{{{i^{\bar\tau} }}}{{{\bar\tau} !}}\exp \bigg(i\frac{{\pi{\delta}t}}{N}\bigg){\bigg(\frac{1}{2}{S_{{k_p}}}(\cos \gamma-1)\bigg)^{\bar\tau} },\label{C11}
\end{align}
where $\bar\tau=\min\left\{|\frac{t}{2}|, N-|\frac{t}{2}|\right\}$. Therefore, (\ref{C5}) can be approximated as
\begin{align}
\tilde{h}_{\textrm{OAM},p}(u,v)
&\approx\eta (p)\frac{{{N}}}{{{2^{\bar\tau} }}} \cdot \frac{{{i^{\bar\tau}}}}{{{\bar\tau} !}}{\bigg(\frac{1}{2}{S_{{k_p}}}(\cos \gamma-1)\bigg)^{\bar\tau} }\nonumber\\
&\times\sum\limits_{\delta= 1}^N {\exp\bigg(i\frac{{2\pi \delta}}{N}({\ell _v}+\frac{t}{2}) + \frac{1}{2}i{{S_{{k_p}}}}\cos \frac{{2\pi \delta}}{N}}\nonumber\\
&+{\frac{1}{2}i{{S_{{k_p}}}}\cos \gamma\cos \frac{{2\pi \delta}}{N}\bigg)}.\label{C12}
\end{align}
Then, similar to the approximation of the summation term with Taylor series expansion in \eqref{C3}, \eqref{C12} can be further approximated as
\begin{align}
\tilde{h}_{\textrm{OAM},p}(u,v)
\approx&\eta (p)(-1)^{\bar\tau}\frac{{{N^2}}}{{{4^{({\bar\tau}+\chi)} }}} \cdot \frac{{{i^{({\bar\tau}+\chi)}}}}{{{\bar\tau}!\chi!}}\cdot{S_{{k_p}}}^{({\bar\tau}+\chi)}\nonumber\\
&\times{(1-\cos \gamma)^{\bar\tau} }{(1 + \cos \gamma )^\chi},\label{C13}
\end{align}
where $\chi=\min\left\{|{\ell_v}+\frac{t}{2}|, N-|{\ell_v}+\frac{t}{2}|\right\}$.

In order to analyze how $\textrm{SIR}_{EO}(p,u)$ changes with the increase of $|\gamma|$ when $\psi=0$, we first express the reciprocal of ${{\textrm{SIR}}_{EO}(p,u)}$ when ${\ell_u}\ne 0$ as
\begin{align}
&\frac{1}{{\textrm{SIR}}_{EO}(p,u)}= \frac{\sum\limits_{u \ne v} {{{\left| \tilde{h}_{\textrm{OAM},p}(u,v) \right|}^2}} }{{\left| \tilde{h}_{\textrm{OAM},p}(u,u) \right|}^2}\nonumber\\
=&{\frac{{{\left| \tilde{h}_{\textrm{OAM},p}(u,v) \right|}^2}}{{\left| \tilde{h}_{\textrm{OAM},p}(u,u) \right|}^2}}\Bigg{|}_{\ell _v=0}
+\sum\limits_{{\ell _u}{\ell _v} < 0}\frac{{{\left| \tilde{h}_{\textrm{OAM},p}(u,v) \right|}^2}}{{\left| \tilde{h}_{\textrm{OAM},p}(u,u) \right|}^2}\nonumber\\
+&\sum\limits_{{\ell _u}{\ell _v} > 0\atop u \ne v}\frac{{{\left| \tilde{h}_{\textrm{OAM},p}(u,v) \right|}^2}}{{\left| \tilde{h}_{\textrm{OAM},p}(u,u) \right|}^2}
.\label{C14}
\end{align}

It is obvious that \eqref{C14} consists of three terms and the derivative of \eqref{C14} is the sum of their derivatives. Moreover, the derivative of $\frac{{{\left| \tilde{h}_{\textrm{OAM},p}(u,v) \right|}^2}}{{\left| \tilde{h}_{\textrm{OAM},p}(u,u) \right|}^2}$ is given as
\setcounter{equation}{11}
\begin{align}
\frac{\partial\left(\frac{{\left| \tilde{h}_{\textrm{OAM},p}(u,v) \right|}^2}{{\left| \tilde{h}_{\textrm{OAM},p}(u,u) \right|}^2}\right)}{\partial\gamma }=T(u,v,\gamma)\left(\kappa(u,v,\gamma)+ \bar\kappa(u,v,\gamma)\right),\label{C15}
\end{align}
\setcounter{equation}{12}
where
\begin{align}
T(u,v,\gamma)&=\frac{{\frac{1}{{{4^{2(\bar \tau  + \chi )}}}} \cdot \frac{1}{{{{(\bar \tau !\chi !)}^2}}} \cdot {S_{{k_p}}}^{2(\bar \tau  + \chi )}}\sin \gamma}{{\frac{1}{{{4^{2\tau }}}} \cdot \frac{{{S_{{k_p}}}^{2\tau }}}{{{{(\tau !)}^2}}}}{{(1 + \cos \gamma )}^{4\tau }}},\label{C16}
\end{align}
\begin{align}
\kappa(u,v,\gamma)=&{{(1 - \cos \gamma )}^{2\bar \tau  - 1}}{{(1 + \cos \gamma )}^{2\chi+2\tau - 1}}\nonumber\\
&\times\big(2(\bar \tau  - \chi ) + 2(\bar \tau  + \chi)\cos \gamma\big),\label{C17}
\end{align}
\begin{align}
\bar\kappa(u,v,\gamma)&=2\tau{{{(1 - \cos \gamma )}^{2\bar \tau }}{{(1 + \cos \gamma )}^{2\chi+2\tau-1}}},\label{C18}
\end{align}
$\bar\tau  - \chi =\frac{{|{\ell _u} - {\ell _v}|}}{2}- \frac{{|{\ell _u} + {\ell _v}|}}{2}$ and $\bar\tau  + \chi  = \frac{{|{\ell _u} - {\ell _v}|}}{2} + \frac{{|{\ell _u} + {\ell _v}|}}{2}$.

When $0<\gamma\le\frac{\pi}{2}$, $T(u,v,\gamma)>0$, thus the sign of the derivative of $\frac{{{\left| \tilde{h}_{\textrm{OAM},p}(u,v) \right|}^2}}{{\left| \tilde{h}_{\textrm{OAM},p}(u,u) \right|}^2}$ depends on $\kappa(u,v,\gamma)$ and $\bar\kappa(u,v,\gamma)$.
We can find that when ${\ell _v}=0$, $\bar\tau  - \chi =0$ and $\bar\tau +\chi = |{\ell _u}|$, thus $\kappa(u,v,\gamma)+\bar\kappa(u,v,\gamma)$ can be written as
\begin{align}
\kappa(u,v,\gamma)+\bar\kappa(u,v,\gamma)=&{(1 - \cos \gamma )}^{2\bar \tau  - 1}{{(1 + \cos \gamma )}^{2\chi+2\tau-1}}\nonumber\\
&\times\big(2(|{\ell _u}|-\tau)\cos\gamma+2\tau\big).\label{C19}
\end{align}
Because $|{\ell _u}| \ge \tau $, $\kappa(u,v,\gamma)+\bar\kappa(u,v,\gamma) > 0$. Hence, the derivative of ${\frac{{{\left| \tilde{h}_{\textrm{OAM},p}(u,v) \right|}^2}}{{\left| \tilde{h}_{\textrm{OAM},p}(u,u) \right|}^2}}\bigg{|}_{\ell _v=0}$ is greater than zero so that ${\frac{{{\left| \tilde{h}_{\textrm{OAM},p}(u,v) \right|}^2}}{{\left| \tilde{h}_{\textrm{OAM},p}(u,u) \right|}^2}}\bigg{|}_{\ell _v=0}$ is increasing with $\gamma$; when ${{\ell _u}{\ell _v} < 0}$, $\bar\tau  - \chi=\min \{ |{\ell _u}|,|{\ell _v}|\}$ and $\bar\tau  + \chi=\max \{ |{\ell _u}|,|{\ell _v}|\}$, thus $\kappa(u,v,\gamma)+\bar\kappa(u,v,\gamma)$ can be written as
\begin{align}
\kappa&(u,v,\gamma)+\bar\kappa(u,v,\gamma)\nonumber\\
=&{(1 - \cos \gamma )}^{2\bar \tau  - 1}{{(1 + \cos \gamma )}^{2\chi+2\tau-1}}\times\bigg(2(\min \{ |{\ell _u}|,|{\ell _v}|\})\nonumber\\
&+2(\max \{ |{\ell _u}|,|{\ell _v}|\})\cos \gamma+ 2\tau{(1 - \cos \gamma )}\bigg).\label{C20}
\end{align}
It can be seen from \eqref{C20} that $\kappa(u,v,\gamma)+\bar\kappa(u,v,\gamma) > 0$. Therefore, the derivatives of every term of $\sum_{{\ell _u}{\ell _v} < 0}\frac{{{\left| \tilde{h}_{\textrm{OAM},p}(u,v) \right|}^2}}{{\left| \tilde{h}_{\textrm{OAM},p}(u,u) \right|}^2}$ are greater than zero so that the term $\sum_{{\ell _u}{\ell _v} < 0}\frac{{{\left| \tilde{h}_{\textrm{OAM},p}(u,v) \right|}^2}}{{\left| \tilde{h}_{\textrm{OAM},p}(u,u) \right|}^2}$ is increasing with $\gamma$; when ${{\ell _u}{\ell _v} >0}$, $\bar\tau  - \chi=-\min \{ |{\ell _u}|,|{\ell _v}|\}$ and $\bar\tau  + \chi=\max \{ |{\ell _u}|,|{\ell _v}|\}$. We can find that when $|{\ell _u}| < |{\ell _v}|$, $\kappa(u,v,\gamma)+\bar\kappa(u,v,\gamma)$ can be expressed as
\begin{align}
&\kappa(u,v,\gamma)+\bar\kappa(u,v,\gamma)\nonumber\\
&=2(1 - \cos \gamma)^{2\bar \tau  - 1}{{(1 + \cos \gamma )}^{2\chi+2\tau-1}}\cos\gamma(|{\ell _v}|-|{\ell_u}|),
\end{align}
when $|{\ell _u}| > |{\ell _v}|$, $\kappa(u,v,\gamma)+\bar\kappa(u,v,\gamma)$ can be expressed as
\begin{align}
&\kappa(u,v,\gamma)+\bar\kappa(u,v,\gamma)\nonumber\\
&=2(1 - \cos \gamma)^{2\bar\tau-1}{{(1 + \cos \gamma )}^{2\chi+2\tau-1}}(|{\ell _u}|-|{\ell _v}|).\label{C23}
\end{align}
Thus, when ${{\ell _u}{\ell _v} >0}$, $\kappa(u,v,\gamma)+\bar\kappa(u,v,\gamma)>0$. Therefore, the derivatives of every term of $\sum_{{\ell _u}{\ell _v} > 0\atop u \ne v}\frac{{{\left| \tilde{h}_{\textrm{OAM},p}(u,v) \right|}^2}}{{\left| \tilde{h}_{\textrm{OAM},p}(u,u) \right|}^2}$ are greater than zero so that $\sum_{{\ell _u}{\ell _v} > 0\atop u \ne v}\frac{{{\left| \tilde{h}_{\textrm{OAM},p}(u,v) \right|}^2}}{{\left| \tilde{h}_{\textrm{OAM},p}(u,u) \right|}^2}$ is increasing with $\gamma$.

When ${\ell_u}=0$, the reciprocal of ${{\textrm{SIR}}_{EO}(p,u)}$ can be written as
\begin{align}
\frac{1}{{\textrm{SIR}}_{EO}(p,u)}=\sum\limits_{{\ell _v} < 0}\frac{{{\left| \tilde{h}_{\textrm{OAM},p}(u,v) \right|}^2}}{{\left| \tilde{h}_{\textrm{OAM},p}(u,u) \right|}^2}+\sum\limits_{{\ell _v} > 0}\frac{{{\left| \tilde{h}_{\textrm{OAM},p}(u,v) \right|}^2}}{{\left| \tilde{h}_{\textrm{OAM},p}(u,u) \right|}^2}.\label{C23}
\end{align}
It is obvious that \eqref{C23} consists of two terms and the derivative of \eqref{C23} is the sum of their derivatives. Moreover, when ${\ell _u}=0$, $\bar\kappa(u,v,\gamma)=0$.
When ${\ell _v}< 0$ or ${\ell _v}> 0$, $\bar\tau  - \chi =0$ and $\bar\tau +\chi = |{\ell _v}|$, thus
\begin{align}
&\kappa(u,v,\gamma)+\bar\kappa(u,v,\gamma)=\kappa(u,v,\gamma)\nonumber\\
&=2|{\ell _v}|\cos \gamma{{(1 - \cos \gamma )}^{2\bar \tau  - 1}}{{(1 + \cos \gamma )}^{2\chi+2\tau - 1}}.\label{C24}
\end{align}
Hence, the derivatives of every term of $\sum_{{\ell _v} < 0}\frac{{{\left| \tilde{h}_{\textrm{OAM},p}(u,v) \right|}^2}}{{\left| \tilde{h}_{\textrm{OAM},p}(u,u) \right|}^2}$ and $\sum_{{\ell _v} > 0}\frac{{{\left| \tilde{h}_{\textrm{OAM},p}(u,v) \right|}^2}}{{\left| \tilde{h}_{\textrm{OAM},p}(u,u) \right|}^2}$ are
greater than zero so that $\sum_{{\ell _v} < 0}\frac{{{\left| \tilde{h}_{\textrm{OAM},p}(u,v) \right|}^2}}{{\left| \tilde{h}_{\textrm{OAM},p}(u,u) \right|}^2}$ and $\sum_{{\ell _v} > 0}\frac{{{\left| \tilde{h}_{\textrm{OAM},p}(u,v) \right|}^2}}{{\left| \tilde{h}_{\textrm{OAM},p}(u,u) \right|}^2}$ are increasing with $\gamma$.

Therefore, we get the conclusion that ${{\textrm{SIR}}_{EO}(p,u)}$ is decreasing with $\gamma$ when $0<\gamma\le\frac{\pi}{2}$. On the contrary, we can prove in the same way that ${{\textrm{SIR}}_{EO}(p,u)}$ is increasing with $\gamma$ when $-\frac{\pi}{2}\le\gamma<0$. Overall, the Proposition 1 that ${{\textrm{SIR}}_{EO}(p,u)}$ is decreasing with $|\gamma|$ is proved.

\bibliographystyle{IEEEtran}
\bibliography{IEEEabrv,desprit}

\begin{thebibliography}{10}
\providecommand{\url}[1]{#1}
\csname url@samestyle\endcsname
\providecommand{\newblock}{\relax}
\providecommand{\bibinfo}[2]{#2}
\providecommand{\BIBentrySTDinterwordspacing}{\spaceskip=0pt\relax}
\providecommand{\BIBentryALTinterwordstretchfactor}{4}
\providecommand{\BIBentryALTinterwordspacing}{\spaceskip=\fontdimen2\font plus
\BIBentryALTinterwordstretchfactor\fontdimen3\font minus
  \fontdimen4\font\relax}
\providecommand{\BIBforeignlanguage}[2]{{%
\expandafter\ifx\csname l@#1\endcsname\relax
\typeout{** WARNING: IEEEtran.bst: No hyphenation pattern has been}%
\typeout{** loaded for the language `#1'. Using the pattern for}%
\typeout{** the default language instead.}%
\else
\language=\csname l@#1\endcsname
\fi
#2}}
\providecommand{\BIBdecl}{\relax}
\BIBdecl

\bibitem{Tataria20216}
H.~Tataria, M.~Shafi, A.~F. Molisch, M.~Dohler, H.~Sj{\"{O}}land, and
  F.~Tufvesson, ``{6G} wireless systems: Vision, requirements, challenges,
  insights, and opportunities,'' \emph{Proc. IEEE}, vol. 109, no.~7, pp.
  1166--1199, 2021.

\bibitem{Allen1992Orbital}
L.~Allen, M.~W. Beijersbergen, R.~J.~C. Spreeuw, and J.~P. Woerdman, ``Orbital
  angular momentum of light and the transformation of {Laguerre-Gaussian} laser
  modes,'' \emph{Phys. Rev. A, Gen. Phys.}, vol.~45, no.~11, pp. 8185--8189,
  Jun. 1992.

\bibitem{Tamburini2012Encoding}
F.~Tamburini, E.~Mari, A.~Sponselli, B.~Thid\'e, A.~Bianchini, and F.~Romanato,
  ``Encoding many channels on the same frequency through radio vorticity:
  {First} experimental test,'' \emph{New J. Phys.}, vol.~14, no.~3, p. 033001,
  2012.

\bibitem{Yan2014High}
Y.~{Yan}{\ }\emph{et\ al}., ``High-capacity millimetre-wave communications with
  orbital angular momentum multiplexing,'' \emph{Nature Commun.}, vol.~5,
  no.~1, p. 4876, Dec. 2014.

\bibitem{Ren2017Line}
Y.~{Ren}{\ }\emph{et\ al}., ``Line-of-sight millimeter-wave communications
  using orbital angular momentum multiplexing combined with conventional
  spatial multiplexing,'' \emph{{IEEE} Trans. Wireless Commun.}, vol.~16,
  no.~5, pp. 3151--3161, May 2017.

\bibitem{Zhang2017Mode}
W.~Zhang, S.~Zheng, X.~Hui, R.~Dong, X.~Jin, H.~Chi, and X.~Zhang, ``Mode
  division multiplexing communication using microwave orbital angular momentum:
  An experimental study,'' \emph{{IEEE} Trans. Wireless Commun.}, vol.~16,
  no.~2, pp. 1308--1318, Feb. 2017.

\bibitem{Chen2018Beam}
R.~Chen, H.~Xu, M.~Moretti, and J.~Li, ``Beam steering for the misalignment in
  {UCA}-based {OAM} communication systems,'' \emph{IEEE Wireless Commun.
  Lett.}, vol.~7, no.~4, pp. 582--585, Aug. 2018.

\bibitem{Chen2018A}
R.~Chen, W.~Yang, H.~Xu, and J.~Li, ``A {2-D} {FFT}-based transceiver
  architecture for {OAM-OFDM} systems with {UCA} antennas,'' \emph{{IEEE}
  Trans. Veh. Technol.}, vol.~67, no.~6, pp. 5481--5485, Jun. 2018.

\bibitem{Zhang2019Orbital}
C.~{Zhang} and Y.~{Zhao}, ``Orbital angular momentum nondegenerate index
  mapping for long distance transmission,'' \emph{{IEEE} Trans. Wireless
  Commun.}, vol.~18, no.~11, pp. 5027--5036, Nov. 2019.

\bibitem{Chen2020Orbital}
R.~{Chen}, H.~{Zhou}, M.~{Moretti}, X.~{Wang}, and J.~{Li}, ``Orbital angular
  momentum waves: Generation, detection and emerging applications,'' \emph{IEEE
  Commun. Surveys Tuts.}, vol.~22, no.~2, pp. 840--868, 2nd Quart., 2020.

\bibitem{Chen2020Multi-mode}
R.~{Chen}, W.-X. Long, X.~{Wang}, and J.~Li, ``Multi-mode {OAM} radio waves:
  Generation, angle of arrival estimation and reception with {UCAs},''
  \emph{{IEEE} Trans. Wireless Commun.}, vol.~19, no.~10, pp. 6932--6947, Oct.
  2020.

\bibitem{Long2021AoA}
W.-X. Long, R.~Chen, M.~Moretti, and J.~Li, ``{AoA} estimation for {OAM}
  communication systems with mode-frequency multi-time {ESPRIT} method,''
  \emph{IEEE Trans. Veh. Technol.}, vol.~99, no.~1, Apr. 2021.

\bibitem{Long2021Joint}
W.-X. Long, R.~Chen, M.~Moretti, J.~Xiong, and J.~Li, ``Joint spatial division
  and coaxial multiplexing for downlink multi-user {OAM} wireless backhaul,''
  \emph{IEEE Trans. Broadcasting}, vol.~67, no.~4, pp. 1--15, Dec. 2021.

\bibitem{Zhang2021Orbital}
X.~{Zhang}{\ }\emph{et\ al}., \emph{Orbital Angular Momentum Based Structured
  Radio Beams and its Applications}.\hskip 1em plus 0.5em minus 0.4em\relax
  John Wiley \& Sons, Ltd, 2021, ch.~9, pp. 269--293.

\bibitem{Xiong2020Performance}
X.~Xiong, S.~Zheng, Z.~Zhu, X.~Yu, X.~Jin, and X.~Zhang, ``Performance analysis
  of plane spiral {OAM} mode-group based {MIMO} system,'' \emph{IEEE Commun.
  Lett.}, vol.~24, no.~7, pp. 1414--1418, 2020.

\bibitem{Trichili2019Communicating}
A.~Trichili, K.-H. Park, M.~Zghal, B.~S. Ooi, and M.-S. Alouini,
  ``Communicating using spatial mode multiplexing: Potentials, challenges, and
  perspectives,'' \emph{IEEE Commun. Surveys Tuts.}, vol.~21, no.~4, pp.
  3175--3203, 2019.

\bibitem{Tian2021Broadband}
Z.~Tian, R.~Chen, W.-X. Long, H.~Zhou, and M.~Moretti, ``Broadband beam
  steering for misaligned multi-mode {OAM} communication systems,'' \emph{J.
  Syst. Eng. Electron.}, vol.~32, no.~4, pp. 779--788, Aug. 2021.

\bibitem{Liang2016Orbital}
J.~{Liang} and S.~{Zhang}, ``Orbital angular momentum {(OAM)} generation by
  cylinder dielectric resonator antenna for future wireless communications,''
  \emph{IEEE Access}, vol.~4, pp. 9570--9574, Apr. 2016.

\bibitem{Zhang2017Four}
W.~Zhang, S.~Zheng, X.~Hui, Y.~Chen, X.~Jin, H.~Chi, and X.~Zhang,
  ``Four-{OAM}-mode antenna with traveling-wave ring-slot structure,''
  \emph{IEEE Antennas Wireless Propag. Lett.}, vol.~16, pp. 194--197, 2017.

\bibitem{Tennant2012Generation}
A.~Tennant and B.~Allen, ``Generation of radio frequency {OAM} radiation modes
  using circular time-switched and phased array antennas,'' in \emph{Proc.
  Loughborough Antennas Propag. Conf. (LAPC)}, Nov. 2012, pp. 1--4.

\bibitem{Mohammadi2010system}
S.~M. Mohammadi, L.~K.~S. Daldorff, J.~E.~S. Bergman, R.~L. Karlsson, B.~Thide,
  K.~Forozesh, T.~D. Carozzi, and B.~Isham, ``Orbital angular momentum in
  radio\rule[1.6pt]{0.3cm}{0.05em}{A} system study,'' \emph{{IEEE} Trans.
  Antennas Propag.}, vol.~58, no.~2, pp. 565--572, Feb. 2010.

\bibitem{Shen2018Generating}
Y.~{Shen}, J.~{Yang}, H.~{Meng}, W.~{Dou}, and S.~{Hu}, ``Generating
  millimeter-wave {Bessel} beam with orbital angular momentum using
  reflective-type metasurface inherently integrated with source,'' \emph{Appl.
  Phys. Lett.}, vol. 112, no.~14, p. 141901, Apr. 2018.

\bibitem{Jian2021Non}
M.~{Jian}, Y.~{Chen}, and G.~{Yu}, ``Non-coaxial {OAM}: Precoding design,
  misaligned parameter estimation and capacity compensation,'' \emph{IEEE
  Access}, vol.~9, pp. 37\,726--37\,738, 2021.

\bibitem{Yagi2021Wireless}
Y.~Yagi, H.~Sasaki, T.~Yamada, and D.~Lee, ``200 {Gb}/s wireless transmission
  using dual-polarized {OAM-MIMO} multiplexing with uniform circular array on
  28 {GHz} band,'' \emph{IEEE Antennas and Wireless Propag. Lett.}, vol.~20,
  no.~5, pp. 833--837, 2021.

\bibitem{Rahmat1979Useful}
Y.~Rahmat-Samii, ``Useful coordinate transformations for antenna
  applications,'' \emph{IEEE Trans. Antennas Propagat.}, vol.~27, no.~4, pp.
  571--574, 1979.

\bibitem{Tajuddin2009TMS320F2812}
M.~F.~N. Tajuddin, N.~H. Ghazali, M.~F. Mohammed, B.~Ismail, Z.~M. Isa, T.~C.
  Siong, and N.~Ghazali, ``{TMS320F2812} digital signal processor {(DSP)}
  implementation of {DPWM},'' in \emph{Proc. IEEE Student Conf. Res. Develop.
  (SCOReD)}, 2009, pp. 142--145.

\bibitem{Pinckney2006Pulse}
N.~Pinckney, ``Pulse-width modulation for microcontroller servo control,''
  \emph{IEEE Potentials}, vol.~25, no.~1, pp. 27--29, 2006.

\bibitem{Xavier2010Coupled}
S.~Xavier-de Souza, J.~A.~K. Suykens, J.~Vandewalle, and D.~Bolle, ``Coupled
  simulated annealing,'' \emph{IEEE Trans. Syst. Man Cybern. B}, vol.~40,
  no.~2, pp. 320--335, 2010.

\bibitem{Abido2000obust}
M.~Abido, ``Robust design of multimachine power system stabilizers using
  simulated annealing,'' \emph{IEEE Trans. Energy conversion}, vol.~15, no.~3,
  pp. 297--304, 2000.

\end{thebibliography}
\begin{IEEEbiography}[{\includegraphics[width=1in,height=1.25in,clip,keepaspectratio]{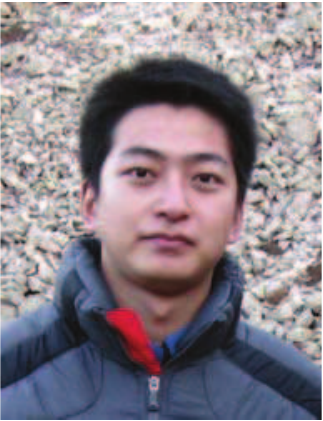}}]{Rui Chen}
(S'08-M'11) received the B.S., M.S. and Ph.D. degrees in Communications and Information Systems from Xidian University, Xi'an, China, in 2005, 2007 and 2011, respectively. From 2014 to 2015, he was a visiting scholar at Columbia University in the City of New York. He is currently an associate professor and Ph.D. supervisor in the school of Telecommunications Engineering at Xidian University. He has published more than 80 papers in international journals and conferences and held 40 patents. He is an Associate Editor for International Journal of Electronics, Communications, and Measurement Engineering (IGI Global). His research interests include broadband wireless communication systems, array signal processing and intelligent transportation systems.
\end{IEEEbiography}

\begin{IEEEbiography}[{\includegraphics[width=1in,height=1.25in,clip,keepaspectratio]{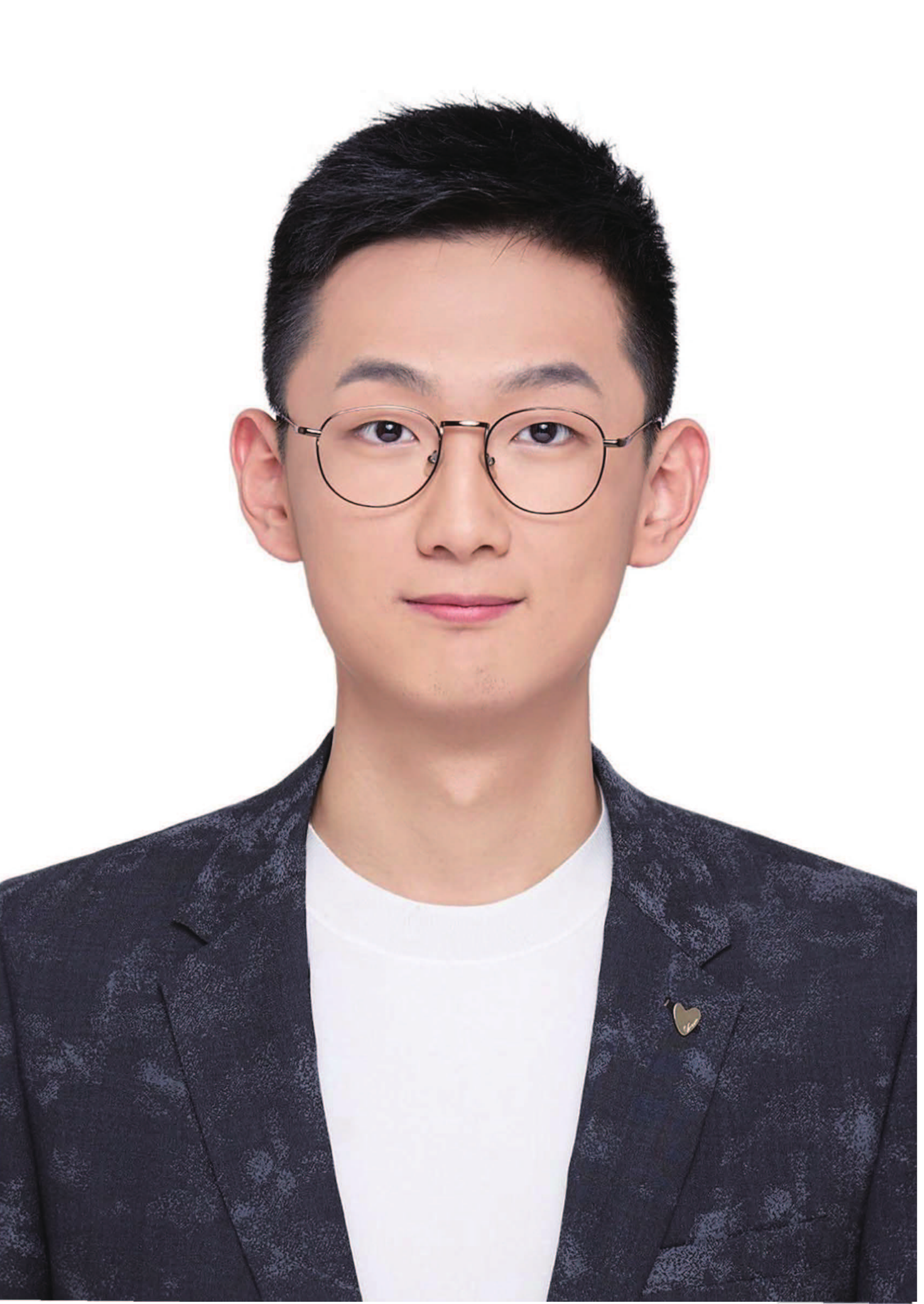}}]{Zhenyang Tian}
(S'18) received the B.S. degree in Electronical Information Science and Technology from the Shanxi University, Taiyuan, China, in 2020. He is currently pursuing the M.S. degree in communications and information systems with Xidian University. His research interests include orbital angular momentum (OAM) communication systems and array signal processing.
\end{IEEEbiography}

\begin{IEEEbiography}[{\includegraphics[width=1in,height=1.25in,clip,keepaspectratio]{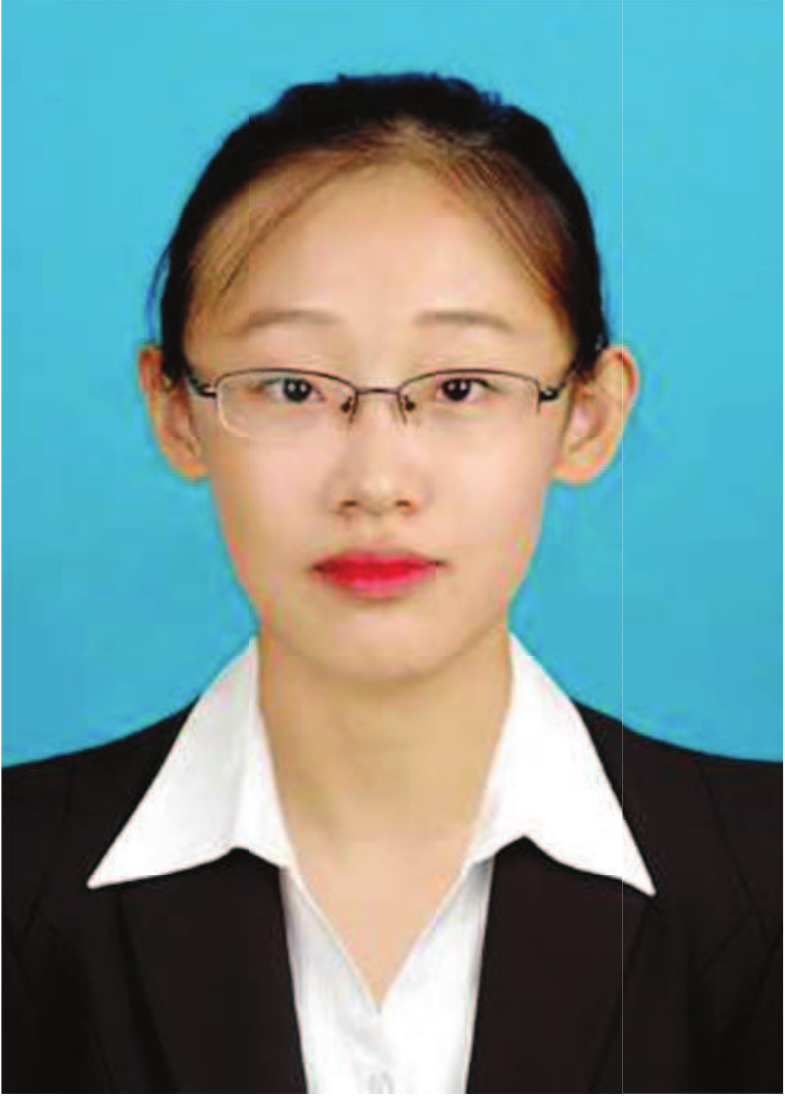}}]{Wen-Xuan Long}
(S'18) received the B.S. degree (with Highest Hons.) in Rail Transit Signal and Control from Dalian Jiaotong University, Dalian, China in 2017. She is currently pursuing a double Ph.D. degree in Communications and Information Systems at Xidian University, China and University of Pisa, Italy. Her research interests include broadband wireless communication systems and array signal processing.
\end{IEEEbiography}

\begin{IEEEbiography}[{\includegraphics[width=1in,height=1.25in,clip,keepaspectratio]{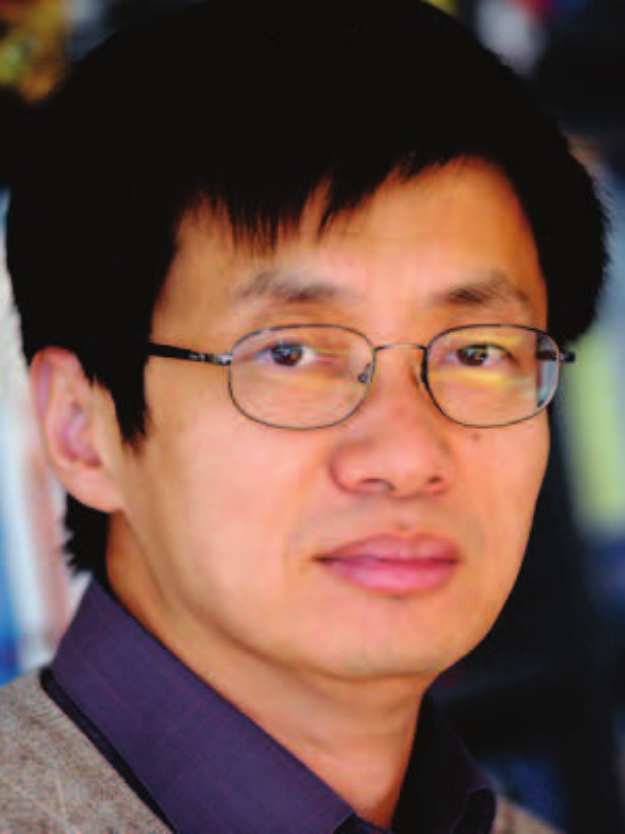}}]{Xiaodong Wang}
(S'98-M'98-SM'04-F'08) received the Ph.D. degree in electrical engineering from Princeton University. He is currently a Professor of electrical engineering with Columbia University, New York, NY, USA. He has authored the book Wireless Communication Systems: Advanced Techniques for Signal Reception (Prentice-Hall, 2003). His research interests include computing, signal processing, and communications, and has published extensively in these areas. His current research interests include wireless communications, statistical signal processing, and genomic signal processing. He was a recipient of the 1999 NSF CAREER Award, the 2001 IEEE Communications Society and Information Theory Society Joint Paper Award, and the 2011 IEEE Communication Society Award for Outstanding Paper on New Communication Topics. He was an Associate Editor for the IEEE TRANSACTIONS ON COMMUNICATIONS, the IEEE TRANSACTIONS ON WIRELESS COMMUNICATIONS, the IEEE TRANSACTIONS ON SIGNAL PROCESSING, and the IEEE TRANSACTIONS ON INFORMATION THEORY. He is listed as an ISI highly cited author.
\end{IEEEbiography}

\begin{IEEEbiography}[{\includegraphics[width=1in,height=1.25in,clip,keepaspectratio]{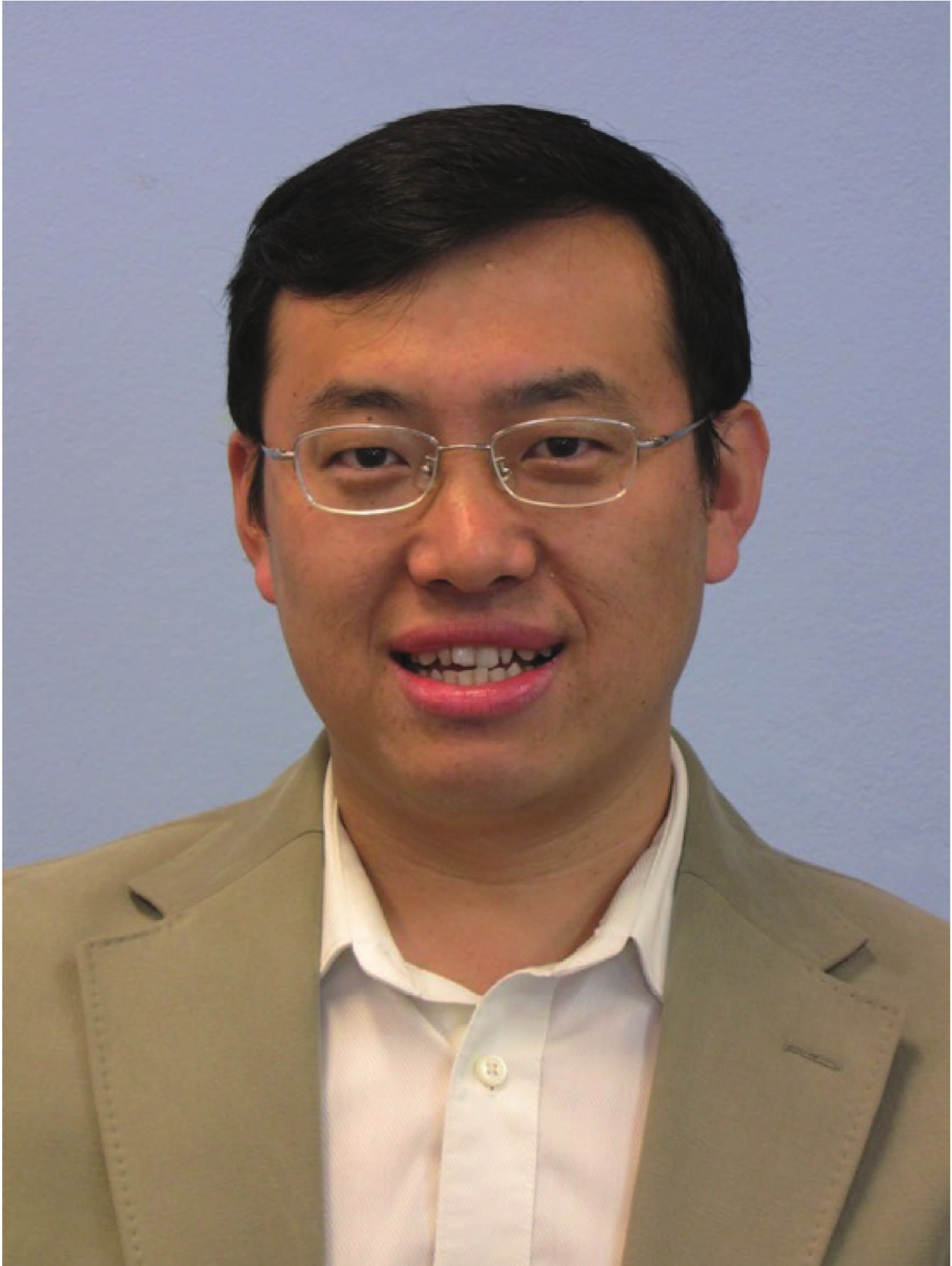}}]{Wei Zhang}
(S'01-M'06-SM'11-F'15) received the Ph.D. degree in Electronic Engineering from the Chinese University of Hong Kong in 2005. Currently, he is Professor at School of Electrical Engineering and Telecommunications, the University of New South Wales, Sydney, Australia. His research interests include space information networks, IoT, and massive MIMO. He has published over 200 papers and holds 5 US patents. He received 5 best paper awards from international conferences, IEEE Communications Society (ComSoc) TCCN Publication Award, and ComSoc Asia Pacific Outstanding Paper Award.

He was ComSoc Distinguished Lecturer in 2016-2017 and the Editor-in-Chief of IEEE Wireless Communications Letters in 2016-2019. Currently, he serves as Area Editor of IEEE Transactions on Wireless Communications and Editor-in-Chief of Journal of Communications and Information Networks. He also serves in Steering Committee of IEEE Networking Letters and IEEE Transactions on Green Communications and Networking.  He is Chair for IEEE Wireless Communications Technical Committee and Vice Director of the IEEE ComSoc Asia Pacific Board. He was a member of Fellow Evaluation Committee of IEEE Vehicular Technology Society in 2016-2019. He served on the organizing committee of the IEEE ICASSP 2016 and the IEEE GLOBECOM 2017. He was TPC Chair of APCC 2017 and ICCC 2019. He is a member of ComSoc IT Committee, TC Recertification Committee, and Finance Standing Committee. He is Member-at-Large of ComSoc in 2018-2020.
\end{IEEEbiography}
\end{document}